\documentclass[11pt]{article}
\usepackage{cite}
\pdfoutput=1
\usepackage{amsmath,amsfonts,amssymb,hyperref}
\usepackage[small,bf,hang]{caption}
\usepackage{slashed}
\input epsf.sty
\usepackage{epsfig}

\def\hybrid{
        \topmargin -20pt
        \oddsidemargin 0pt
        \headheight 0pt \headsep 0pt
        \textwidth 6.55in 
        \textheight 9.5in 
        \marginparwidth .875in
        \parskip 5pt plus 1pt \jot = 1.5ex}

\hybrid

 \csname
@addtoreset\endcsname{equation}{section}

\newcommand{\sectiono}[1]{\section{#1}\setcounter{equation}{0}}

\def\moth{\mathsurround=0pt}
\newdimen\zo \zo=0pt

\def\tick{\leaders\hrule height 0.5ex depth 0pt \hskip 0.5pt}
\def\upboxfill{$\moth \setbox\zo\hbox{\tick}%
  \hskip 3pt\hbox to 0pt{$\tick$\hss}\hrulefill \hbox to 7.5pt{$\tick$\hss}$}

\def\dtick{\leaders\hrule height .34pt depth 0.5ex \hskip 0.5pt}
\def\downboxfill{$\moth \setbox\zo\hbox{\dtick}%
  \hskip 2pt\hbox to 0pt{$\dtick$\hss}\hrulefill \hbox to 2pt{$\dtick$\hss}$}

\DeclareMathOperator{\area}{area}
\DeclareMathOperator{\flux}{flux}

\def\bec{\begin{center}}
\def\ec{\end{center}}

\def\be{\begin{equation}}
\def\ee{\end{equation}}
\def\bea{\begin{eqnarray}}
\def\eea{\end{eqnarray}}
\def\ba{\begin{array}}
\def\ea{\end{array}}

\begin{document}

\begin{titlepage}

\rightline{\tt BRX-TH-6328} 
\rightline{\tt MIT-CTP-4896} 
\begin{center}
\vskip 2.5cm

{\Large \bf {Convex programs for minimal-area problems }}   

\vskip 0.5cm

  \vskip 1.0cm
 {\large {Matthew Headrick}}
 
{\em  \hskip -.1truecm Martin Fisher School of Physics \\
Brandeis University\\
Waltham MA 02143, USA\\
\tt headrick@brandeis.edu}

 \vskip 0.5cm

 {\large {and}}

  \vskip 0.5cm
 {\large {Barton Zwiebach}}

{\em  \hskip -.1truecm 
Center for Theoretical Physics \\
Massachusetts Institute of Technology\\
Cambridge MA 02139, USA\\
\tt zwiebach@mit.edu \vskip 5pt }

\vskip 1.5cm  
{\bf Abstract}

\end{center}

\noindent
\begin{narrower}

\baselineskip15pt

The minimal-area problem that defines string diagrams 
in closed string field theory
asks for the metric of least area on a Riemann surface with the condition
that all non-contractible closed curves have length
at least  $2\pi$. 
This is an extremal length  problem in conformal geometry 
 as well as a problem in systolic geometry.  
 We consider the analogous minimal-area problem 
for homology classes of curves and, with the aid of calibrations
and the max flow-min cut theorem, 
formulate it as a local convex program.  
We derive an equivalent dual program involving 
maximization of a concave functional.
These two programs give new insights into the form of the minimal-area metric and are amenable to numerical 
solution.  We explain how the homology problem 
can be modified to provide the solution to the original homotopy problem.

\end{narrower}

\end{titlepage}

\baselineskip11pt

\tableofcontents


\baselineskip15pt

\section{Introduction and summary}

Progress in our mastery of string theory often 
requires finding answers to certain mathematical questions.  
This was the situation
with string field theory, whose  formulation needs concrete
ways of constructing the moduli spaces of Riemann surfaces with marked points,  and finding a way to assign local complex coordinates at the marked points for each surface in each moduli space\cite{Witten:1985cc,Giddings:1986wp,Zwiebach:1992ie}. 

While the formulation of open string field theory required 
some results on the moduli spaces of Riemann
surfaces with boundaries, most of the needed facts were
known in the literature through work of Strebel~\cite{strebel}.
The case of closed string field theory proved much more
challenging.  After some work elucidating the construction
of the moduli spaces of {\em spheres} with marked points~\cite{Saadi:1989tb,Kugo:1989aa}---a construction relevant to classical 
closed string field theory---the general construction of all moduli spaces required for
the full quantum action was claimed
to arise via the following minimal-area problem 
\cite{Zwiebach:1990ni,Zwiebach:1990nh}: 

\begin{quote}
{\em \underline{Minimal-Area Problem:} \ Given a genus $g$ Riemann surface with $n \geq 0$ marked points
($n \geq 2$ for $g = 0$), find the metric of minimal (reduced) area under the condition that the length of any noncontractible closed curve be greater than or equal to $2\pi$.}
\end{quote}

The marked points are the locations where 
string vertex operators are inserted. The metric's Weyl class is fixed by the complex structure: the length
element is
$ds = \rho |dz|$, where $z = x+ iy$ is a complex coordinate,  
and the area element is
$dA = \rho^2 dx \wedge dy$.  The homotopy of curves is relative
to the punctures, and therefore curves surrounding each puncture are considered noncontractible. 
All noncontractible closed  curves are required to 
have length at least $2\pi$.  That number, while convenient for
closed string theory, is just conventional; it could be set equal to
1 or to any other number.  In the extremal metric, that 
number  happens to be the  {\em systole} of the surface, that is, the length of the shortest noncontractible closed geodesic.  
For $g\geq 1$ and $n=0$, the area of the extremal metric
is finite.  When $n \geq 1$, however, one finds that 
in some open neighborhood of each of the  punctures
the extremal metric is a semi-infinite 
cylinder of circumference $2\pi$.  As a result, the area is infinite
and one must consider a regularized ``reduced area'' for 
minimization~\cite{Zwiebach:1990nh}. 
Briefly stated, the above minimal-area problem gives a string field theory because the plumbing
of extremal metrics gives extremal metrics.

For a minimal-area problem where the length condition is applied
to a set of homotopy classes of curves that have non-intersecting 
representatives, the extremal metric is known and arises as
the norm of a Jenkins-Strebel
quadratic differential~\cite{strebel}. In our problem, we
constrain {\em all} homotopy classes of curves. It turns out that for genus-zero surfaces with marked points,
the extremal metric nonetheless arises from quadratic
differentials. This is not the case, however, 
for large classes of surfaces in each of the moduli spaces of $g\geq 1$ Riemann surfaces.
Generically,  the extremal metric does not
arise from a quadratic differential, there is no general method
to find it and, in fact, the solution is unknown.  The
length constraint, which applies to an infinite set of curves, makes 
even a numerical analysis of the problem very challenging.

The minimal-area problem is a particularly hard case of the problem of finding the {\em extremal length} conformal invariant~\cite{ahlfors}.  Moreover, the problem can also be considered as one in systolic geometry \cite{m_katz,mberger,guth}. Gromov \cite{gromov} studied in detail the 
problem of finding the minimal-volume Riemannian
metric on an $n$-dimensional manifold $M$ with a given systole. The specialization of this question to two-dimensional manifolds with 
{\em fixed}    complex structure \cite{gromov,bavard} is the same problem posed above.  For two-dimensional manifolds but without fixing the Weyl class of the metric, a set of results was derived by Calabi~\cite{calabi}.  The case of metrics that are, additionally, of non-positive curvature is also of interest~\cite{katz-sabourau}.  

The extremal metric for our minimal-area problem is expected
to have systole-length geodesics that cover the surface.  
For brevity we will use the term {\em systolic geodesic} for
a systole-length geodesic. Near the marked points these geodesics are circles foliating
a flat semi-infinite cylinder, with the marked point at infinite distance. 
Through each point on that cylinder there is just one systolic
geodesic.   On the rest of the surface there can exist
regions with one and only one 
systolic geodesic going through each point,
but also regions where more than one such geodesic
goes through each point.  In general the surface is covered by 
multiple {\em bands} of intersecting and non-intersecting 
systolic geodesics. 
The metric is not known as soon as systolic bands intersect.
This will happen for surfaces over finite subsets of each
moduli space ${\cal M}_{g,n}$ of Riemann surfaces of
genus $g\geq 1$ with $n$ marked points.

This paper reports on progress on this minimal-area problem. The advances rely, to a large degree, on the geometrical application of tools from the subject of {\em convex optimization}~\cite{boyd}. A basic example of such an application, which we will use extensively in this paper, is the max flow-min cut theorem on manifolds, which is proved with the help of strong duality of convex programs \cite{headrick-hubeny-refs}. A review of this material 
can be found in the recent paper~\cite{Headrick:2017ucz} of Hubeny and Headrick. (See also \cite{Freedman:2016zud} for a different physical application of the max flow-min cut theorem on manifolds, in the context of holographic entanglement entropy.)

To make progress on the minimal-area problem, we begin by modifying it.  As stated, the length constraints apply to {\em homotopy} classes of curves.  Instead, we consider length constraints applied to {\em homology} classes of curves.  Any closed curve that is homologically nontrivial is also homotopically nontrivial, but there are homotopically nontrivial curves that are homologically trivial.  The length of those curves are
not constrained in the homology problem.  These curves, however,
can be dealt with by 
passing to a suitable covering space where they 
become homologically nontrivial.  We discuss in detail how a suitably extended homology problem provides a solution
of the original minimal-area problem.

In the general homology minimal-area problem, we fix a set $\{C_\alpha\}$ of homology cycles and numbers $\{\ell_\alpha\}$ and require the length of every curve in the class $C_\alpha$ to have length at least $\ell_\alpha$. 
In this paper we present two main advances on this problem:
\begin{enumerate} 
\item The first advance is to replace the length constraints 
by the existence of calibrations: closed one-forms with norm everywhere less than or equal to one.  A calibration with period $\ell_\alpha$ on a 
homology cycle 
$C_\alpha$ constrains the lengths of all curves in the corresponding homology class to be greater than or equal to 
$\ell_\alpha$. Requiring the existence of a calibration places
no additional constraints because the
max flow-min cut theorem guarantees the converse: if all curves in 
the homology class 
have length at least $\ell_\alpha$ then there is a calibration with period $\ell_\alpha$.
The length constraint, a nonlocal condition on the metric on the surface,   
is implemented as
 a local condition on the one-form calibration,  a significant simplification.  
 The calibrations can be written in terms of abelian differentials consistent with the requisite periods plus exact one-forms. Minimizing the area subject to the calibration constraint is a convex optimization problem, or convex program, henceforth called the {\em primal program}.  This immediately  guarantees that any local minimum is a global minimum (i.e.\ there are 
 no false minima). Furthermore, it allows the application of powerful numerical methods for solving convex programs.

\item The second advance follows from
applying Lagrangian duality to the primal.  This means adding Lagrange multipliers to impose the constraints and then solving for 
the original variables of the primal.  The result is a 
{\em dual program} in which
we {\em maximize} a functional of the Lagrange multipliers. 
The dual program variables are a collection of functions $\varphi^\alpha$ 
and  numbers $\nu^\alpha$ 
associated with the 
homology classes $C_\alpha$. 
 The value
of $\nu^\alpha$ is the discontinuity of $\varphi^\alpha$ across an arbitrarily chosen representative of the class $C_\alpha$.   At the optimum, the curves of constant $\varphi^\alpha$ are the 
systolic geodesics in the class $C_\alpha$ and $\nu^\alpha$ is, roughly, 
the height of the annular domain 
covered by those geodesics.   Because of convexity of the primal and an easily satisfied
technical condition, the property of 
{\em strong duality} holds: the minimum in the
primal is guaranteed to coincide with the maximum of the dual. 
\end{enumerate}

\begin{figure}[!ht]
\leavevmode
\begin{center}
\epsfysize=5cm
\epsfbox{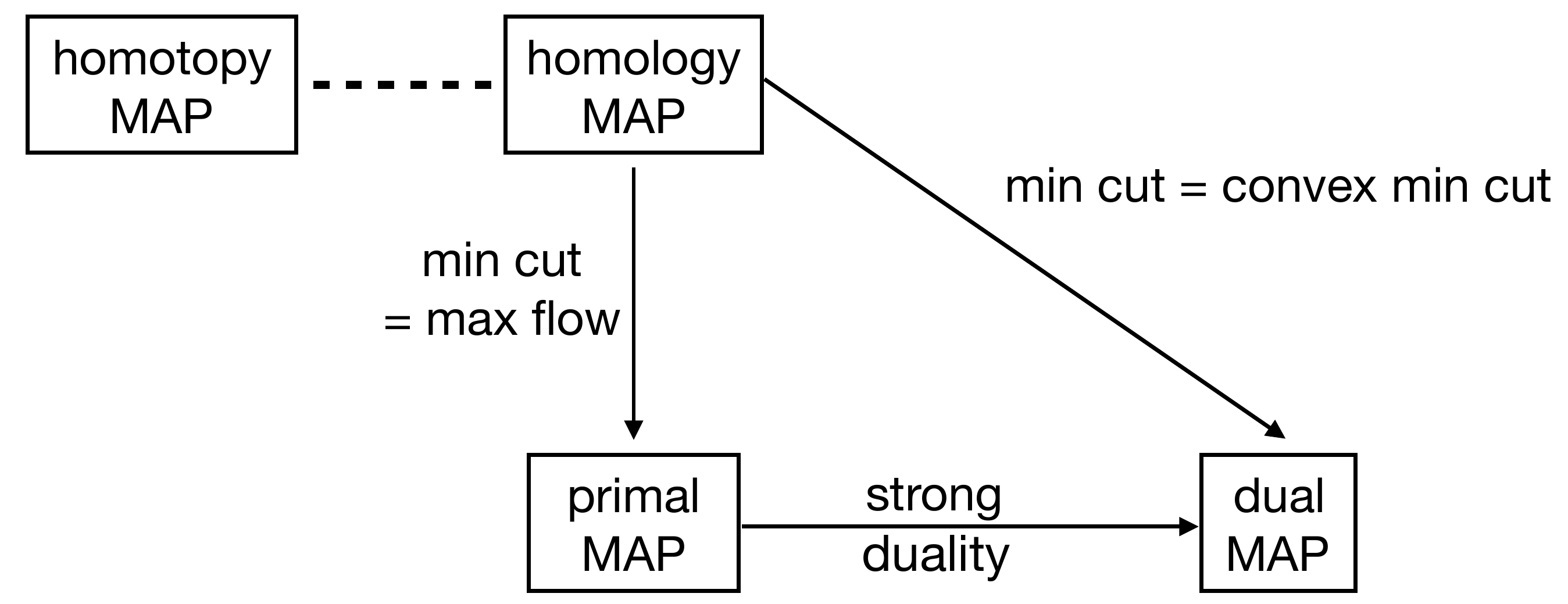}
\end{center}
\caption{\small  
Relations among the minimal-area programs (MAPs) discussed in this paper. The homotopy MAP is given in (\ref{homotopy-MAP}), the  homology MAP  in (\ref{firstprogram}), the primal MAP in 
(\ref{secondprogram}), and the dual program
 in (\ref{thirddual}). 
The dotted line indicates that the programs are 
related but not equivalent.  The arrows relate equivalent 
programs, and the text on the arrows indicate the main
tool used to prove the equivalence. }
\label{vmbtvmvt1}
\end{figure}

The relations between the various programs are illustrated in Figure~\ref{vmbtvmvt1}.  The homotopy minimal area problem (MAP) suggests the analogous homology problem, the two corresponding 
boxes joined by a dotted line. Our work with covering spaces in
section~\ref{sec:homotopy} shows how to use the homology version
of the problem to solve the closed string 
field theory (homotopy) problem.  
The passage from homology MAP to primal MAP was explained in item 1
above, and the dualization in item 2.   An alternative derivation, shown by the diagonal arrow in the diagram, provides additional
insight into the dual and the max flow-min cut theorem.  

In an accompanying paper \cite{headrick-zwiebach2}, 
we illustrate the programs described here by numerically solving for the 
{\em unknown} extremal metric of a relatively simple
Riemann surface,  the square torus with one
boundary.  The extremal metric on this surface is also the
extremal metric on the so-called Swiss cross.  
The extremal metric displays crossing bands of geodesics
and non-zero Gaussian curvature.  We also find the previously
unknown metric for the once-punctured square torus.  
This metric is needed for the definition of closed string field 
theory to one loop.  We will demonstrate that the programs are easily solved numerically.   
Having a minimization program and a maximization
program with the same solution proves very helpful.    
The methods of the present paper can also be applied to the study
of conformal metrics of non-positive curvature on Riemann surfaces.  These metrics are also of interest for string field theory
and exhibit surprising features revealed by the numerical study:  not all of the Riemann surface is necessarily covered by saturating geodesics~\cite{headrick-zwiebach3}\footnote{We have learned of work by Katz and Sabourau~\cite{k-s},
dealing with systolic geometry of surfaces with Riemannian metrics of non-positive curvature. Regions without systolic geodesics play an important role
in this work.}.

In addition to being readily amenable to numerical solution, the programs above offer novel
theoretical insights into the minimal-area metrics. Indeed, strong duality implies
\emph{complementary slackness}, which relates the two programs, giving information about the solution.
In particular:   
\begin{enumerate}
\item[(i)]  
We  find a remarkably simple
relation, (\ref{area-heights}),
between the extremal area $A$ of the surface, 
the heights $\nu^\alpha$ of  
geodesic bands, and the minimum lengths $\ell_\alpha$:
\be
\label{area-height-vm}
A = \sum_\alpha  \nu^\alpha \ell_\alpha \,. 
\ee
Thus,
even when bands cross and there may be curvature, the total area of the 
surface is the sum of that of flat cylinders of height 
$\nu^\alpha$ and circumference
$\ell_\alpha$.   
\item[(ii)]  The dual program allows the definition of 
the local {\em density} $\rho_\alpha$ of geodesics in the $\alpha$-band at any point on the surface.
We find that the pointwise sum of the densities
of all the bands is constant over the entire surface, (\ref{density}).
Equivalently, 
 a sum rule constrains the metric 
at every point of
the surface ((\ref{solution2}) and (\ref{solution3})).
 The sum rule implies that
a region foliated by a single band of systole-length geodesics must
be flat, a fact previously established by a more involved argument 
\cite{Ranganathan:1991qd,Wolf:1992bk}. 
\end{enumerate} 

We have tried to understand in what way the minimal-area metrics generalize those that arise as the norm of Jenkins-Strebel quadratic differentials.
We have also tried to develop intuition about the somewhat
mysterious dual program.  Here are a few remarks:
\begin{enumerate}
\item[(a)] The systolic geodesics in the extremal metrics
are the generalization of the horizontal trajectories of 
these quadratic differentials.   The heights $\nu^\alpha$ of 
the dual program generalize the heights of the ring domains
in the quadratic differentials.  It is striking
that in the systolic problem the area formula (\ref{area-height-vm})
takes exactly the same form as for quadratic differentials, where
the metric is indeed that defining flat cylinders of prescribed circumference.

\item[(b)]  When constraining homotopy classes of curves with representatives that do not intersect, quadratic differentials arise as the problem of maximizing the
weighted sum $\sum \ell_\alpha^2 M_\alpha$ where $M_\alpha$ are the moduli of non-overlapping ring domains $R_\alpha$ in the homotopy classes constrained with $\ell_\alpha$.  We argue that the dual functional is a generalization of such a weighted sum.  
For this we see that the curves of constant $\varphi^\alpha$ (the precursors of systolic geodesics) segregate and form ring domains (subsection \ref{sec:prop-sol-dual}).  Moreover,  the dual program applied to a ring domain gives precisely $\ell_\alpha^2 M_\alpha$ as the optimum.

\end{enumerate} 

This paper is organized as follows.  
In the section \ref{sec:csft} we go
over the basics of the minimal-area problem and its relevance
to closed string field theory.  In 
section \ref{sec:convex-optimization} we begin by giving 
a brief review of the relevant aspects of convex optimization. We then examine and discuss in detail the max flow-min cut theorem, giving
a proof based on duality that brings to the fore a ``convex min cut'' program
that arises by convex relaxation.   In section \ref{sec:homology}, we introduce the 
homology minimal-area problem and reformulate it in terms of calibrations.  
The dual minimal-area program, given in various equivalent forms,
is the subject of  section \ref{sec:dual}.  
We derive it by several routes, including directly from the convex min-cut program and by  direct dualization of the primal.   In section \ref{sec:examples}  we  work through the solvable examples
of the cylinder and torus to give some intuition for the primal and dual programs.
 In section \ref{sec:optimal} we use both programs to learn about
the extremal metrics, and then turn to the dual, focusing on 
its properties and its relation to quadratic differentials.
 Finally, in section \ref{sec:homotopy} we return to the homotopy
problem that motivated this paper and show how the use 
of certain   covering spaces makes it possible to solve the homotopy problem
in terms of the homology problem.

\section{Review of closed string field theory and minimal-area metrics}
\label{sec:csft}

String perturbation theory constructs scattering amplitudes
by integration of conformal field theory (CFT) correlators of vertex operators
over moduli spaces of Riemann surfaces.   Any construction of a closed 
string field theory must find a way to generate the moduli spaces 
of Riemann surfaces.  More precisely, it must generate the moduli spaces ${\cal M}_{g,n}$ of Riemann surfaces of genus $g$ with
$n$ marked points.  The marked points, or punctures, are the places where
vertex operators are inserted. 

In this section we review the basic ideas that demonstrate
that the minimal area problem 
stated in the introduction allows for the formulation
of a closed string field theory.  We also discuss how the minimal
area problem fits in the context of conformal geometry
and systolic geometry. 

\subsection{Closed string field theory}

 The classical closed string field theory
must generate the genus zero moduli spaces ${\cal M}_{0,n}$
with $n\geq 3$.  The full quantum theory must also generate
the $g\geq 1$ moduli spaces with $n\geq 1$, although the
case $n=0$ is also of interest.  
Similar remarks apply for open string field theory and open-closed
string field theory in which the Riemann surfaces can also have
boundary components and marked points on those boundaries.
The string field theory generates the moduli spaces of 
Riemann surfaces via its Feynman diagrams, or string diagrams.
The Feynman rules effectively become a prescription for building Riemann surfaces algorithmically 
and must construct each surface exactly once, thus generating the relevant moduli spaces.  The Feynman rules
use vertices and propagators. 

As the string field theory builds surfaces  it must also provide 
a  local coordinate around each marked point. Such a coordinate is needed to insert vertex operators that are not dimension-zero primaries, as required for off-shell amplitudes.   It turns out that the local coordinate need only be defined up to an overall phase. 

In summary, 
the string field
theory must provide:
\begin{enumerate}

\item  A construction of the moduli spaces ${\cal M}_{g,n}$ of
Riemann surfaces.

\item  For each surface, a choice of local coordinates (up to phases) at the marked points.  The choice must be
continuous over the moduli space.

\item  The construction of the moduli spaces must be generated
through Feynman rules, using vertices and propagators.

\end{enumerate}

In order to satisfy such requirements canonically, 
an organizing principle is needed.   In light-cone string field theory, for example,  the string diagrams
are the solution to the following problem~\cite{DHoker:1987hzc}: 
 Given a Riemann  surface find the (unique) abelian differential with purely imaginary periods.  This abelian differential provides a metric
on the Riemann surface and allows one to visualize the surface
concretely.  The result is the familiar light-cone string
diagrams. 

For {\em covariant} closed string field theory the minimal-area problem stated
in the introduction provides the organizing principle~\cite{Zwiebach:1990nh}. 
 One could
say closed string field theory raises that minimal-area problem.
Since the minimal-area metric is unique, if 
known for all surfaces,  
it provides a concrete construction of the moduli spaces of Riemann surfaces.   This is requirement (1) above.

\begin{figure}[!ht]
\leavevmode
\begin{center}
\epsfysize=4.0cm
\epsfbox{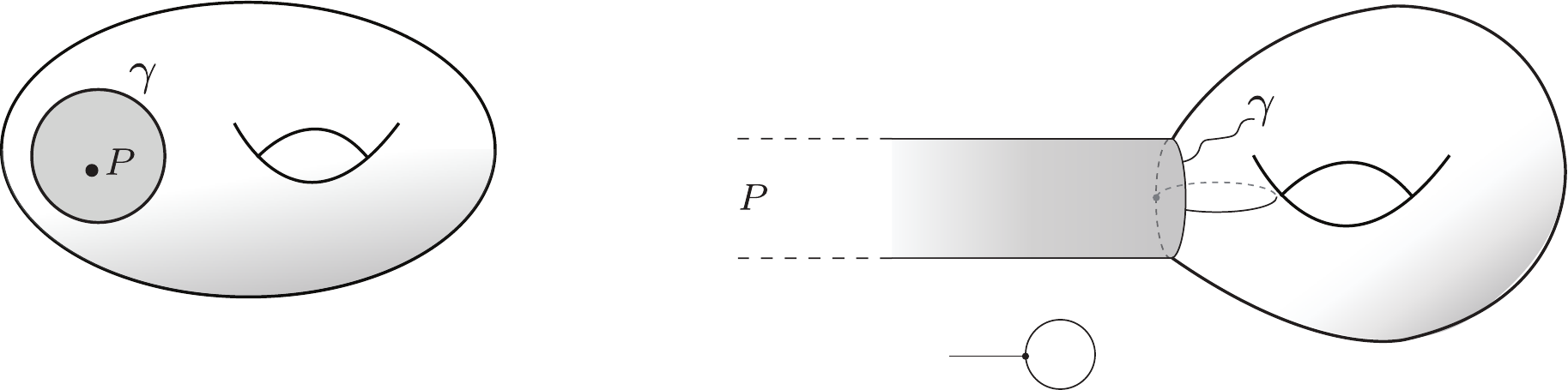}
\end{center}
\caption{\small  A torus with a marked point $P$ (left) and the same surface equipped with its minimal-area metric (right). The curve $\gamma$ is the last systolic geodesic  on the semi-infinite cylinder.  That curve is identified as the
$|z|=1$ curve of a local coordinate that vanishes at $P$. In this way the minimal area metric fixes a local coordinate at the marked point. Once the surface is equipped with a minimal-area metric, the Feynman graph that would produce it is manifest:  in this case the surface arises as a one-loop tadpole.}
\label{fhsv1}
\end{figure}

In this minimal-area problem, the homotopy of closed curves is defined relative to the marked points, or punctures.  
Curves surrounding marked points
are homotopically nontrivial and must satisfy the length condition.
This, in fact, guarantees that (if the minimal-area metric exists) there is a neighborhood of the marked point that is isometric to a flat semi-infinite cylinder of circumference equal to the systole.\footnote{The infinite area requires the definition of a 
finite regularized {\em reduced area} obtained with a subtraction
that uses an arbitrary choice of local coordinates at the marked points.}
This can be used to define a local coordinate $z$ around each marked point:  one simply defines the last systolic geodesic on the cylinder
to be the curve $|z|=1$ (see Figure~\ref{fhsv1}). With $z=0$ the location of the marked point, the Riemann mapping theorem guarantees that there is
a map from the coordinate disk $|z|\leq 1$ to the semi-infinite
cylinder, defined uniquely up to a phase.  This is requirement (2) above.

\begin{figure}[!ht]
\leavevmode
\begin{center}
\epsfysize=8.0cm
\epsfbox{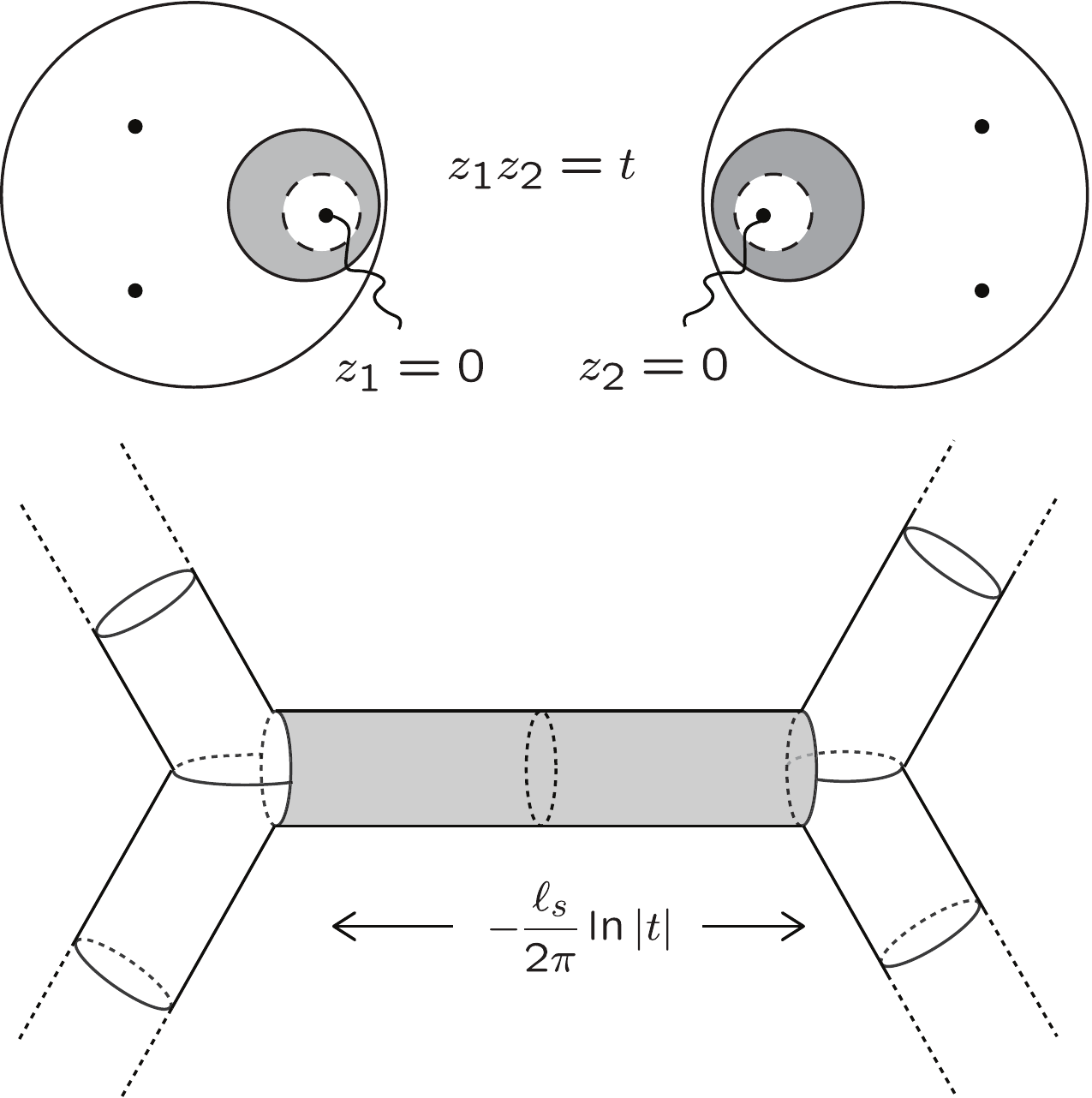}
\end{center}
\caption{\small  Two three-punctured spheres are glued by the identification
$z_1 z_2= t$ where $|t| \leq 1$, and $z_1$ and $z_2$ are, respectively, coordinates around
a marked point on the left and on the right sphere.  If the spheres are equipped
with minimal area metrics each one looks like three semi-infinite cylinders
coming together.  In this case the cylinders with $z_1$ and $z_2$ coordinates
are partially amputated and  glued so that  a cylinder of length
$- {\ell_s\over 2\pi} \ln |t|$ remains (shown shaded). The resulting four-punctured sphere is equipped with a minimal area metric.}
\label{ffsvt1}
\end{figure}

\begin{figure}[!ht]
\leavevmode
\begin{center}
\epsfysize=8.0cm
\epsfbox{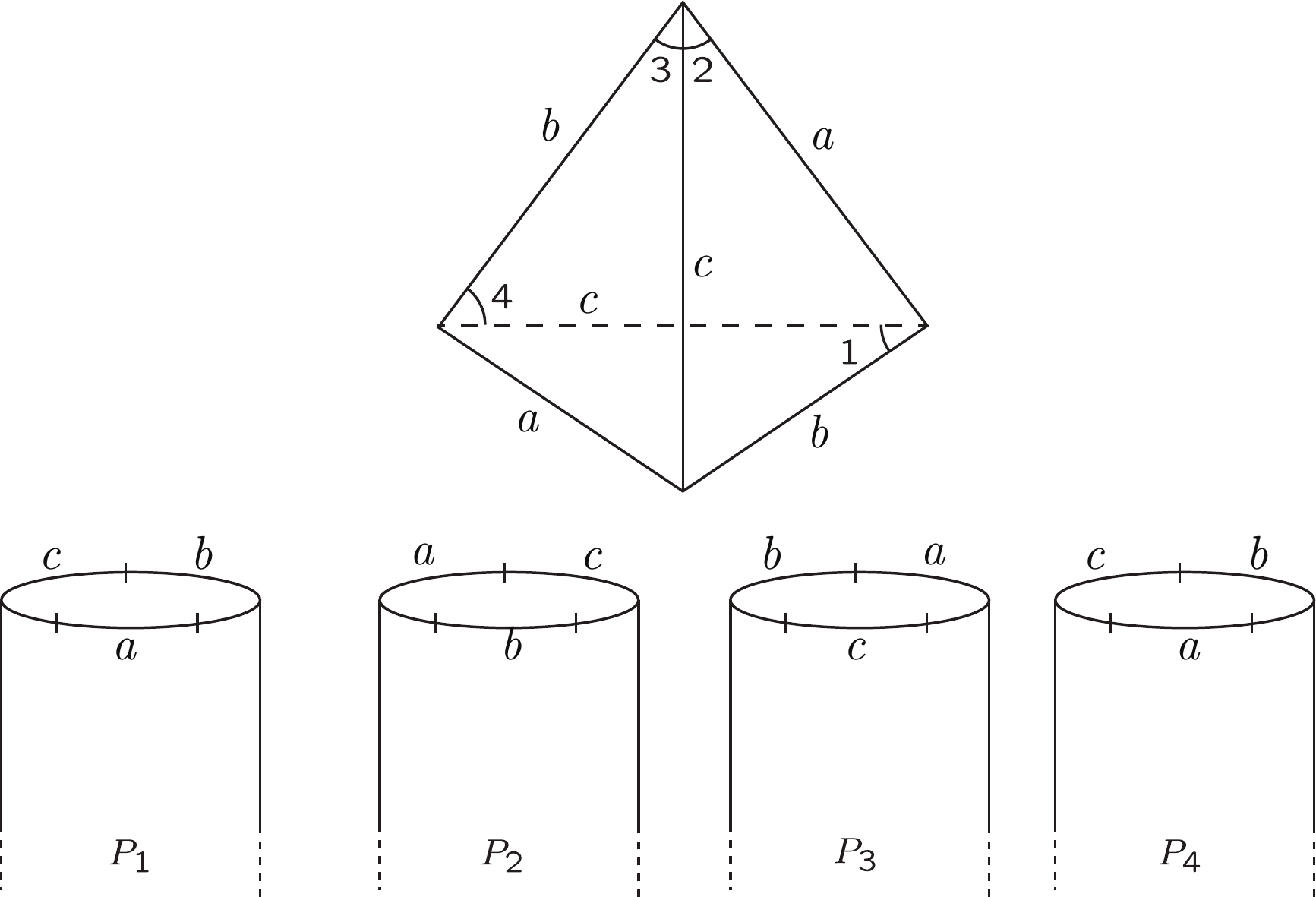}
\end{center}
\caption{\small  For some four-punctured spheres the minimal 
area metric arises by gluing the edges of 
four semi-infinite cylinders on the faces of a tetrahedron graph. 
Here $a,b,c$ are length parameters on the graph and $a+b+c= \ell_s$. 
In order not to have closed curves that violate
the length conditions one must also have $a+b$, $b+c$, and $c+a$ 
all greater than or equal to $\ell_s/2$. }
\label{ffsvt231}
\end{figure}

The most nontrivial constraint is in fact (3). Minimal 
area metrics, however,  are naturally built with
the use of vertices and propagators thus satisfying this 
requirement.  Indeed, once a surface is equipped with its minimal
area metric the associated Feynman graph that would produce it is 
obvious (see Figure~\ref{fhsv1}, right).  The propagator can be viewed as a sewing prescription.  Two local coordinates $z_1, z_2$ defined around
two marked points (on two surfaces or on the same surface)
allow sewing via $z_1 z_2 = t$ with $t$ a complex number
 $|t| \leq 1$.  As indicated in Figure~\ref{ffsvt1} the effect of sewing  minimal area metrics with some
fixed $t$ is that of partially amputating the semi-infinite cylinders associated with the
$z_1$ and $z_2$ coordinates and gluing their boundaries, leaving a finite intermediate cylinder of length $-{\ell_s\over 2\pi} \ln |t|$ and circumference equal to the systole $\ell_s$.   The new surface can be shown
to inherit a minimal-area metric through this sewing operation,
guaranteeing consistency.  The basic intuition proves correct:
the new metric satisfies all length conditions\footnote{\label{foot3}In order to construct higher genus diagrams without violating the length conditions one must redefine all the
vertices by introducing stubs.  This effectively means that
the local coordinates are defined using the geodesic on the
cylinder a distance
$\ell_s/2$ away from the last geodesic.  Thus any gluing operation will automatically introduce a cylinder $\ell_s$ long, preventing the appearance of new short geodesics. 
A Riemann surface whose minimal-area metric does \emph{not} contain an internal cylinder of length at least $\ell_s$ is considered a vertex, while one that \emph{does} contain such a cylinder is built as a Feynman diagram out of vertices and propagators. For more details see~\cite{Zwiebach:1992ie}.}  
 and its area cannot
be lowered because it would imply that the constituent surfaces
would admit  metrics with lower area than that of their minimal-area metrics. The surface built in Figure~\ref{ffsvt1} is a four-punctured sphere with a minimal
area metric. 
 It corresponds to a Feynman graph with an internal  propagator; 
 this diagram, as the length and twist 
of the intermediate cylinder 
are varied, covers some region of the moduli space.
Part of the moduli space is covered by surfaces that represent ``vertex" contributions and are shown in Figure~\ref{ffsvt231}.   Each  
surface in the vertex  
is an elementary
four-point interaction and is constructed by gluing four semi-infinite cylinders
on a tetrahedron graph.  The perimeter of each of the faces of the tetrahedron
must equal the systole length: $a+b+c= \ell_s$.  By varying the length
parameters $a,b,$ and $c$, while guaranteeing that no curve becomes too short,
one covers the  
vertex region ${\cal V}_{0,4}$ of the moduli space ${\cal M}_{0,4}$.  Together, the two types of Feynman
diagrams construct all surfaces in the moduli space of four punctured spheres. 
 
\begin{figure}[!ht]
\leavevmode
\begin{center}
\epsfysize=12.0cm
\epsfbox{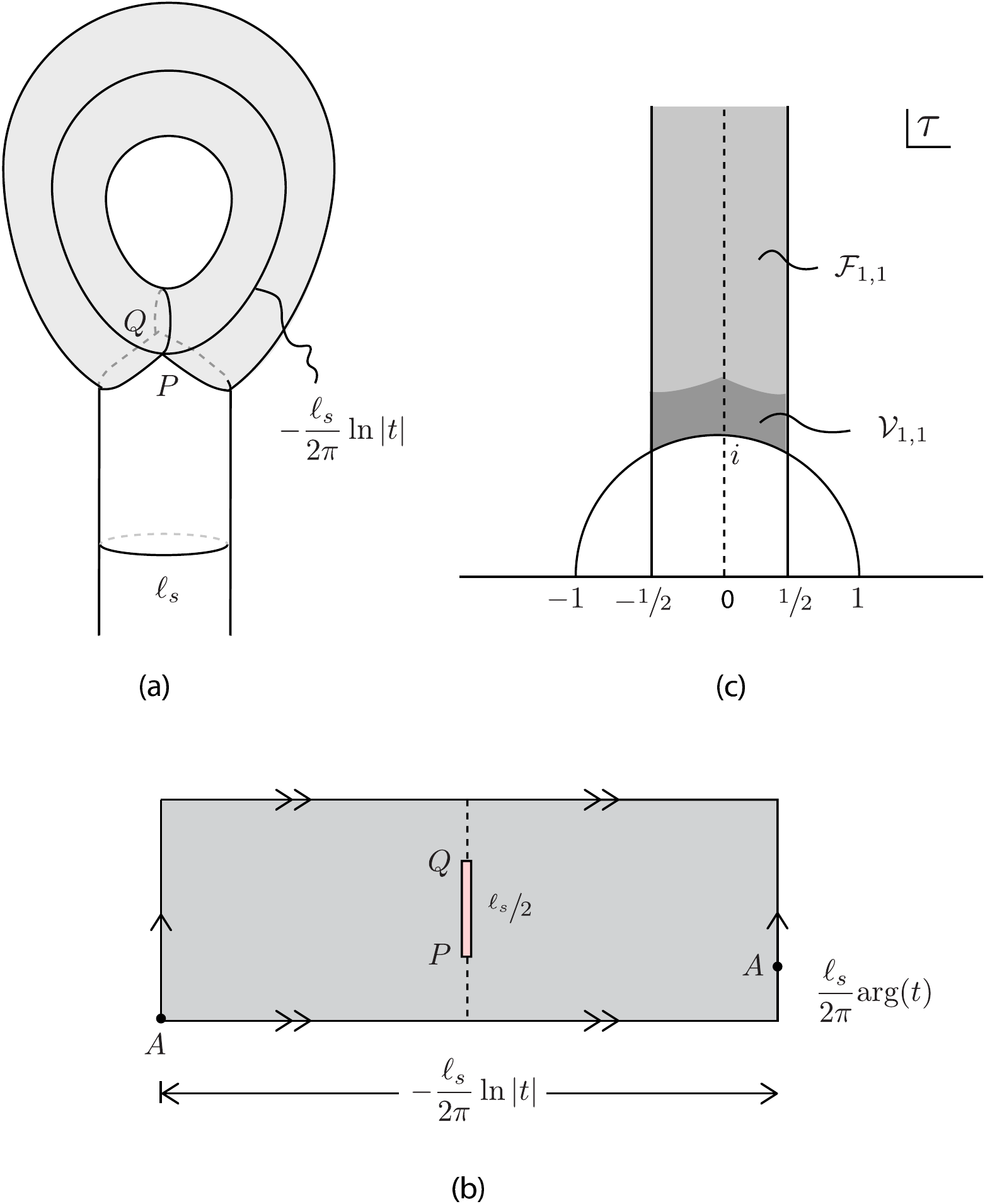}
\end{center}
\caption{\small  (a) The Feynman graph obtained by gluing a
cylinder on the three-string vertex through the identification 
$z_1 z_2 = t$, with $|t| < e^{-2\pi} $.   
This effectively
introduces a cylinder of systolic circumference and length $- {\ell_s\over 2\pi} \ln |t|$ greater than $\ell_s$.  
  (b) The cylinder region shown in detail, including the effect of the twist 
angle arg($t$).  The semi-infinite cylinder representing the puncture is 
attached at the slit $PQ$.  (c) The moduli space ${\cal M}_{1,1}$ of once-punctured
tori (shaded) 
and a qualitative sketch of the Feynman region ${\cal F}_{1,1}$ and the 
(darker)   
vertex region ${\cal V}_{1,1}$.  }
\label{tori-once-mod.sp}
\end{figure}

For further illustration consider the moduli space ${\cal M}_{1,1}$
of once-punctured tori, relevant to one-loop tadpoles mentioned
in Figure~\ref{fhsv1}.   This tadpole amplitude computes
one-point functions of vertex operators and is composed 
of a Feynman
diagram and a vertex ${\cal V}_{1,1}$.  The Feynman diagram uses
the three-string vertex---that is, the three-punctured sphere with the
minimal area metric---and glues two of the punctures 
together via the identification
$z_1 z_2 = t$ of the local coordinates with $|t| < 1$, which as explained above introduces
a cylinder of length $-{\ell_s\over 2\pi} \ln |t|$ and circumference equal to the systole $\ell_s$ (see Figure~\ref{tori-once-mod.sp}).  For small enough $|t|$ 
this operation produces metrics of minimal area with systole $\ell_s$. 
It is clear, however, that as $|t|\to 1$ 
gluing will fail to produce surfaces with an 
allowed minimal area metric: as the length of the
cylinder goes to zero, noncontractible closed curves shorter than $\ell_s$ will
appear.  It is simple to figure out the subset of $|t| <1$ that generates
minimal area metrics, but quite a bit harder~\cite{Zemba:1988rf}  to determine the part of the moduli space ${\cal M}_{1,1}$ that these 
surfaces cover.  They certainly fail to cover it all:  not all minimal
area metrics on ${\cal M}_{1,1}$ arise from a Jenkins-Strebel (JS)
quadratic differential.  The stubs that must be included in the three-string vertex 
(see footnote \ref{foot3}) imply that effectively the Feynman
region ${\cal F}_{1,1}$ is generated by a cutoff region $|t| < e^{-2\pi}$.
All the surfaces in ${\cal F}_{1,1}$ have minimal area metrics arising from
a JS quadratic differential.   
But, unavoidably, not all of the fundamental domain is covered.   
The missing region corresponds
to the string vertex ${\cal V}_{1,1}$, shown with darker shade  
in Figure~\ref{tori-once-mod.sp}(c).  
The minimal area metrics on these conformal structures
are not known, but they can be constructed with the methods of this paper.
In particular, in the companion paper~\cite{headrick-zwiebach2} we 
construct the minimal area metric on the 
simplest surface in ${\cal V}_{1,1}$, the square torus $\tau = i$.

The minimal-area metric is known for all genus-zero surfaces with 
$n\geq 3$ marked points~\cite{Zwiebach:1992ie}.  In this case the metric is locally 
flat, has negative curvature conical singularities at some
special points, and arises from a Jenkins-Strebel quadratic differential.  Let {\em regular} points be the points on the surface that are not conical singularities nor marked points.
The key property of these metrics is that going 
through every regular point  there is a unique systolic geodesic.
These are the so-called horizontal trajectories of the quadratic
differential.  At any conical singularity more than one saturating
geodesic goes through.    This knowledge of the genus-zero
minimal-area metrics suffices to construct the classical closed string field theory and determines all the string vertices of the classical theory.

Equipped with the classical string vertices and the propagator,
the Feynman rules allow us to build loop diagrams.  These diagrams
will build some but {\em not all} of the Riemann surfaces in the moduli spaces ${\cal M}_{g,n}$, with $g\geq 1$.  All the 
built surfaces will share
the properties of the genus-zero metrics:  the metrics are flat
except for conical singularities and there is a single systolic
geodesic through each regular point.  The missing surfaces, in general, will have a different kind of minimal-area metric:  
we expect them to have regions where
through every point there is more than one systolic geodesic. 
These metrics are unknown 
and have not yet been proven to exist. The purpose of this paper is to 
develop methods to find these metrics and learn about their properties. We will not attempt here
a proof of existence, but such a proof would be important
progress as it would
establish the existence of a closed string field theory!  Indeed, 
even if we did not know the metrics, if they exist and satisfy 
some weak conditions, they would fulfill the
three requirements above and thus ensure that a string field theory exists.   

It is prudent to note that there are non-canonical
proposals that possibly yield 
string field theories.  One of them, based on
symmetric, factorizable quadratic differentials was
formulated by Sonoda and Zwiebach~\cite{Sonoda:1989wa}.  The construction is not 
without some (weak) mathematical assumptions and may
require an adjustment of the propagator at each order
of perturbation theory.  Another option, recently investigated
by Moosavian and Pius~\cite{Moosavian:2017qsp},
is to use hyperbolic metrics.  There is an obvious 
challenge here: hyperbolic metrics are not consistent with sewing.  Nevertheless,
it seems possible, by non-canonical deformations and a recursive
procedure, to adjust the
local coordinates to obtain consistency with sewing.  In this paper we will not discuss these attempts. 

Suppose we knew the minimal-area metrics exist, but their explicit
form for each surface is complicated.  It seems quite likely that
the minimal-area metrics display some regularities or patterns regarding systolic geodesics.  Let $U_n$ denote  
the subset of points in a Riemann surface where 
exactly $n$ systolic geodesics go through.  
We expect that each minimal-area surface will have regions $U_1, U_2, \ldots, U_k$ where $k$ depends on 
the genus and the number of marked points.  Systole-length geodesics in the same homotopy class form bands that foliate
the surface.  
The parameters associated with these bands include a suitably defined height, and perhaps angles formed by different intersecting bands.  Such parameters could help construct a new decomposition of the moduli spaces of Riemann surfaces.  It is also plausible that a number of questions in string field theory would only require
partial information about the minimal-area metrics. 

Of course, it would be ideal to know the 
 explicit form of all minimal-area metrics.  
 In principle, however, 
 for any fixed surface the exact form of the metric
 is not needed on much of the surface.  Arbitrary
 off-shell computations and the definition of the
 string field theory action  {\em only} require
 the explicit form of the local 
 coordinates at the marked points, and that requires finding
 the last systolic geodesic homotopic to each of the marked
 points on the surface.  
 The value of the metric outside the coordinate disks does not affect anything.

\subsection{Mathematical context}

Minimal-area problems
have a long history in the theory of Riemann surfaces.   
They led to the definition of the  
 {\em extremal length} 
conformal invariant $\lambda$ associated to a collection of
curves $\Gamma$ on a surface $M$ (see \cite{ahlfors}). 
To compute $\lambda$ one   
minimizes the 
area of $M$ over the conformal metric 
while keeping all curves in $\Gamma$ longer 
than a constant $\ell_s$ identified as the systole.  
The conformal invariant $\lambda(\Gamma)$ is then given by the
ratio of the squared systole and the minimal area.  A useful test by 
Beurling~\cite{ahlfors} 
can be used to tell if a candidate metric is of minimal area, and a partial converse was given in \cite{Wolf:1992bk}.  The specification of $\Gamma$ is usually done by including some homotopy classes of curves on it.

The minimal-area metric can be shown to exist and arises 
from Jenkins-Strebel quadratic differentials whenever the 
chosen homotopy classes of curves in $\Gamma$ have
representatives that {\em do not} intersect~\cite{strebel}.  The extremal
metric is known even when different length conditions 
are placed on the various homotopy classes in $\Gamma$.
The homotopy classes can include those homotopic
to the marked points and the required regularization of the area 
is inspired by Teichm\"uller's treatment of the modulus of
a punctured disk \cite{strebel}.  The minimal-area problem for closed
string field theory is much harder because $\Gamma$ includes
all homotopy classes of curves, and therefore there are infinite
instances of pairs of homotopy classes where representatives
always intersect.  As soon as such pairs exist, the minimal-area metric can easily fail to arise from a quadratic differential.
In the minimal-area metric there are systole-length curves for
some subset of the homotopy classes of curves.  If these ``realized'' homotopy classes have representatives that do not intersect, then
the minimal-area metric will arise from a quadratic differential.
If realized homotopy classes intersect, the metric will almost always fail to arise from a quadratic differential.
For genus zero, the minimal-area metric arises from
a quadratic differential because realized homotopy classes
cannot intersect.  If two systole-length curves 
in different  homotopy classes intersect at one point, they 
would have to intersect in a second point because each geodesic cuts a genus zero surface into two pieces. 
But one can prove that two non-homotopic systoles 
cannot generally intersect at more than one 
point~\cite{Zwiebach:1992ie}.  

Another perspective on the minimal-area problem arises from
systolic geometry~\cite{m_katz}.  
Gromov introduced the notion of {\em essential}
$n$-dimensional manifolds $M$ as those for which the 
{\em systolic} volume 
\be
{\hbox{Vol} (M, g)\over (\hbox{sys} (M,g))^n}
\ee
has a lower bound on the space of Riemannian metrics $g$ on $M$
and the bound only depends on the topology of $M$~\cite{gromov}.
When the manifold is two dimensional, the above ratio coincides
with the quantity that one considers for the calculation of the
extremal length.   In the extremal length problem, the conformal
structure of the surface is fixed and the metrics are restricted to
be conformal. 
In minimizing the above ratio, the conformal structure is not kept fixed as one works with all Riemannian metrics
and there is just one minimum ratio for each genus.  Of course, 
in systolic geometry one can study ``conformal systolic inequalities'' \cite{bavard}
and this is exactly the same subject considered in the theory of Riemann surfaces.    The highest lower bound for the systolic ratio is
known for genus one, and it is attained for the torus $\tau= e^{i\pi/3}$ equipped with a flat metric~\cite{berger}.  The highest lower bound for the systolic ratio is not known for genus two, or in fact for any $g \geq 2$.    For surfaces, and allowing the complex structure to vary, Calabi~\cite{calabi} showed that there is no 
region $U_1$ with just one band of systole-length geodesics and, 
 in $U_2$ the two bands of systolic geodesics are
perpendicular everywhere and the metric is flat.  
He described the foliations using calibrations, as elaborated  
by Bryant~\cite{bryant}.  Calabi's results, however,  
do not apply for the conformal case.  There are many
examples with regions $U_1$, and the proof of flatness for $U_2$  
does not extend to the conformal case.

\section{Review of convex programming and max flow-min cut}
\label{sec:convex-optimization}

In this section we review basic facts about convex programs
and their duals that will be used in the rest of the paper.
We will state the relevant results mostly avoiding proofs.
For a somewhat more detailed review of similar facts,
the reader should consult~\cite{Headrick:2017ucz}.  For 
more details as well as proofs, the book by 
 S. Boyd and L. Vandenberghe 
\cite{boyd} is highly recommended.  
We also discuss
the max flow-min cut theorem as applied to calibrations
and curves in homology, both as an illustration of these ideas and because it is the starting point for our investigations.

\subsection{Convex programs}\label{sec:convex}

In discussing convex programs we need the terminology of 
convex and affine sets, and convex and affine functions.
Consider a set $C\subseteq \mathbb{R}^n$. 
We will denote a point in $\mathbb{R}^n$ as $y$. 
The set $C$ is
affine if the full line going through any two distinct points in $C$ lies in $C$.  A set $C$ is convex if the line {\em segment} joining any
two distinct points in $C$ lies in $C$.   A function $f: C \to \mathbb{R}$ is convex if its domain $C$ is a convex set and for all
$y_0,y_1$ in that domain
\be
\label{convex-functions}
f( t y_0 + (1-t) y_1 )  \leq  t f(y_0) + (1-t) f(y_1) \,   \ \ \forall  t\in [0,1] \,. 
\ee
A function $f$ is said to be concave if
$-f$ is convex.   
A function $h: \mathbb{R}^n \to \mathbb{R}$ is affine if it is
the sum of a linear function and a constant function. Affine functions
satisfy equation (\ref{convex-functions}) with the inequality
replaced by equality.  Thus affine functions are convex and 
also concave.   A function is strictly convex if the strict inequality
holds for $t\in (0,1)$.  It is natural to search for the minimum
of a convex function and for the maximum of a concave function. 

Some operations preserve convexity. The non-negative weighted
sum of a set of convex functions is itself a convex function.    
If $f_1$ and $f_2$ are convex functions, their {\em pointwise maximum}  $f$, defined for each $y$ by
$f(y) =  \max \{ f_1(y), f_2(y)\}$,
is a convex function whose domain is the intersection of the domains of
$f_1$ and $f_2$. 
Similarly, the pointwise minimum of two concave functions is concave. 
These statements obviously generalize to the pointwise maximum (minimum) of any number of convex (concave) functions. A special case we will use below involves a continuous family of affine functions of $y$ indexed by a variable $w\in\mathcal{B}$: If $f(w,y)$ is affine in $y$ for all $w\in\mathcal{B}$, then
\be
g(y) \ := \inf_{w\in {\cal B}}  \, f(w,y)  \,
\ee
is a concave function of $y$ on the domain where the infimum is finite.  

Now consider the following optimization problem presented in standard form:   \be
\label{convex-program-standard}
\begin{split}
\hbox{Minimize}  \qquad   & f_0(y) \,  \quad \hbox{over} \  y\in 
{\cal D}  \\
\hbox{subject to} \qquad & f_i(y) \leq 0 \, \ \ i = 1 , \ldots, m \,, \\
& h_j(y) = 0 \ \ j = 1 , \ldots, p \,,
\end{split}
\ee
In this program $y \in \mathbb{R}^n$ is the variable we minimize over.  Here $f_0(y)$ is the {\em objective function},  the $f_i$'s with $i\geq 1$ define the inequality constraints, and the $h_j$'s define the equality constraints.
The {\em domain} ${\cal D}$ of the problem
is a subset of $\mathbb{R}^n$
defined as the common domain of the functions $f_i$ ($i=0, \ldots, m$) and  $h_j$ ($j=1, \ldots , p$).   
The {\em feasible} set ${\cal F}\subseteq {\cal D}$ is defined to be the subset of the domain in which
the constraints hold:
\be
{\cal F}  \ = \ \{ y  \ |    y \in {\cal D} ,  \ f_i( y) \leq 0\ \forall i \,, \ 
h_j (y) = 0\ \forall j \, \} \,.
\ee
The {\em optimal value} $p^*$ is the minimum value of the objective over the feasible set ${\cal F}$:     
\be
p^* \ = \  \inf_{y \in {\cal F}}\   f_0 (y) \,. 
\ee
A point $y^*\in {\cal F}$ is said to be an {\em optimal point} if $f_0(y^*)= p^*$. Since, in much of this paper, we will be working in function spaces rather than in $\mathbb{R}^n$, we will also use the terms \emph{optimal configuration} and \emph{solution}, to avoid confusion with the points of the underlying manifold. (In the optimization literature, ``solution'' generally refers to the optimal \emph{value} $p^*$; however, it seems more natural in the physics context to use 
the word to refer to the optimal point or configuration.) Note that, in general, an optimal point need not exist, and if one exists it need not be unique.

The above problem is a {\em convex program}
if the objective $f_0$ and inequality constraint functions $f_i$ are convex functions over a convex domain ${\cal D}$, and the $h_i$ are affine functions. 
It follows that the feasible set ${\cal F}$ is also convex.  
For a convex program, any local minimum is in fact a global
minimum. If $f_0$ is strictly convex, then the optimal point $y^*$ (if it exists) is unique, otherwise it need not be. However, the set of optimal points $\{y^*\in\mathcal{F}|f(y^*)=p^*\}$ is necessarily convex.

We will also refer to the maximization problem
\be\label{convex-program-standard2}
\begin{split}
\hbox{Maximize}  \qquad   & g_0(y) \,  \quad \hbox{over} \  y\in 
{\cal D}  \\
\hbox{subject to} \qquad & g_i(y) \ge 0 \, \ \ i = 1 , \ldots, m \,, \\
& h_j(y) = 0 \ \ j = 1 , \ldots, p\,,
\end{split}
\ee
where the functions $g_i$ ($i=0,\ldots,m$) are \emph{concave}, as a convex program, since it is obviously related to the convex program \eqref{convex-program-standard} just by taking $f_i(y)=-g_i(y)$ 
($i=0,\ldots,m$).

Symmetries of a convex program can greatly aid their solution. Suppose that a convex program is invariant under a finite or compact group $G$ of affine transformations of $\mathbb{R}^n$; that is, the domain $\mathcal{D}$, the objective $f_0$, the set of inequality constraint functions $\{f_i\}$, and the set of equality constraint functions $\{h_j\}$ are all invariant under the action of $G$. Then, acting on any feasible point $y$ by any element $g$ of $G$ produces a point $g(y)$ that is also feasible and has the same value of the objective~$f_0$. Furthermore, by the convexity of the feasible set, 
the average $\langle y\rangle_G$ over the $G$ action is also feasible, and by the convexity of $f_0$, the value of $f_0$ for the average is no larger than for $y$:
\begin{equation}
f_0(\langle y\rangle_G)\le f_0(y)\,.
\end{equation}
This average $\langle y\rangle_G$  
 is clearly a fixed point of $G$. Therefore, when 
 minimizing 
 we can impose the symmetry at the outset without changing the value of $p^*$.
In other words,  we can reduce the domain $\mathcal{D}$ to the 
  locus  of $G$-fixed points within $\mathcal{D}$. 
  Since $G$ acts affinely, this locus is a convex set. Such a reduction by a symmetry can make it much easier to solve the problem, as we will see at several points in this paper and the sequel~\cite{headrick-zwiebach2}.

\subsection{Lagrangian duality}
\label{prim-dual-opt}

\emph{Lagrangian duality} is a method for transforming a constrained optimization problem into another one, 
written in terms of a different set of variables, that gives information about the solution to the first one. In fact, under fairly general circumstances, the two problems are equivalent, in the sense that they have the same optimal value. The method involves introducing Lagrange multipliers for the constraints and then solving for the original variables, leaving a program expressed in terms of the Lagrange multipliers. In the process, a program involving minimizing a convex objective, such as \eqref{convex-program-standard}, becomes a program involving maximizing a concave objective, and vice versa. In this context, the original program is referred to as the \emph{primal} and the new one as the \emph{dual}. We will now describe the method in more detail, taking \eqref{convex-program-standard} as our primal.

The {\em Lagrangian function} $L(x, \lambda, \nu)$ 
is built by adding to the objective of the primal a sum of terms with Lagrange multipliers 
$\lambda_i$ and $\nu_j$ multiplying the
constraint functions:
\be
L(y, \lambda, \nu) \ := \ f_0(y)  + \sum_{i=1}^m \lambda_i  f_i(y)  +  \sum_{j=1}^p \nu_j h_j(y) \,.
\ee
Since $L(y, \lambda, \nu)$ is, for any 
fixed $y$, 
an affine function of $\lambda$
and $\nu$, minimizing over $y$ yields a
concave function of $\lambda$ and $\nu$ called the \emph{Lagrange dual function} or {\em dual objective} $g_0(\lambda, \nu)$:
\be
g_0(\lambda, \nu ) \ := \  \inf_{y \in {\cal D}}  \, L( y, \lambda , \nu)  \,.
\ee
Its domain $\mathcal{D}'$ is the set of points $(\lambda,\nu)$ on which the infimum is finite. One can prove that for any $\nu$ and any $\lambda\ge0$ (i.e.\ $\lambda_i\ge0$ for all $i$) the dual objective provides a lower bound for the optimal value $p^*$ of the primal:
\be
\label{main-result-1}
g_0( \lambda\ge0, \nu ) \ \leq \ p^*  \,. \ 
\ee

Since $g_0(\lambda, \nu)$ is a concave function of $\lambda$ and $\nu$, it makes sense to look for its maximum. 
Indeed, one  defines the {\em dual program}:
\be  
\begin{split}
\hbox{Maximize}  \qquad   & g_0(\lambda , \nu ) \,  \quad 
\hbox{over} \ \ (\lambda,\nu)\in\mathcal{D}'
 \,,  \\
\hbox{subject to} \qquad & \lambda_i \geq 0 \,, \, \ \ i = 1 , \ldots, m \,. 
\end{split}
\ee
This is a convex optimization problem of the form \eqref{convex-program-standard2} (regardless of whether the primal is convex). The dual feasible set is the set
\begin{equation}
\mathcal{F}' := \{(\lambda,\nu)\in\mathcal{D}'|\lambda\ge0\}\,.
\end{equation}
We have a {\em dual optimal value}
$d^*$ and \emph{dual optimal point}
(or \emph{configuration})
$(\lambda^*, \nu^*)\in\mathcal{F}'$ if
\be
d^* \ := \  \sup_{(\lambda, \nu)\in\mathcal{F}'}  g_0(\lambda ,  \nu) \ = \ 
g_0(\lambda^*, \nu^*) \,. 
\ee 
It follows from (\ref{main-result-1}) that
\be
\label{orderingxx}
d^*  \leq \, p^* \,;
\ee
this relation is called \emph{weak duality}. We have {\em strong duality} if the primal and dual optimum
values are the same: $d^* = p^*$.  Strong duality holds in 
a variety of situations but can be guaranteed when the 
primal is convex and the {\em Slater condition} is satisfied:  
there exists a feasible point $x$ in the interior
of ${\cal D}$ where the inequality constraints are
strictly satisfied 
($f_i(y)<0$ for $i=1,\ldots,m$, $h_j(y)=0$ for $j=1,\ldots,p$).

Suppose strong duality holds, and let $y^*$ and $(\lambda^*, \nu^*)$ be 
primal and dual optimal points respectively.  The inequality
constraints and their Lagrange multipliers then satisfy
{\em complementary slackness}: 
\be
\label{complementary-slackness}
 \lambda_i^*  f_i (y^*) \ = \ 0   \ \ \hbox{for each } \ i \,. \ 
\ee
Note that the primal and dual optimal points may not be unique. However, since \eqref{complementary-slackness} holds for \emph{any} primal and dual optimal points, we have the following:
\begin{enumerate}
\item If there exists a primal optimal point $y^*$ such that $f_i(y^*)<0$, then for all dual optimal points $\lambda^*_i=0$ (slack in the constraint implies no slack in the multiplier).
\item \label{slackness2} If there exists a dual optimal point $(\lambda^*,\nu^*)$ such that $\lambda^*_i>0$, then for all primal optimal points $f_i(y^*)=0$ (slack in the multiplier implies no slack in the constraint).
\end{enumerate}
In the 
case $f_i(y^*)=0$, one says that the constraint is {\em active} (or saturated).   
One can also show that, for any dual optimal $(\lambda^*,\nu^*)$, any primal optimal $y^*$ minimizes $L(y, \lambda^*, \nu^*)$:
\be\label{dualmin}
\inf_{y \in {\cal D}} L (y, \lambda^*, \nu^*)  = L (y^*, \lambda^*, \nu^*)\,.
\ee
This follows from strong duality as follows:
\begin{equation}
\inf_{y \in {\cal D}} L (y, \lambda^*, \nu^*)  = g_0(\lambda^*,\nu^*) = d^* = p^* = f_0(y^*) = L(y^*,\lambda^*,\nu^*)\,.
\end{equation}

In the rest of this paper we will be working mostly not in $\mathbb{R}^n$ but in infinite-dimensional function spaces. In doing so, we will simply assume that the above results carry over, without worrying too much about the functional analysis that would be necessary to prove our claims rigorously.

\subsection{Calibrations and the max flow-min cut theorem}\label{max-flow-section}

In network theory one often considers a source node $s$ and a sink node $t$ connected by a graph with some number of nodes and edges.  Each  edge is
assigned  a capacity, the maximum flux it can handle.  With this
graph one can pose two different
 problems that are nontrivially related.   
The first is a {\em max flow} problem.  A flow is an assignment of fluxes to each edge of the graph consistent with their capacity. The assignment must respect conservation (flux in $=$ flux out) at each node except $s$ and $t$.
In the max flow problem one searches for the flow with maximum
flux from the source $s$ to the sink~$t$.  
The second problem is a {\em min cut}
problem.  A cut is a partition of the nodes into two sets $S$ and $T$, with $S$ containing $s$ and $T$ containing $t$.  The capacity of the
cut is the sum of capacities of the edges connecting $S$ and $T$.  The min cut problem asks for the cut of minimum capacity.  The max flow-min cut (MFMC) theorem states that for any network the flux of the max flow equals the capacity of the min cut.  These two problems can be stated as linear programs related by strong duality.

In this paper we will make extensive use of a closely analogous theorem in the setting of Riemannian manifolds. In this theorem, the role of a \emph{flow} is played by a norm-bounded divergenceless vector field and the role of a \emph{cut} is played by a hypersurface in a specified homology class. The proof of the theorem, which hinges on strong duality of convex programs, can be found in the references \cite{headrick-hubeny-refs}. Here we will explain the setup, state the theorem, and sketch its proof. Our sketch focuses on the aspects of the proof involving convex optimization and skips over those involving geometric measure theory (properly defining minimal surfaces, proving their existence, etc.). Several of the techniques and intermediate results in the proof will be used in the rest of the paper. In reference \cite{Headrick:2017ucz} one can find 
a detailed  discussion of the MFMC theorem
focused on the case when the homology class is defined by a region of the manifold's boundary, the situation relevant to the study of holographic entanglement entropy.

\subsubsection{Setup and statement}

Let $M$ be a connected Riemannian manifold, possibly with boundary. In what follows $M$ and all submanifolds  are assumed compact and oriented. While in most of this paper $M$ is two-dimensional, in this subsection $M$ may have arbitrary dimension $d$. We define a \emph{flow} as a vector field $v$ obeying
\begin{equation}
\nabla_\mu v^\mu = 0\,,\qquad |v|\le1
\end{equation}
everywhere. Let $C$ be an integral $d-1$ homology class, $C\in H_{d-1}(M,\mathbb{Z})$.\footnote{With minor modifications, everything in this subsection can be generalized to the case where $C$ is a homology class relative to some subset of $M$.} A \emph{cut} is a hypersurface $m$ in the class $C$. The volume form $\omega=\sqrt g\hskip1pt d^dx$ on $M$ can be pulled back onto $m$ to give its area form $\omega_{\parallel}$, whose integral is the total area of $m$:
\footnote{In the case $d=2$ relevant to the rest of the paper,
a ``hypersurface'' is a curve, and its ``area''in the sense of \eqref{areadef} is its length.}
\begin{equation}\label{areadef}
\area(m) := \int_m\omega_\parallel\,.
\end{equation}
We now consider the flux of a flow $v$ through a cut $m$,
\begin{equation}
\flux_m(v) := \int_m\omega_\parallel\,n_\mu v^\mu\,, 
\end{equation}
where $n_\mu$ is the unit normal to $m$. 
By virtue of the divergenceless condition, this flux is independent of $m$, so we can write instead $\flux_C(v)$. On the other hand, the 
norm bound $|v| \leq 1$ 
implies $n_\mu v^\mu\le1$; integrating this inequality over $m$ shows that the flux is bounded above by the area of $m$. All in all, for any flow $v$ and any cut $m$ we have
\begin{equation}
\flux_C(v) \le \area(m)\,.
\end{equation}
The MFMC theorem asserts that this inequality is tight:
\begin{equation}
\sup_v\flux_C(v) = \inf_{m\in C}\area(m)\,.
\end{equation}
A flow $v^*$ that achieves the supremum is called a max flow and a cut $m^*$ that achieves the infimum is called a min cut. Under suitable regularity conditions, 
$v^*$ and $m^*$ 
exist; however, they are not necessarily unique.

An equivalent, and in some cases more convenient, language for talking about flows is that of calibrations 
\cite{harvey-lawson}. A $p$-\emph{calibration} is a $p$-form $u$ that is closed, $du=0$, and has norm $|u|\le1$. The norm bound
implies that for any $p$-dimensional submanifold $m$ the 
period of $u$ along $m$ is bounded by the area of $m$:
\begin{equation}\label{calbound}
\int_mu \le \int_m\omega_\parallel\,.
\end{equation}
If $u=\omega_\parallel$ everywhere on $m$
we say that $u$ \emph{calibrates} $m$.   Clearly, if $u$ calibrates $m$ then the bound \eqref{calbound} is saturated.  
Conversely, if $u$ is a calibration and \eqref{calbound} is saturated, then $u$ calibrates $m$. 

The relation
\begin{equation}\label{flowcalibration}   
u=*(g_{\mu\nu}v^\mu dx^\nu)\,,
\end{equation}
with $*$ the Hodge star,
gives a one-to-one map between flows $v$ and $(d-1)$-calibrations $u$; $\nabla_\mu v^\mu=0$ is equivalent to $du=0$, and $|v|\le1$ is equivalent to $|u|\le1$. 
While the condition $\nabla_\mu v^\mu=0$ involves the metric via the Christoffel symbol, the condition $du=0$ does not. In the rest of the paper we will be varying the metric. 
It will therefore be more convenient for us to 
use the language of calibrations, where the metric only appears in the constraint $|u|\le1$. When the metric is fixed, however, the two notions are interchangeable.

Under the relation \eqref{flowcalibration}, we have
\begin{equation}
\flux_C(v) = \int_Cu\,,
\end{equation}
where by the closedness of $u$ we needn't specify on which representative of $C$ we evaluate the period of $u$. Therefore, in the language of calibrations, the MFMC theorem reads
\be\label{MFMC}
\sup_{u} \int_{C}u = 
\inf_{m\in C}\area(m)\,.
\ee
A calibration $u^*$ that achieves the supremum is called a max calibration. Like the max flow $v^*$, it is not necessarily unique. However, any max calibration must calibrate every min cut $m^*$, and therefore is fully determined on the locus of min cuts. Elsewhere, however, it is 
underdetermined. In particular, for any point $x$ \emph{not} on  
any min cut, there exists a max calibration $u^*$ such that $|u^*(x)|<1$. We will take this fact as intuitively clear (although the reader can find an argument in subsection 3.4 of \cite{Headrick:2017ucz}). This will become important below when we use max calibrations to find min cuts.

\subsubsection{Proof}\label{sec:MFMCproof}

We will now sketch 
the proof  of \eqref{MFMC}. Here is the general outline. First, we note that the conditions defining a calibration, $du=0$, $|u|\le1$, are convex, and the functional $\int_Cu$ is linear in~$u$. Therefore we can write a convex program, which we call the \emph{max flow} program, whose optimal value is the left-hand side of \eqref{MFMC}. On the other hand, for the right-hand side, which we call the \emph{min cut} program, the objects being minimized over---hypersurfaces in the homology class $C$---do not naturally form a convex set. We therefore rewrite the problem as a minimization over representatives of the Poincar\'e dual \emph{cohomology} class, which are closed one-forms. These \emph{do} form a convex set, on which we can define a convex objective that is a generalized ``area'' functional. We will call this the \emph{convex min cut} program. A non-trivial step is to show that it is equivalent to the original min cut problem. Furthermore, it is related by strong duality to the max flow program. We will thus show
\begin{equation}\label{twoequalities}
\text{max flow} = \text{convex min cut} = \text{min cut} \,.  
\end{equation}
Here = means the programs are equivalent, 
i.e.\ have the same optimal value. 
We will go from left to right in this equation: write down the max flow program, dualize it in order to get the convex min cut program, and finally show that the latter reduces to the original min cut program.

We first consider the case where $M$ is closed, in order to eliminate certain complications that 
would distract from the main story. At the end we will fill in the generalization to manifolds with boundary.

\paragraph{From max flow to convex min cut:} To write the left-hand side of \eqref{MFMC} in terms of a convex program, we let the domain be the set of $(d-1)$-forms on $M$, which is clearly a convex set, and we impose the definition of a calibration, $du=0$ and $|u|\le1$, as explicit constraints. The first is a linear equality constraint, while the second is a convex inequality constraint. For the objective, we would like to write $\int_Cu$. However, this is only well-defined if $u$ is closed---otherwise the integral depends on the choice of representative of $C$---and we need an objective that is well-defined on the entire domain, i.e.\ before imposing the constraints. We could resolve this problem by picking an arbitrary representative of $C$. However, it will turn out to be more convenient to appeal to Poincar\'e duality and instead choose a representative $\eta_0$ of the dual \emph{cohomology} class $\tilde C\in H^1(M)$. This means that, for any closed $(d-1)$-form $u$,
\begin{equation}
\int_Cu = \int_Mu\wedge\eta_0\,.
\end{equation}
The right-hand side provides an objective that is well-defined for any $(d-1)$-form and equals the one we want for closed ones. The max flow program 
then reads 
\be\label{maxflow}
\begin{split}
\hbox{\bf Max flow:} \ \ \ \ 
\hbox{Maximize}  \qquad   & \int_Mu\wedge\eta_0 \,  \quad 
\text{over $u$ ($(d-1)$-form)}  \\
\   \ \hbox{subject to}\ \   \quad & 1-|u|\ge0\,,  \\
&    \qquad  du=0\,,\quad\forall x\in M\,.
\end{split}
\ee
We will follow the convention of indicating the type of each variable in a program (constant, function, form, etc.) in parentheses after its name. The position $x$ in the manifold is playing the role of the indices $i$ and $j$ in the equality and inequality constraints; henceforth we will not write $\forall x\in M$ in each program.

We now dualize the max flow program. Since $du$ is a top-form, it has only one independent component. As a result, a single Lagrange multiplier $\psi$, a scalar
function on $M$, suffices for the constraint $du=0$. 
We introduce a second scalar Lagrange multiplier $\phi\ge0$ for the inequality constraint  $1-|u|\ge0$. 
The Lagrangian functional is
\begin{equation}\label{Lag1}
L[u,\psi,\phi] =  \int_M u\wedge\eta_0 +\int_M\left((-1)^d\psi \,du+\omega\phi(1-|u|)\right),
\end{equation}
where again $\omega$ is the volume form and the factor $(-1)^d$ is included to simplify future equations.  We now wish to maximize $L$ with respect to $u$, where $u$ is now an arbitrary $(d-1)$-form. We can do the maximization pointwise on $M$ if we first integrate the $\psi\, du$ term by parts to get the derivative off of $u$:
\begin{equation}\label{Lag2}
L[u,\psi,\phi] = \int_M\left( u\wedge(\eta_0+d\psi)-\omega\phi|u|+\omega\phi\right)=\int_M\omega(I_u+\phi)\,,
\end{equation}
where
\begin{equation}
I_u=(-1)^{d-1}v^\mu\eta_\mu-\phi|v|\,, \qquad
\eta:=\eta_0+d\psi\,, 
\end{equation}
and the vector $v$ is defined 
in terms of $u$ by \eqref{flowcalibration}. Maximizing $L$ with respect to $u$ is equivalent to maximizing $I_u$ with respect to $v$ at each point. One can convince oneself that $I_u$ is bounded above if and only if $\phi\ge|\eta|$, and in this case its maximum is zero. This inequality on $\phi$ allows us to drop the earlier constraint $\phi \ge0$. So the dual program is 
\be\label{dual1}
\begin{split}
&\hbox{Minimize}  \quad   \  \int_M\omega\phi
\,  \quad 
\text{over $\psi$, $\phi$ (functions)}\,, \\
& \text{subject to}\ \ \  \phi\ge|\eta_0+d\psi| \, .  
\end{split}
\ee
Clearly the minimum with respect to $\phi$ is achieved by setting $\phi=|\eta_0+d\psi|$. This leaves the program
\be\label{mincut}
\begin{split}
\hbox{Minimize}  \qquad   & \int_M\omega\, |\eta_0+d\psi|
\,  \quad 
\text{over $\psi$ (function)}
\,.
\end{split}
\ee
As expected, the objective is convex in $\psi$.
Moreover, there are 
no explicit constraints.

For any function $\psi$ the one-form $\eta_0+d\psi$ is in the class $\tilde C$, and conversely any one-form $\eta$ in that class can be written as $\eta_0+d\psi$ for some function $\psi$; in other words the map $\psi\mapsto\eta_0+d\psi$ from functions to $\tilde C$ is surjective. It is also affine. So we can change variables to $\eta$ and write \eqref{mincut} as
\be\label{mincut2}
\begin{split}
\hbox{\bf Convex min cut:} \ \ \ \ 
\hbox{Minimize}  \quad   & \int_M\omega|\eta|
\,  \quad 
\hbox{over $\eta\in\tilde C$}
\,.
\end{split}
\ee
(The reason for the name will be explained shortly.) Strong duality holds because Slater's condition, stated below \eqref{orderingxx}, holds: the $d-1$ form $u=0$ is feasible and strictly satisfies the inequality constraint $|u|\le1$. Therefore, the convex min cut program is equivalent to the max flow program.  
This establishes the first equality in \eqref{twoequalities}.

\paragraph{From convex min cut to min cut:} The min cut, by definition, is the solution to the following program:
\be\label{mincut0}
\begin{split}
\hbox{\bf Min cut:} \ \ \ \ 
\hbox{Minimize}  \quad   & \hbox{area} (m)  \quad 
 \hbox{over $\ m\in  C$}
\,.
\end{split}
\ee
Since the homology class $C$ is not a convex set, this is not a convex program. However, it is related to the \emph{convex min cut} program \eqref{mincut2} by convex relaxation. Convex relaxation essentially means turning an optimization problem with a non-convex feasible set into a convex program by embedding the feasible set into a larger convex set and extending the definition of the objective to a convex function on the new feasible set. In this case, we will use a map $\eta_m$ 
defined below to embed 
the non-convex set $C$ into the convex set $\tilde C$, and replace the area functional with the convex functional $\int_M\omega|\eta|$.
In general, relaxation can result in a program with a lower optimal value than the original one. We will check, however, that this is not the case here: the convex min cut and original min cut programs have the same optimal values.

To relate the one-forms appearing in the convex min cut program to the hypersurfaces appearing in the min cut program, we recall that there is a map taking an arbitrary hypersurface $m$ to a one-form $\eta_m$ such that, for any (not necessarily closed) $(d-1)$-form $u$,
\begin{equation}
\label{def-eta-m}
\int_mu = \int_Mu\wedge\eta_m\,.
\end{equation}
In a local coordinate system $x^\mu$ ($\mu=1,\ldots,d$) in which $m$ is at $x^d=0$ and has orientation form 
$dx^1\wedge\cdots\wedge dx^{d-1}$, 
the one-form $\eta_m$ 
is given by\footnote{In \eqref{bumpform}, we assume that $m$ has unit multiplicity at $x^d=0$. More generally, if it has multiplicity $n_+$ with orientation $dx^1\wedge\cdots\wedge dx^{d-1}$ and $n_-$ with opposite orientation, then $\eta_m = (n_+-n_-)\delta(x^d)dx^d$.}
\begin{equation}\label{bumpform}
\eta_m =\delta(x^d)dx^d\,.
\end{equation}
$\eta_m$ is sometimes called a ``bump form'' (see \cite{bott-tu}, section 6 for a more detailed discussion). It is clear that if $m$ is closed then $\eta_m$ is closed, and furthermore if $m\in C$ then $\eta_m\in\tilde C$. Working in the above local coordinate system, one can also show that $\omega|\eta_m| = \omega_\parallel\wedge\eta_m$, so
\begin{equation}
\int_M\omega|\eta_m| = \area(m)\,.
\end{equation}
Taking the infimum over $m$ we get
\begin{equation}\label{oneway}
\inf_{m\in C}\area(m) =\inf_{m\in C}\int_M\omega|\eta_m|\ge \inf_{\eta\in\tilde C}\int_M\omega|\eta|\,.
\end{equation}

We would have gotten an equality in \eqref{oneway} 
if  $m\mapsto\eta_m$  
were a surjective map to $\tilde C$. 
It is not, however: most one-forms in $\tilde C$ are not the bump form of any hypersurface in $C$. However, as we will now show, any $\eta\in\tilde C$ corresponds in some sense to a convex combination of hypersurfaces in $C$. 
By this we mean the following.  
Since $\eta$ is closed, by the Frobenius theorem it is hypersurface-orthogonal. We will construct those hypersurfaces, 
which we will call $m_\eta(t)$,
 and show that 
 they are in the class $C$, are parametrized by points on the circle $\mathbb{R}/\mathbb{Z}$, and have average area equal to $\int\omega|\eta|$. This will establish equality between the left- and right-hand sides of \eqref{oneway}.

Let $\eta$ be a continuous one-form in $\tilde C$; we will treat one-forms that are discontinuous or have delta functions as limits of continuous ones. Since $\tilde C$ is the Poincar\'e dual of $C$, which is an integral homology class, the integral of $\eta$ over any closed curve $c$ equals the intersection number $\#(c,C)$, which is an integer. Therefore, given two points $x_{1,2}\in M$, the integral of $\eta$ along a path from $x_1$ to $x_2$ is independent of the choice of path up to an integer. Fixing a base point $x_0$, we define the continuous function $\Psi_\eta:M\to\mathbb{R}/\mathbb{Z}$ by
\begin{equation}
\Psi_\eta(x) = \int_{x_0}^x\eta\,.
\end{equation}
It is clear that $d\Psi_\eta = \eta$. 
(In fact, we could equivalently define $\Psi_\eta$ as the solution to $d\Psi_\eta=\eta$ with $\Psi_\eta(x_0)=0$.) Changing the base point $x_0$ just shifts $\Psi_\eta$ by an unimportant constant, so we will not indicate $x_0$ explicitly. Given a closed curve $c:\mathbb{R}/\mathbb{Z}\to M$, the  map 
\be
\Psi_\eta\circ c:\mathbb{R}/\mathbb{Z}\to\mathbb{R}/\mathbb{Z}
\ee
has winding number
\be
\int_cd\Psi_\eta=\int_c\eta=\#(c,C)  \,. 
 \ee
We will use this fact below.

Any connected 
subset of $M$ on which $\eta$ vanishes maps under $\Psi$ to a single point of $\mathbb{R}/\mathbb{Z}$. Therefore the locus $\{x\in M|\eta(x)=0\}$ maps to a set of isolated points in $\mathbb{R}/\mathbb{Z}$. Let $S\subseteq\mathbb{R}/\mathbb{Z}$ be the complement of this set. 
Since the isolated points form a set 
  of measure zero, in integrals we will ignore the difference between $\mathbb{R}/\mathbb{Z}$ and $S$. For a point $t\in S$, define the \emph{level set} $m_\eta(t)$ as the inverse image of~$\Psi$:
\begin{equation}\label{metadef}
m_\eta(t) := \Psi_\eta^{-1}(t)\,.
\end{equation}
This level set is a closed hypersurface on which $\eta\neq0$; 
we fix its orientation by pulling the $(d-1)$-form $*\eta$ back to $m_\eta(t)$. 
The intersection number of any closed curve $c$ with $m_\eta(t)$ is the number of times (counted with signs) that $\Psi_\eta\circ c$ intersects $t$, which is the winding number of $\Psi_\eta\circ c$, which in turn as shown above is $\#(c,C)$. The intersection numbers of a hypersurface with a complete set of closed curves (representing a basis for $H_1(M,\mathbb{Z})$) fully determine its homology class; since these intersection numbers are the same for $m_\eta(t)$ as for $C$, $m_\eta(t)$ must be in the class $C$.

Finally, by the coarea formula,
\begin{equation}\label{coarea}
\int_M\omega|\eta| = \int_M\omega|d\Psi_\eta| = \int_0^1dt\area(m_\eta(t))\,,
\end{equation}
which implies
\begin{equation}\label{etaarea}
\int_M\omega|\eta| \ge\inf_{m\in C}\area(m)\,.
\end{equation}  

Taking the infimum of \eqref{etaarea} over $\eta$,
\begin{equation}\label{secondway}
\inf_{\eta\in\tilde C}\int_M\omega|\eta| \ge\inf_{m\in C}\area(m)\,.
\end{equation}
Together with \eqref{oneway}, which is the above inequality
in the opposite direction, this establishes the equivalence of the min cut and convex min cut programs:
\be
\label{min-cut=convex-min-cut}
\inf_{m\in C}\area(m)\ = \
\inf_{\eta\in\tilde C}\int_M\omega|\eta| \,.  
\ee

\paragraph{Manifolds with boundary:} The generalization of the proof of MFMC to manifolds with boundary requires only a minor modifications of the above, essentially just imposing certain boundary conditions.

On a manifold with boundary we must replace Poincar\'e duality with Poincar\'e-Lefschetz duality, which establishes an isomorphism between the homology group $H_{d-1}(M)$ and the cohomology group relative to the boundary $H^1(M,\partial M)$. Relative cohomology means that all forms are subject to the boundary condition that their pullback onto $\partial M$ vanishes. In particular, for us it means that every element $\eta$ of the dual class $\tilde C$ is normal to $\partial M$, and any two elements differ by an exact one-form $d\phi$ where $\phi$ vanishes on $\partial M$. 

The max flow program \eqref{maxflow} and the Lagrangian \eqref{Lag1} are unchanged. However, when we integrate by parts, there is a new surface term $(-1)^d\int_{\partial M}\psi u$ which must be added to the right-hand side of \eqref{Lag2}. When we then maximize with respect to $u$ (which, recall, is unconstrained at this stage), this term is bounded above if and only if $\psi$ vanishes on $\partial M$, in which case the term vanishes. So the only change to the programs \eqref{dual1} and \eqref{mincut} is the addition of a boundary condition:
\begin{equation}\label{psibc}
\left.\psi\right|_{\partial M} = 0\,.
\end{equation}
This boundary condition is precisely what is required to ensure that $\eta:=\eta_0+d\psi$ is an element of $\tilde C$, and conversely that every element of $\tilde C$ can be written as $\eta_0+d\psi$. So the convex min cut program written in the form \eqref{mincut2} remains unchanged. The rest of the proof goes through as before.

\sectiono{The homotopy and homology problems}
\label{sec:homology}

In this section we discuss minimal-area metrics on
a compact Riemann surface $M$, possibly with boundary.
After fixing some notation we state the closed string 
field theory
minimal-area problem as a convex program.
This program is not amenable to numerical solution
as stated because it involves a 
functional set of nonlocal constraints: 
 the lengths of an infinite number of homotopically nontrivial closed curves
are constrained, and each length constraint imposes nonlocal conditions
on the metric. 
  The minimal area problem can be modified
to consider constraints on curves in a set of homotopy classes,
with different length conditions on the various classes.

We examine then a more significant modification where
we work with {\em homology} classes of curves.  This
time we consider  a collection of non-trivial
integral homology 1-cycles  $C_\alpha \in H_1 (M,\mathbb{Z})$, with $\alpha \in J$ an index labeling the cycles.  
We aim to 
minimize the total area of $M$ subject to the constraint that, for each $\alpha$, all representatives of $C_\alpha$ have length at least $\ell_\alpha$, 
where $\ell_\alpha >0$ is a constant. 
In subsection \ref{sec:homotopicMAP} we write this as a convex program. This program, just like the homotopy one, is not very practical as it also involves a functional constraint set. Therefore,  in subsection \ref{sec:localprimal} we use calibrations and the max flow-min cut theorem to rewrite the problem as a convex program with a {\em finite} number of {\em local}    
 constraints. This form of the program allows one to obtain rigorous upper bounds on the
minimal area and is also practical for numerical minimization.

We restrict ourselves in this paper to homology classes of closed curves, both for the sake of concreteness and because this is the case relevant to closed string field theory. Our considerations, however, carry over almost without modification to open curves, with the homology class defined relative to the endpoints of the curve. More generally, one can consider classes defined relative to other sets, such as the boundary of $M$; this requires imposing appropriate boundary conditions on the calibrations $u^\alpha$ and scalar fields $\varphi^\alpha$ defined below
and in the following section.

\subsection{Notation}

Let us fix some notation to get started.  We denote by 
$g^0_{\mu\nu}$ a fiducial conformal metric on the Riemann surface $M$. This means that for any local complex coordinate $z= x^1+ ix^2$ on the surface, the fiducial metric takes the form 
\be
ds^2 = 2g^0_{z\bar z}dzd\bar z=
2g^0_{z\bar z}\left((dx^1)^2 + (dx^2)^2\right).  
\ee
The conformal metrics $g_{\mu\nu}$ on $M$ are in one-to-one correspondence with functions $\Omega \geq 0$  on $M$ via
\begin{equation}
g_{\mu\nu}\, =\, \Omega \, g^0_{\mu\nu}\,.
\end{equation}
We will also write 
\be
\Omega=\rho^2\,,  
\ee 
where $\rho$ is a 
non-negative function on $M$:\footnote{In the literature on Riemann surfaces, the metric is often written $ds^2=\rho^2dzd\bar z$. The reader familiar with that literature should be aware that our use of $\rho$ is different.}
\begin{equation}
g_{\mu\nu}\, =\, \rho^2 \, g^0_{\mu\nu}\,.
\end{equation}
Since the relation between $\Omega$ and $\rho$ is non-linear, it does not automatically preserve convexity properties. Given that convexity plays a central role in our analysis, we will take care in each instance to specify whether we are using $\Omega$ or $\rho$ as our variable.

The area of $M$ is
\begin{equation}
\text{area} (M) = \int_M d^2x \sqrt g = \int_M d^2 x \sqrt{g^0}\,\Omega =\int_Md^2\sqrt{g^0}\,\rho^2\,.
\end{equation}
We also write
$$\area(M) = \int_M \omega = \int_M \omega_0 \,\Omega =\int_M\omega_0\,\rho^2\,, $$
where we define the volume form $\omega$ and the 
fiducial volume form $\omega_0$ as
\be
\omega \, = \, d^2x \sqrt{g}\,, \quad  \omega_0 \, = \, d^2 x \sqrt{g_0} \,. 
\ee

Consider now the lengths of curves on the surface. 
Let $\gamma$ be a closed curve 
defined by a map $x^\mu(t): [0,1] \to M$. 
We then have
\begin{equation}
\begin{split}
\text{length} (\gamma)   := & \   \int_0^1 dt\,   | \dot x_i  (t)| 
= \int_0^1  dt\sqrt{g_{\mu\nu}\, \dot x^\mu\, \dot x^\nu} \\
=  & \ \int_0^1 dt \sqrt{\Omega \, g^0_{\mu\nu}\,
  \dot x^\mu\, \dot x^\nu} 
  = \int_0^1 dt \sqrt{\Omega} \,  |\dot x |_{{}_0}
= \int_0^1 dt \,\rho \,  |\dot x |_{{}_0} \, ,
\end{split}
    \end{equation} 
with $t$ derivatives denoted by dots, $| \cdot|$ denoting
norm in the metric $g$ and $| \cdot|_{{}_0}$ denoting
norm in the fiducial metric $g_0$ (see Appendix \ref{sec:notation} for more details on
notation).  For brevity we will write, symbolically, 
 \begin{equation}
\text{length} (\gamma) 
 \  = \  \int_\gamma  \sqrt{\Omega} \,  |\dot x |_{{}_0}   \,.
   \end{equation}
We can also consider  lengths of representatives of non-trivial homology 1-cycles. Such a representative $m$ is a set of $k$ closed oriented curves  $\gamma_i(m)$, with $i= 1 , \ldots, k$. We parameterize each of them with $t\in [0,1]$ 
using maps $x_i^\mu (t):  [0,1] \to M$.  The length of $m$ is then defined as 
\begin{equation}
\text{length} (m)  \ := \  \sum_{i=1}^k \int_0^1 dt\,   | \dot x_i  (t)| 
  \ = \ 
  \sum_{i=1}^k\int_0^1 dt \, \sqrt{\Omega } \,  \bigl| \dot x_i \bigr|_{{}_0} \,.  
   \end{equation}
For brevity we will write, symbolically
 \begin{equation}
\text{length} (m) 
 \  = \  \int_m  \sqrt{\Omega} \,  |\dot x |_{{}_0}   \,.
   \end{equation}
Here we are adding  lengths regardless of orientation of the curves.
If a representative contains both a curve $\gamma$ and the oppositely
  oriented curve $-\gamma$, which 
  cancel in homology,
  the length functional adds the two lengths.

\subsection{Homotopic minimal-area programs}
\label{sec:homotopicMAP}

The closed string field theory minimal-area problem (MAP) 
stated in the introduction and reviewed in section~\ref{sec:csft}
 is now stated as a convex program.  We let $\ell_s$ 
 denote the systole and $\Gamma$ the set of 
 non-contractible closed curves on $M$, and write
\begin{equation}
\label{csft-MAP}
\begin{split}
 \text{\bf Closed string field theory MAP: }\ \ \  & \text{Minimize } \,  
\int_M\omega_0 \,\Omega \ \  
\ \ \text{over }\Omega\ge0 \text{ (function)}\\
& \hbox{subject to} \ \ \ \ 
\, \ell_s-\int_\gamma\sqrt{\Omega} \, |\dot x |_{{}_0} \le0\,, \ \ 
  \forall\,   \gamma \in \Gamma \,. \ \ 
  \end{split} 
\end{equation}
The notation here means that $\Omega\ge0$ is an {\em implicit} constraint:  it defines the domain on which the objective and other constraint functions are defined.  We have allowed the scale factor $\Omega$ to vanish, since this may occur at points on $M$ on the minimal-area metric (e.g.\ if it has conical singularities). The other constraint is explicit:  
\be
\ell_s -\int_\gamma\sqrt{\Omega} \, |\dot x |_{{}_0} \, \le \, 0\, .
\ee
In a well-defined convex program, the objective and all explicit constraint functions must be convex on the domain defined by the implicit constraints, which must itself be a convex subset of an affine space. 
We can check that \eqref{csft-MAP} is indeed a convex program: The domain, consisting of non-negative functions $\Omega$ on $M$, is clearly convex. The objective is a linear functional of $\Omega$ and therefore convex. And, for each curve $\gamma$, the constraint function $\ell_s-\int_\gamma\sqrt\Omega|\dot x|_0$, is an affine functional, with positive coefficients, of $-\sqrt\Omega$, which is itself a convex function of $\Omega$ for $\Omega\ge0$. This will be discussed more explicitly below in the context of the homology version.

We can also write the program in terms of $\rho$:
\begin{equation}
\label{csft-MAPrho}
\begin{split}  
& \text{Minimize } \,  
\int_M\omega_0 \,\rho^2 \ \  
\ \ \text{over }\rho
\ge0 
\text{ (function)}\\
& \hbox{subject to} \ \ \ \ 
\, \ell_s-\int_\gamma\rho \, |\dot x |_{{}_0} \le0\,, \ \ 
  \forall\,   \gamma \in \Gamma \,. \ \ 
  \end{split} 
\end{equation}
This is also a convex program: The domain is again the set of 
non-negative functions on $M$, the objective is a convex functional, and the constraint function is an affine, hence convex, functional. (See \cite{Zwiebach:1992ie} for further discussion.) Since the objective in \eqref{csft-MAPrho} is \emph{strictly} convex, this way of writing the program has the advantage of showing that the solution $\rho^*$ is \emph{unique}, which implies that the solution $\Omega^*$ of \eqref{csft-MAP} is also unique.

A generalized version of this problem imposes 
different conditions on different homotopy classes of closed curves. Let $D_\beta\in \pi_1(M)$, with $\beta \in K$ an
index labeling the various homotopy classes.  
To the curves on $D_\beta$
we associate the length constraint $\ell_\beta$.  We then have
\begin{equation}
\label{homotopy-MAP}
\begin{split}
 \text{\bf Homotopy MAP: }\ \ \  & \text{Minimize } \,  
\int_M\omega_0 \,\Omega \ \   
\ \ \text{over }\Omega\ge0 \text{ (function)}\\
& \hbox{subject to} \ \ \ \ 
\, \ell_\beta-\int_\gamma\sqrt{\Omega} \, |\dot x |_{{}_0} \le0\,, \ \ 
  \forall\,   \gamma \in D_\beta, \ \ \forall \beta \in K \,. 
  \end{split} 
\end{equation}

\subsection{Homological minimal-area 
program}\label{sec:isosystolic}
  
Again, let $C_\alpha$ with $\alpha \in J$ be
a set of non-trivial homology cycles on $M$. 
   We look for the minimal
  area (conformal) metric 
  under the condition that, for any $\alpha\in J$, the length of any representative 
  of $C_\alpha$  be greater than or equal to 
  $\ell_\alpha$.  In analogy to the homotopy MAP above, we write  
\begin{equation}
\label{firstprogram}
\begin{split}
 \text{\bf Homology MAP: }\ \ \  & \text{Minimize } \,  
\int_M\omega_0 \,\Omega \ \   
\ \ \text{over }\Omega\ge0 \text{ (function)}\\
& \hbox{subject to} \ \ \ \ 
\, \ell_\alpha-\int_m\sqrt{\Omega} \, |\dot x |_{{}_0} \le0\,, \ \ 
\forall \,  m\in C_\alpha\,, \  \forall\,   \alpha \in J \,.
  \end{split} 
\end{equation}
 
 Let us make explicit the fact that we have a convex program
 by showing that by discretization we get a program of the
 form (\ref{convex-program-standard}).
 Our variable $\Omega$ on the surface $M$ 
 is the analog of the vector $y\in \mathbb{R}^n$
 in (\ref{convex-program-standard}).  We think of $\Omega$
 as a vector $\vec{\Omega}\in \mathbb{R}^N$, whose components are the values $\Omega[i]$ over a discretization of the surface $M$ into small
 plaquettes indexed by $i=1, \cdots N$, for some large value of $N$. 
 The implicit constraint $\Omega[i] \geq 0$ for all $i$, indeed
 defines a convex subspace of the affine space $\mathbb{R}^N$
 where $\vec{\Omega}$ lives.

  The objective $f_0$ in  (\ref{convex-program-standard})
 is now a sum of the discrete values $\Omega[i]$ multiplied by 
 the quantity $\omega_0[i]\geq 0$   representing the value of 
 $d^2x\sqrt{g^0}$ for the $i$-th plaquette:  
 \be
 f_0 (\vec{\Omega})\, =  \, \sum_{i=1}^N  \Omega(i)  \, \omega_0(i)\,. 
 \ee
Written this way, the objective is clearly an affine function
of $\vec{\Omega}$, and thus it is convex.  
Consider now the constraint on the length.  Take some fixed
cycle $m$, parameterized as $y(t)$.  The cycle will go through some 
plaquettes whose values of $i$ are in a set 
$S_m$. For each $i\in S_m$ there is positive number
$\chi_m[i]\geq 0$ capturing the way the curve crosses
the  $i$-th plaquette  so that the length constraint takes
the form
\be
\ell_\alpha - \sum_{i\in S_m}  \sqrt{\Omega[i]} \, \chi_m[i]  \, \leq \, 0 \,. 
\ee
Since the function $(-\sqrt{x})$ is convex for $x\ge0$, the
left-side of this inequality, being a non-negatively weighted
sum of convex functions,  is a convex function of $\vec{\Omega}$ over the domain $\vec{\Omega} \geq 0$.
 
Note that even as we discretize the program (\ref{firstprogram}),
there are an infinite number of constraints, since there
are an infinite number of curves whose length
must be constrained.  In the program each cycle 
is represented by a set $S_m$ and the values
of the (continuous) coefficients $\chi_m[i]$.
An infinite number of constraints is difficult to handle.
Of course, one gets a finite number of constraints if
one decides to constrain the length of a finite number of curves,
but it is unclear when the chosen set of curves suffices
to get a good approximation. 

There is no need to discretize to ascertain convexity 
 if we are willing to consider
a generalized form of the standard program (\ref{convex-program-standard}) where the variable $x$ is replaced 
by a function, both the
objective and the inequality constraints are defined by 
convex functionals, and the equality constraints are
affine functionals.  Letting $\Omega$ denote the function variable
such program reads   
\be
\label{convex-program-functional}
\begin{split}
\hbox{Minimize}  \qquad   & F_0[\Omega] \,  \ \ 
\hbox{over} \  \Omega \in {\cal D}  \\
\ \hbox{subject to} \hskip-3pt 
\qquad & F_i[\Omega] \leq 0 \, \ \ i = 1 ,2 \ldots, \,, \\
& H_j[\Omega ] = 0 \ \ j = 1 , \ldots  \,.  
\end{split}
\ee 
Here $F_0$ and $F_i$ are convex functionals
over some domain ${\cal D}$ for the functions $\Omega$,
where ${\cal D}$ is a convex subset of an affine space.  
Moreover, the $H_j$ are affine functionals. 
The affine space
in our case is the space of all functions $\Omega$ and the
convex subset is the set of all $\Omega \geq 0$.
A functional
$\Phi$ is said to be convex if for functions $\Omega_1$ and
$\Omega_2$ in ${\cal D}$ we have 
\be
\Phi \, [\, t\Omega_1 + (1-t) \Omega_2\, ] \, \leq \,  t\, \Phi\, [\Omega_1]
+ (1-t) \, \Phi \, [\Omega_2] \,,  \quad  t\in [0,1] \,. 
\ee	
The area functional 
\be
F_0[\Omega] \ = \ \int_M  \omega_0\, \Omega \,,  
\ee 
 is manifestly an affine, and thus convex functional.
 We have no equality constraints.  
 The inequality constraints are the length constraints. 
 Applied to the
 cycle $m\in C_\alpha$ it takes the form
 \be
 F_m[\Omega] \ = \ \ell_\alpha  - \int_m \sqrt{\Omega} \,  | \dot y |_0 \,, 
 \ee
 and $F_m$ is quickly shown to be a convex functional using the convexity
 of $-\sqrt{x}$, for $x>0$.   Each constraint is non-local
 on the surface $M$, involving the scale factor on curves
 that stretch along $M$.  Moreover we really have a
 functional constraint set:  each constraint is associated
 to a function, the embedding map that defines the cycle
 on the surface.   

We can also write the homology program in terms of $\rho$:
\begin{equation}\label{firstprogramrho}
\begin{split}
& \text{Minimize } \,  
\int_M\omega_0 \,\rho^2 \ \   
\ \ \text{over }\rho
\ge0 \text{ (function)}\\
& \hbox{subject to} \ \ \ \ 
\, \ell_\alpha-\int_m\rho \, |\dot x |_{{}_0} \le0\,, \ \ 
 \forall \,  m\in C_\alpha\,, \  \forall\,   \alpha \in J \,.
  \end{split} 
\end{equation}
As for \eqref{csft-MAPrho}, in this form the objective is strictly convex, which shows that the solution $\rho^*$ is unique. This implies in turn that the program \eqref{firstprogram} also has a unique solution $\Omega^*$.

The advantage of the homology MAP over the homotopy one is that, as we will show in the next subsection, by using calibrations it can be converted into a program with a finite number of local constraints. Furthermore, as we will show in section \ref{sec:homotopy}, 
the closed string field theory homotopy program
can be written as a homology program by the trick of passing to a covering space.

\subsection{Reformulation as a local problem using calibrations}\label{sec:localprimal}

In this subsection 
we will use the device of calibrations to reformulate the homological minimal-area problem \eqref{firstprogram} as a convex program with constraints that are almost entirely \emph{local} on $M$. For this we will rely crucially on the max flow-min cut (MFMC) theorem. Calibrations and MFMC were described 
in subsection \ref{max-flow-section}.  
Here we will specialize to the two-dimensional case and give further details.

We define a  \emph{1-calibration}  
on a manifold as a closed one-form $u$ obeying $|u|\le1$ everywhere:
\begin{equation}
\text{Calibration:} \ \   du = 0 \,,  \ \ | u| \leq 1 \,. 
\end{equation}
There is no boundary condition on $u$. Since $u$ is closed, its integral over a cycle 
in a homology class $C$ is independent of the representative on which it is integrated, so we write 
\be
\int_Cu\,, 
\ee
for the period of the calibration.  As is well known, by virtue of the constraint $|u|\le1$, this integral gives
a lower bound for the length of any representative $m\in C$:
\begin{equation}\label{calibration}
\biggl| \int_{C}u \, \biggr| \,  \leq \, \text{length}(m) \,  \,.
\end{equation}
To verify this we write $m$ as a set of curves $x_i^\mu(t)$.  Since 
$m$ belongs to
the homology class $C$ and the form $u$ is closed we have
\begin{equation}
\int_{C}u  =   \int_{m}u   =  \ \sum_i\int_0^1 dt \,  u (\dot x_i(t))  =  \   \  \sum_i\int_0^1 dt \,  \langle \hat u , \dot x_i(t) \rangle \,.
   \end{equation}
Here the vector $\hat u$ is obtained by raising the index on the
one-form $u$ with the help of the metric, and $\langle \cdot\, , \, \cdot\rangle$ is the inner product on the space of vectors
(see Appendix A for notation and identities).  Taking 
absolute values we have
\begin{equation}
\biggl|\int_{C}u \, \biggr| \  =    \  \biggl|\sum_i\int_0^1 dt \,  \langle \hat u , \dot x_i(t) \rangle\biggr|  \leq  \, \sum_i \biggl|\int_0^1 dt \,  \langle \hat u , \dot x_i(t) \rangle\biggr|   \leq  \,  \sum_i \int_0^1 dt \, \bigl|  \langle \hat u , \dot x_i(t) \rangle\bigr|  \,.
   \end{equation}
The inequalities above are all saturated if the integrand $\langle \hat u , \dot y_i(t) \rangle$ is everywhere positive. 
Using the Schwarz inequality 
$| \langle v, v'\rangle | \leq |v| |v'|$ and  $|\hat u| = |u| \leq 1$ we now get
 \begin{equation}
\biggl|\, \int_{C}u  \, \biggr|   \   \leq  \ 
\sum_i \int_0^1 dt \,  | u |  |\dot x_i(t)| \leq \int_0^1 dt \,  |\dot x(t)| = \text{length} (m) \,,
  \end{equation}
  which is what we wanted to prove. 
It is also clear what we need for the length inequality
to be saturated: 
\begin{equation}
\label{systole-u}
\text{length}(m) = \biggl|\, \int_{C}u  \, \biggr|  \quad \text{requires}  \ \   \hat u \propto \dot x    \ 
\  \text{and} \ \ 
|u| = 1 \, \ \text{on} \, \, m \,. 
\end{equation}
Furthermore, the sign of the proportionality between $\hat u$ and $\dot x$ should be constant on $m$.

We have seen that if there exists a 
calibration with period $\ell$ on $C$ then all representatives of $C$ have length at least $\ell$. Furthermore, if the minimal-length representative has length $\ell'\ge \ell$, then, by the MFMC theorem, 
there exists a calibration $u'$ with period $\ell'$.  
Then $u:=(\ell/\ell')u'$ is also a calibration, and has period $\ell$.
In summary, there exists a 
calibration with period $\ell$ {\em if and only if} all representatives of $C$ have length at least $\ell$. In this way, we can replace the length condition on every representative of a cycle with the existence of a calibration of a given period.  Except for the  period condition, all conditions on a calibration $u$ are local on the surface.

For each cycle $C_\alpha$, $\alpha \in J$, on which we  impose the length constraint, we demand the existence of a corresponding calibration $u^\alpha$. The constraints on $u^\alpha$ are:\begin{equation}
\label{calib-u}     
du^\alpha=0\,,\quad 
|u^\alpha|\le1 \,, \quad
\int_{m_\alpha}u^\alpha=\ell_\alpha\,,
\end{equation}
where $m_\alpha$ is an arbitrary representative of $C_\alpha$. 
It is necessary to choose a representative since the period $\int_{C_\alpha}u^\alpha$ is not well-defined before the constraint $du^\alpha=0$ is imposed. The first and third constraints are affine in the variable $u^\alpha$ and don't depend on the scale factor $\Omega$.  
We can rewrite the second constraint as follows:
\begin{equation}\label{flowconstraint}
 |u^\alpha|_0^2-\Omega\le0\,.
\end{equation}
As written, the constraint \eqref{flowconstraint} is convex in $u^\alpha$ and affine in $\Omega$. 
Viewed as  $\Omega \geq |u^\alpha|_0^2$, 
we see that the fiducial  norm of the calibration ``props up'' the scale factor $\Omega$. We can now drop the implicit constraint $\Omega\ge0$ that we had in our first program (\ref{firstprogram}), which simplifies the subsequent analysis a bit. We now have the following convex program:
\begin{equation}\label{secondprogram}
\begin{split}
\text{\bf Primal MAP v1:} \ \   
\ \ &\text{Minimize }\ \, \int_M \omega_0 \,\Omega \quad 
\text{ over 
$\Omega$ (function), $u^\alpha$ (one-forms)}  \ \  \\
&   \hbox{subject to}  \hskip25pt |u^\alpha|_0^2-\Omega\le0\,,\\
&  \hskip99pt
  du^\alpha=0\, , \  \\
& \hskip66pt  \ell_\alpha-\int_{m_\alpha}\hskip-8pt u^\alpha=0\,, 
\ \forall \alpha \in J  \,.  \phantom{{x}_A}  \  \\
\end{split}  
\end{equation}
There is no boundary condition on $u^\alpha$.  Any {\em feasible} $u^\alpha,\Omega$ provides a rigorous upper bound on the value of the minimum. 
We call this program ``primal'' because in the next section we will derive a second minimal-area program, related to this one by Lagrangian duality, which we will call the \emph{dual minimal-area program}.

As for the homotopy and homology programs, we can make the change of variables from $\Omega$ to $\rho$ in \eqref{secondprogram}, and the program remains convex:
\begin{equation}\label{secondprogramrho}
\begin{split}
\ \ &\text{Minimize }\ \, \int_M \omega_0 \,\rho^2 \quad 
\text{ over 
$\rho$ (function), $u^\alpha$ (one-forms)}  \ \  \\
&   \hbox{subject to}  \hskip25pt |u^\alpha|_0-\rho\le0\,,\\
&  \hskip99pt
  du^\alpha=0\, , \  \\
& \hskip66pt  \ell_\alpha-\int_{m_\alpha}\hskip-8pt u^\alpha=0\,, 
\ \forall \alpha \in J  \,.  \phantom{{x}_A}  \  \\
\end{split}  
\end{equation}
Again, since the objective is strictly convex in $\rho$, it is clear that $\rho$ takes a unique value $\rho^*$ in the solution. On the other hand, the one-forms $u^\alpha$ do not appear in the objective, so they are not necessarily unique. 
To see the non-uniqueness we solve for $\rho$:   
\begin{equation}
\label{vm-find-omega}   
\rho \ = \ \max_\alpha|u^\alpha|_0\,, \quad \text{at every point on $M$.}
\end{equation}
If we
use this solution, the program is now:
\begin{equation}\label{solveOmega}
\begin{split}
&\text{Minimize }\ \int_M\omega_0 \,\max_\alpha|u^\alpha|_0^2\ \
\text{ over }
u^\alpha\text{ (one-forms)} \, \\
& \hbox{subject to} \quad  \, du^\alpha=0\,,\\
&  \hskip62pt  \ell_\alpha- \int_{m_\alpha}u^\alpha=0\, ,  \ \ 
\forall \alpha \in J
\,.
\end{split}
\end{equation}
The objective is  convex in the space of $u$'s:  
it is the pointwise maximum of
a set of convex functions.  While \eqref{solveOmega} is simpler-looking,
we have found it easier in practice to work with the  form~\eqref{secondprogram}.
Returning to the non-uniqueness of $u^\alpha$, it is clear from
(\ref{vm-find-omega}) that we can change any calibration $u^\alpha$
over a region $R$ where $|u^\alpha|_0 < \rho$ without changing the
value of the objective while keeping $u^\alpha$ closed and its periods
unchanged.  For this we simply let $u^\alpha \to u^\alpha + d\chi$, where 
$\chi$ is a smooth function, chosen to be sufficiently small and supported only
inside the region~$R$.   

The program \eqref{secondprogram} still contains 
a finite number of non-local constraints, namely the period conditions 
$\int_{m_\alpha}u^\alpha=\ell_\alpha$,
as compared to the functional infinity of constraints appearing in \eqref{firstprogram}. 
By using a basis of real  closed one-forms, however, 
we can rewrite these constraints as algebraic conditions, 
eliminating the need to do integrals on $M$ to check that the constraints are satisfied.  Moreover, the algebraic conditions can sometimes be solved  directly.
Let  $\omega^i$ be a 
real basis for $H^1(M)$,\footnote{We work with real one-forms throughout, as the calibrations are defined to be real.} closed one-forms with periods:
\be
\omega^i_\alpha \ := \ \int_{C_\alpha}\omega^i\,. 
\ee
(Do not confuse the basis one forms $\omega^i$ with the
volume forms $\omega$ and $\omega_0$.) 
We can then write $u^\alpha$ as as a linear combination
of the basis one-forms plus an exact one-form $d\phi^\alpha$
where $\phi^\alpha$ is a function on $M$:
\begin{equation}
u^\alpha = \sum_i  c_i^\alpha \omega^i+d\phi^\alpha\,.  
\end{equation}
The constants $c_i^\alpha$  are constrained by the period
conditions $\int_{C_\alpha} u^\alpha = \ell_\alpha$:
\begin{equation}\label{cconstraint}
\sum_i c_i^\alpha\omega^i_\alpha =\ell_\alpha\,. \end{equation}
If the number of  constraints here (equal to the number of
homology classes) is no more than the number of independent cycles, then by an appropriate choice of one-forms $\omega^i$ the constraint \eqref{cconstraint} can be solved directly.
Having replaced the calibrations 
by the constants $c_i^\alpha$ and the functions $\phi^\alpha$,
where the zero mode of $\phi^\alpha$ drops out,
the program reads:
\begin{equation}\label{thirdprogram}
\begin{split}
\text{\bf Primal MAP v2:} \  \ 
\ &\text{Minimize }\, \int_M \omega_0\,\Omega\, \quad  \ 
\text{over $c^\alpha_i$ (constants), $\Omega$, $\phi^\alpha$ (functions)} \\
& \, \hbox{subject to} \ \ \ 
\Bigl|\sum_ic_i^\alpha \omega^i+d\phi^\alpha\Bigr|_0^2-\Omega\le0\,,\\
& \hskip99pt   \ell_\alpha - \sum_ic^\alpha_i\omega^i_\alpha=0\,, \qquad\qquad\forall \alpha \in J\,.
\end{split}  
\end{equation}
We will see in~\cite{headrick-zwiebach2} that with a suitable discretization 
the program  \eqref{thirdprogram} is straightforward to solve numerically.

\section{An equivalent dual problem}\label{sec:dual}

\begin{figure}[!ht]
\leavevmode
\begin{center}
\epsfysize=3.5cm
\epsfbox{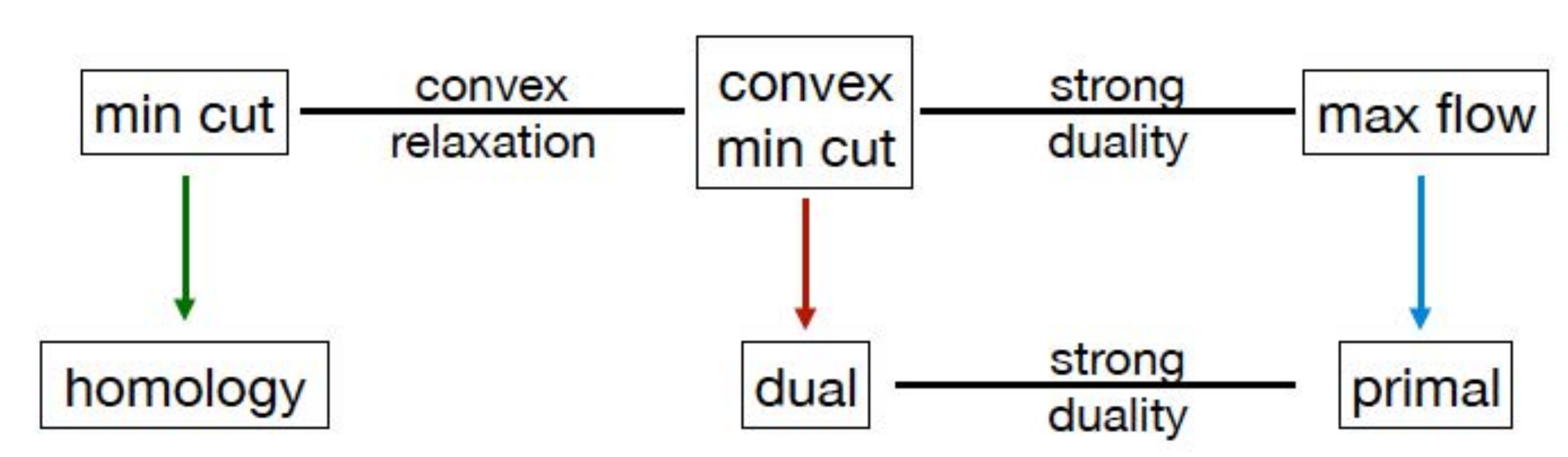}
\end{center}
\caption{\small Illustration of the relations among the various programs in this paper. The top line gives the programs for finding the minimal length in a given homology class. Min cut is given in 
(\ref{mincut0}),  convex min cut  in (\ref{mincut2}), and max flow  in (\ref{maxflow}). The bottom line 
gives the minimal-area programs. The homology minimal 
area program (MAP) is given in subsection \ref{sec:isosystolic}, the dual program in \ref{sec:dualstate}, and the primal in 
\ref{sec:localprimal}. 
Each one is connected by an arrow to the minimal-length program employed in its derivation. The green arrow (left) 
is direct. The blue arrow (right) 
 requires a 
short argument, appearing between 
(\ref{systole-u}) and (\ref{calib-u}). The derivations of the dual MAP using convex min cut (red arrow) are given in subsections \ref{sec:maximin} and \ref{sec:directdual}.
}
\label{fig:road-map}
\end{figure}

In this section we will derive another convex program that
solves the homological minimal-area problem described in subsection \ref{sec:isosystolic}. Like the primal minimal-area program \eqref{secondprogram}, it involves only local variables and constraints, and is amenable to numerical solution. Perhaps surprisingly, it is a \emph{maximization} problem and does not involve the metric at all. It involves instead a function $\varphi^\alpha$ and a constant $\nu^\alpha$ for each 
 homology class 
 $C_\alpha$ constrained with length parameter $\ell_\alpha$. 
However, as we will explain in section \ref{sec:saturating}, it is straightforward 
to use a solution to this program to read off both the minimal-area metric and the curves saturating the length constraint (if any) in each class.  We will call this program, in its various simply related versions, the {\em dual minimal-area program}, since its derivation involves a dualization, starting either from the homology minimal-area program \eqref{firstprogram} or the so-called primal one \eqref{secondprogram}. The program will be stated and briefly discussed in subsection \ref{sec:dualstate}, and then derived in three different ways in \ref{sec:dualderive}. Figure \ref{fig:road-map} illustrates the relations among the various programs in this paper.

\subsection{Statement}\label{sec:dualstate}

We start with the same data defining the homological minimal-area problem (\ref{firstprogram})  described in the previous section: a Riemann surface $M$, possibly with boundary, 
 equipped with a fiducial metric $g^0_{\mu\nu}$, a set of non-trivial integral homology classes $C_\alpha\in H_1(M,\mathbb{Z})$, and for each one a 
positive number~$\ell_\alpha$. We then require that, for all $\alpha$, the length of every representative of $C_\alpha$ is at least $\ell_\alpha$, and we wish to find the smallest area among metrics in the Weyl class of $g^0_{\mu\nu}$ satisfying this constraint.

Let $\tilde C^\alpha\in H^1(M)$ be the 1-cohomology class that is Poincar\'e 
dual  to $C_\alpha$ or, in the presence of a boundary, 
the Poincar\'e-Lefschetz dual (see above \eqref{psibc}).
We will denote by $\eta^\alpha$ a general representative of $\tilde C^\alpha$. We claim that the above problem is solved by the following convex program, henceforth
called the first dual version:
\begin{equation}
\label{dualeta}
\begin{split}
\hbox{\bf Dual MAP v1:} \ \ & \ \  \text{Maximize}\quad
2\sum_\alpha \nu^\alpha \ell_\alpha -\int_M\omega_0
\Bigl(\sum_\alpha\nu^\alpha|\eta^\alpha|_0\Bigr)^2
\quad\text{over $\nu^\alpha$ (constants), $\eta^\alpha\in\tilde C^\alpha$} \\
&\ \  \text{subject to} \qquad   \,\nu^\alpha\ge0\,, \quad\forall\alpha\in J \,.
\end{split}
\end{equation}
Here $\omega_0=\sqrt{g^0}d^2x$ and $|\cdot|_0$ are, respectively,  the area form and norm with respect to the fiducial metric. The program in fact is independent of the choice of fiducial metric within a given Weyl class, since under a Weyl transformation $g^0_{\mu\nu}\to\hat\rho^2g^0_{\mu\nu}$, the area form and norm transform as $\omega_0\to\hat\rho^2\omega_0$ and $|\cdot|_0\to\hat\rho^{-1}|\cdot|_0$, so the objective is unchanged. It is a straightforward exercise to show that the objective is 
a concave functional of $\nu^\alpha$ and $\varphi^\alpha$, and the constraint functions are linear, hence concave, so 
(\ref{dualeta})  is indeed a convex program. We will derive this program by three different routes in subsection \ref{sec:dualderive}.

A solution to \eqref{dualeta}  
gives us not only the value of the minimal area 
but also the minimizing metric and curves saturating 
the length constraint.   Given a solution $(\nu^{\alpha*},\eta^{\alpha*})$, the metric is
\begin{equation}\label{solution}
g_{\mu\nu} = \Omega^*g^0_{\mu\nu}\,,\qquad\Omega^* = (\rho^*)^2\,,\qquad
\rho^* = \sum_\alpha\nu^{\alpha*}|\eta^{\alpha*}|_0\,.
\end{equation}
Furthermore, for any $\alpha$ such that $\nu^{\alpha*}>0$, the level sets $m_{\eta^{\alpha*}}(t)$ defined in \eqref{metadef} saturate the length constraint. We will derive these facts in subsection \ref{sec:saturating}.

We can write the program \eqref{dualeta} in a more concrete form by choosing, for each $\alpha$, a fiducial representative $\eta^\alpha_0$ of $\tilde C^\alpha$. Then any $\eta\in\tilde C^\alpha$ can be written $\eta^\alpha = \eta_0^\alpha+d\psi^\alpha$ for some function $\psi^\alpha$ satisfying $\psi^\alpha|_{\partial M} = 0$.  Defining $\varphi^\alpha$ by   
\begin{equation}\label{psiphirelation} 
 \varphi^\alpha 
 :=\nu^\alpha\psi^\alpha \,,
\end{equation}
and recalling that $\nu^\alpha\ge0$, we have
\begin{equation}
\nu^\alpha|\eta^\alpha|_0 = |\nu^\alpha\eta^\alpha_0+d\varphi^\alpha|_0\,.
\end{equation}
This change of variables puts the program into the following form:
\begin{equation}
\label{seconddual}
\begin{split}
&\hbox{\bf Dual MAP v2:}\\
 & \ \ \text{Maximize}\quad
2\sum_\alpha \nu^\alpha \ell_\alpha -
\int_M\hskip-3pt\omega_0\Bigl(\sum_\alpha
\left|\nu^\alpha \eta_0^\alpha+d\varphi^\alpha\right|_0\Bigr)^2\  
\text{ over $\nu^\alpha$ (constants), $\varphi^\alpha$ (functions)}  \\
& \ \ \text{subject to} \qquad   \, \left.\varphi^\alpha\right|_{\partial M}=0 \quad\forall\alpha\in J \,  .
\end{split}
\end{equation}
In going from \eqref{dualeta} to \eqref{seconddual}, we actually dropped two constraints. First, according to \eqref{psiphirelation}, $\nu^\alpha=0$ implies $\varphi^\alpha=0$, but in \eqref{seconddual} we do not impose this constraint. However, if $\nu^\alpha=0$, then the objective will in any case be maximized by setting $d\varphi^\alpha=0$. If $M$ has a boundary, the boundary condition then implies $\varphi^\alpha=0$. If not, the program \eqref{seconddual} is invariant under $\varphi^\alpha\to\varphi^\alpha+$ constant, so we are free to set $\varphi^\alpha=0$. Second, we dropped the constraint $\nu^\alpha\ge0$. However, if $\nu^\alpha<0$ then the objective will be increased by taking $\nu^\alpha\to-\nu^\alpha$ and $\varphi^\alpha\to-\varphi^\alpha$. So neither of these ``relaxations'' changes the solution, and \eqref{seconddual} is indeed equivalent to \eqref{dualeta}.

We will now write \eqref{seconddual} in an even more concrete third form by making a specific choice of fiducial one-forms $\eta_0^\alpha$, namely the delta-function ``bump'' forms defined in \eqref{bumpform}. Choose a representative $m_\alpha$ of each homology class $C_\alpha$ and set $\eta_0^\alpha = \eta_{m_\alpha}$, which equals $\delta(x^2)dx^2$ in a local coordinate system in which $m_\alpha$ is at $x^2=0$. 
The second term in the objective \eqref{seconddual} contains in the integrand the square of a sum of norms  of one-forms. It therefore contains the square of a delta function, giving a divergent integral, unless $d\varphi^\alpha$ has a compensating delta function,
\begin{equation}
d\varphi^\alpha = -\nu^\alpha\delta(x^2)dx^2+\text{regular}\,.
\end{equation}
Thus the objective is finite only if $\varphi^\alpha$ jumps by $-\nu^\alpha$ along $m_\alpha$. We can make the constraint $\Delta\varphi^\alpha|_{m_\alpha}=-\nu^\alpha$ (which is linear in $\varphi^\alpha$ and $\nu^\alpha$) explicit. 
Away from $m_\alpha$, $\eta_{m_\alpha}$ vanishes. In this form, the program becomes
\begin{equation}\label{thirddual}
\begin{split} 
&\hbox{\bf Dual MAP v3:} \\
 &\quad\text{Maximize}\ \  
  \, 2\sum_\alpha \nu^\alpha \ell_\alpha-\int_{M'}
\omega_0\Bigl(\sum_\alpha\left|d\varphi^\alpha\right|_0\Bigr)^2\   \ \     \hbox{over $\nu^\alpha$ (constants), $\varphi^\alpha$ (functions)}
\\
&\quad \hbox{subject to} \ \ \ \ 
 \Delta\varphi^\alpha|_{m_\alpha}=-\nu^\alpha\,, \\
 &  \hskip70pt  \varphi^\alpha|_{\partial M}=\ 0\, ,  \quad \forall \alpha\in J \,. 
\end{split} 
\end{equation}
Here $M'$ is the manifold $M$ with the chosen representative curves $m_\alpha$ removed:
 \be
 M' = M\setminus \cup_{\alpha\in J}  m_\alpha\,. 
 \ee
 This is just a way to tell us that in
 calculating the objective with our necessarily 
 discontinuous $\varphi^\alpha$ there is no delta-function 
 contribution in $d\varphi^\alpha$ on $m_\alpha$.
  Note also that we don't need to be careful about the orientation of $m_\alpha$ or the sign of the jump, since the objective is invariant under $\varphi^\alpha\to-\varphi^\alpha$. 
  
The program (\ref{thirddual}) is the final form of our dual program. 
In trying to maximize the objective the first term tries to make 
$\nu^\alpha$ large; however, a non-zero jump forces $\varphi^\alpha$ to have a non-zero gradient somewhere, which makes the second term more negative. The former is linear while the latter is quadratic, so we 
expect a maximum to exist. Note that the second term in the objective
is a little unusual:  it is the square of a sum of norms rather than
the more familiar (to physicists) sum of norm-squared terms.  
On the maximum not all homology classes
may be active: $\nu^\alpha$ and $\varphi^\alpha$ may vanish for some $\alpha$.

An interesting special case occurs when, for some $\alpha$, a subset $m'_\alpha$ of the full boundary $\partial M$ is a representative of the class $C_\alpha$. 
In this case the representative $m_\alpha$ where $\varphi^\alpha$ jumps
can be pushed all the way until it reaches the curve $m'_\alpha$ in the boundary.  
Since the value of $\varphi^\alpha$ at the boundary is supposed to be zero, placing
the discontinuity  at $m'_\alpha$ effectively sets the value of $\varphi^\alpha$ 
on $m'_\alpha$  equal to $\nu^\alpha$. The value of $\varphi^\alpha$ on the rest
of the $\partial M$ remains zero. In other words, for that $\alpha$ we replace the constraints in \eqref{thirddual} by the following boundary conditions:
\begin{equation}\label{boundaryrep}
\left.\varphi^\alpha\right|_{\partial M\setminus m'_\alpha}=0\,,\qquad
\left.\varphi^\alpha\right|_{m'_\alpha}=\nu^\alpha\,.
\end{equation}

Like the primal \eqref{thirdprogram},  the dual program \eqref{thirddual} is straightforward to solve numerically. 
Furthermore, any trial values for $\nu^\alpha,\varphi^\alpha$ give a rigorous lower bound on the maximum. Using both the primal~\eqref{thirdprogram} and the dual~\eqref{thirddual}, we can thus bound the minimum area both above and below.
As in program Dual MAP v1, a solution $( \nu^{\alpha*}, d\varphi^{\alpha*})$ of 
Dual MAP v3 also gives us the
minimizing metric $\rho^*$
\be
\label{minmetric-v3}
\rho^*  = \sum_\alpha | d\varphi^{\alpha*} |_0 \,. 
\ee

We noted that the solution of the primal problem was not unique because
the calibrations $u^\alpha$ are ambiguous in the regions where their 
fiducial norm squared does not equal the Weyl factor $\Omega$ and they  
do not appear explicitly in the objective. 
For the dual program (\ref{thirddual}),
one can find fine-tuned examples for which the solution is not unique (such as the square torus; see section \ref{sec:torus}). 
We believe, however, that the solution
is generically unique, up to shifts of the functions $\varphi^\alpha$ 
by constants.  This intuition is supported by the identification
of saturating geodesics from the functions $\varphi^\alpha$,
to be discussed in section \ref{sec:saturating}. 

\paragraph{Solving for the $\boldsymbol{\nu}$'s:}   The dual programs
presented above involve maximization over some 
non-negative constants $\nu^\alpha$ and some one-forms,
or functions, depending on the version.  
As mentioned below \eqref{solution}, on the solution, $\nu^{\alpha*}>0$ for any $\alpha$ such that the length constraint in that homology class is active, i.e.\ such that there exist representatives $m\in C_\alpha$ saturating the length constraint. If we happen to know which homology classes are active, then we can ignore the $\nu^\alpha\ge0$ constraint for those classes, and simply  drop the other ones altogether from the program. This makes it possible 
to perform the maximization over the $\nu$'s.  We will do
so, although it is not clear if the resulting objective, while simple looking, is
particularly useful.  Take, for example,
version 1 of the dual program (\ref{dualeta}), whose objective 
$O_1$ is 
\begin{equation}
\label{dualeta-vv}
O_1 =  
2\sum_\alpha \nu^\alpha \ell_\alpha - \sum_{\alpha, \beta} 
 \nu^\alpha \nu^\beta  M_{\alpha \beta} \,,  \quad
 M_{\alpha\beta} := \int_M\omega_0
|\eta^\alpha|_0 |\eta^\alpha|_0 \geq 0 \,. 
\end{equation}
 Using vector notation $\boldsymbol{\nu} = (\nu^1, \ldots ), \,  {\boldsymbol{\ell}} = (\ell_1, \cdots )$ and letting ${\bf M}$ denote
the symmetric matrix with components $M_{\alpha\beta} \geq 0$, the objective is
\be
O_1 =  2\,  \boldsymbol{\nu}^T  \boldsymbol{\ell } - \boldsymbol{\nu}^T \, {\bf {M}} \,   \boldsymbol{\nu}\,.
\ee
Generically the matrix $\bf M$ is invertible (otherwise a slightly more involved treatment is required). 
Maximization over $\boldsymbol{\nu}$ leads to 
$\boldsymbol{\ell}= {\bf M } \, \boldsymbol{\nu}$ and therefore   
$\boldsymbol{\nu}= {\bf M }^{-1} \, \boldsymbol{\ell}$.  Putting it back into
the objective we get the new objective
\be
O'_1 =  \,   \boldsymbol{\ell }^T  {\bf M}^{-1}  \boldsymbol{\ell}\,.
\ee
It remains, of course to maximize $O'_1$ over the one-forms $\eta$ entering into the definition of the ${\bf M}$ matrix elements.  
It is noteworthy
that the Schwarz inequality on the surface implies that
\be
M_{\alpha \beta} \leq  \sqrt{ M_{\alpha\alpha} M_{\beta\beta}} \,, 
\ee
where repeated indices are not summed.  Note that this result
can easily be adapted for version 3 of the dual program.  In this
case one would define tilde functions $\varphi^\alpha = \nu^\alpha \tilde \varphi^\alpha$ with unit discontinuities $\Delta \tilde \varphi^\alpha = -1$ and a similar writing of the objective
is possible.

\subsection{Derivations}\label{sec:dualderive}

In this subsection, we will derive the dual minimal-area program. We will in fact present three different derivations. 
Each of these derivations 
illuminates the relation between the dual program and the original homological minimal-area program \eqref{firstprogram}.  
This is valuable given that, at first blush, these two problems appear to be totally unrelated. The derivations 
also give insight into the nature of the solutions. 
Lagrangian duality plays a central role
in each one, justifying the name ``dual minimal-area program''. The first derivation is perhaps the most intuitive, the second is the shortest, and the last one highlights the relationship between the dual and the primal minimal-area program \eqref{secondprogram}.

\subsubsection{Maximin}
\label{sec:maximin}

In proving the MFMC theorem, we established the equality (\ref{min-cut=convex-min-cut}) of the min cut and convex min cut programs.  In our present context, where 
$M$ is two-dimensional, 
we write 
\begin{equation}\label{metaequiv}
\inf_{m\in C}\text{length}(m) = \inf_{\eta\in\tilde C}\int_M\omega|\eta|\,.
\end{equation}
Given the homological minimal area problem,
a given metric $\rho$ is feasible if and only if, for all $\alpha$, every representative $m_\alpha$ of the class $C_\alpha$ has length at least $\ell_\alpha$. Using \eqref{metaequiv}, 
this is equivalent to the condition that, for all $\alpha$ and all $\eta^\alpha\in\tilde C^\alpha$,
\begin{equation}\label{etabound}
\int_M\omega|\eta^\alpha| \ge \ell_\alpha\,.
\end{equation}
Making the dependence on $\rho$ explicit by substituting $\omega=\omega_0\rho^2$ and $|\eta^\alpha|=\rho^{-1}|\eta^\alpha|_0$, \eqref{etabound} becomes
\begin{equation}\label{etabound2}
\int_M\omega_0\, \rho\, |\eta^\alpha|_0 \ge \ell_\alpha\,.
\end{equation}
This provides an equivalent reformulation of the length conditions in the homological minimal area problem (\ref{firstprogram}).  

Any given $\eta^\alpha\in\tilde C^\alpha$ imposes, via \eqref{etabound2}, a constraint on $\rho$. Rather than considering all of these constraints at the same time, we will pick just {\em one}  $\eta^\alpha$ for each $\alpha$, and minimize the area $\int_M\omega_0\rho^2$ subject only to 
the corresponding \emph{finite} number of constraints. 
The resulting metric only gives a lower bound on the area, for \eqref{etabound2} may be violated for some other choice of $\eta^\alpha$s.  
However, we will argue that, 
among those choices,  the metric giving the \emph{greatest} lower bound in fact obeys all the constraints in \eqref{etabound2}. The proof of the last statement hinges crucially on the convexity of the space of $\eta^\alpha$s; it would not work if we tried to follow the same procedure using instead the homology representatives $m_\alpha$, which do not form a convex set.

Let us then fix an $\eta^\alpha$ from each class $\tilde C^\alpha$. Denote this set $\boldsymbol{\eta}=(\eta^\alpha)_{\alpha\in J}$. The following program minimizes the area over $\rho$ subject to the constraints \eqref{etabound2}:
\begin{equation}\label{rhominimize}
\begin{split}
&\text{Minimize }\ \int_M\omega_0\rho^2
\quad\text{over $\rho$ (function) } \\
& \hbox{subject to} \quad  \,
\ell_\alpha - \int_M\omega_0\rho|\eta^\alpha|_0\le0\quad
\forall \alpha \in J
\,.
\end{split}
\end{equation}
Let $\rho[\boldsymbol{\eta}]$ 
denote the minimizing metric  and 
$A[\boldsymbol{\eta}]$ its 
area. (According to the notation of subsection \ref{sec:convex}, we should write these as $\rho^*[\boldsymbol{\eta}]$ and $A^*[\boldsymbol{\eta}]$, but to avoid cluttering the notation we drop the stars.) We now dualize this program. 
Associated to the length constraints we introduce Lagrange multipliers $\nu^\alpha\ge 0$, assembled into a vector $\boldsymbol{\nu}$. The Lagrangian functional is
\begin{equation}
L= 2\sum_\alpha\nu^\alpha\ell_\alpha+\int_M\omega_0\Bigl(\rho^2-2\rho\sum_\alpha\nu^\alpha|\eta^\alpha|_0\Bigr).
\end{equation}
We can easily minimize $L$ pointwise with respect to $\rho$. The minimum is at
\begin{equation}\label{rhomin}
\rho = \sum_\alpha\nu^\alpha|\eta^\alpha|_0\,,
\end{equation}
and we are left with the following program:
\begin{equation}\label{numaximize}
\begin{split}
&\text{Maximize }\ g_{\boldsymbol{\eta}}[\,\boldsymbol{\nu}]:=2\sum_\alpha\nu^\alpha\ell_\alpha-\int_M\omega_0\Bigl(\, \sum_\alpha\nu^\alpha|\eta^\alpha|_0\Bigr)^2
\quad\text{over $\boldsymbol{\nu}$  (constants) } \\
& \hbox{subject to} \quad  \,
\nu^\alpha\ge0\,,  \quad\forall \alpha \in J 
\,.
\end{split}
\end{equation}
Strong duality holds for the convex program \eqref{rhominimize}: we can always make $\rho$ large enough that the constraint is strictly obeyed, so Slater's condition is satisfied. Let $\boldsymbol{\nu}[\boldsymbol{\eta}]$ be the maximizing value of $\boldsymbol{\nu}$ in the program \eqref{numaximize}. 
According to \eqref{dualmin}, the solution to \eqref{rhominimize} is given by \eqref{rhomin} with $\boldsymbol{\nu}$ set to~$\boldsymbol{\nu}[\boldsymbol{\eta}]$:
\begin{equation}\label{rhomin2}
\rho[\boldsymbol{\eta}] = \sum_\alpha\nu^\alpha[\boldsymbol{\eta}]\, |\eta^\alpha|_0\,.
\end{equation}

Let $A^*$ be the minimal area 
subject to the constraint \eqref{etabound2} 
for \emph{all} possible choices of $\boldsymbol{\eta}$;
this is the optimal value of the homology minimal-area program of section \ref{sec:homology}.  
As we discussed above,   
for any given 
$\boldsymbol{\eta}$ we have
\begin{equation}
A[\boldsymbol{\eta}] \le A^*\,,
\end{equation}
so
\begin{equation}\label{suplower}
\sup_{\boldsymbol{\eta}}A[\boldsymbol{\eta}]\le A^*\,.
\end{equation}
Let $\boldsymbol{\eta}^*$ be the maximizer of $A[\boldsymbol{\eta}]$. 
We will show below that the corresponding metric $\rho[\boldsymbol{\eta}^*]$ is feasible, i.e.\ obeys \eqref{etabound2} for all $\boldsymbol{\eta}$. 
Assuming this for the moment,  we now find 
\begin{equation}\label{supupper}
\sup_{\boldsymbol{\eta}}A[\boldsymbol{\eta}] = A[\boldsymbol{\eta}^*] \ge A^*\,,
\end{equation}
because any feasible metric must have area greater than or equal to the minimum. 
Combining the last two inequalities 
gives
\begin{equation}\label{suplower-vm}
\sup_{\boldsymbol{\eta}}A[\boldsymbol{\eta}]=A^*\,.
\end{equation}
This shows that the primal minimal-area 
program, having $A^*$ as optimum, is equivalent to the program (\ref{numaximize}) supplemented by
maximization over $\boldsymbol{\eta}$.   But such a program is
in fact \eqref{dualeta}, thereby proven equivalent to the 
homological  minimal-area program~(\ref{firstprogram}).

It remains to show that $\rho[\boldsymbol{\eta}^*]$ is feasible. We will proceed by contradiction, showing that any $\boldsymbol{\eta}$ for which $\rho[\boldsymbol{\eta}]$ is \emph{infeasible} does \emph{not} maximize $A[\boldsymbol{\eta}]$. Suppose that for some $\beta\in J$ and some $\hat\eta^\beta\in\tilde C^\beta$,  the length condition 
\eqref{etabound2} is violated:
\begin{equation}\label{etaboundviolate}
\int_M\omega_0\,
\rho[\boldsymbol{\eta}]  \,|\hat\eta^\beta|_0<\ell_\beta\,.
\end{equation}
We will construct an $\boldsymbol{\eta}'$ and a $\boldsymbol{\nu}'$ such that 
\be
\label{last-step}
g_{\boldsymbol{\eta}'}[\, \boldsymbol{\nu}']>g_{\boldsymbol{\eta}} [\, 
\boldsymbol{\nu}[\boldsymbol{\eta}]]\, ,
\ee
where $\boldsymbol{\nu}[\boldsymbol{\eta}]$ denotes the maximizing value of $\boldsymbol{\nu}$ in the program (\ref{numaximize}).
It will then follow that 
\begin{equation} 
A[\boldsymbol{\eta}'] = \sup_{\boldsymbol{\nu}}g_{\boldsymbol{\eta}'}[\, \boldsymbol{\nu}] \ge g_{\boldsymbol{\eta}'}[\, \boldsymbol{\nu}'] > g_{\boldsymbol{\eta}}[\, \boldsymbol{\nu}[\boldsymbol{\eta}]]=A[\boldsymbol{\eta}]\,,
\end{equation}
showing that $\boldsymbol{\eta}$ is not maximal, and completing the proof. 

It remains to construct an $\boldsymbol{\eta}'$ and a $\boldsymbol{\nu}'$ so that  (\ref{last-step}) holds.  The two cases $\nu^\beta[\boldsymbol{\eta}]>0$ and $\nu^\beta[\boldsymbol{\eta}]=0$ require separate analyses. For $\nu^\beta[\boldsymbol{\eta}]>0$, we know by complementary slackness (see (\ref{complementary-slackness}))   that the corresponding constraint in \eqref{rhominimize} is saturated, i.e.
\begin{equation}\label{etasaturate}
\int_M\omega_0\, \rho[\boldsymbol{\eta}]\, |\eta^\beta|_0 = \ell_\beta\,.
\end{equation}
We use the length-violating one-form $\hat \eta^\beta \in \tilde C$ to 
define $\boldsymbol{\eta}'$ and $\boldsymbol{\nu}'$ by
\begin{equation}
\eta'^{\alpha} = \eta^\alpha \quad\text{for }\alpha\neq\beta\,,\qquad
\eta'^{\beta} = (1-\epsilon)\eta^\beta+ \epsilon \hat\eta^\beta\,,\qquad\nu'^{\alpha} = \nu^\alpha[\boldsymbol{\eta}]\quad\text{for all }\alpha\,,
\end{equation}
where $0<\epsilon<1$. By the convexity of the norm 
we have 
\begin{equation}
|\eta'^{\beta}|_0 \le (1-\epsilon)|\eta^\beta|_0+\epsilon|\hat\eta^\beta|_0\,,
\end{equation}
and therefore  
\begin{equation}
\sum_\alpha   \nu'^{\alpha}|\eta'^{\alpha}|_0 \le 
\sum_\alpha\nu^\alpha[\boldsymbol{\eta}]\, |\eta^\alpha|_0  
+\epsilon\nu^\beta[\boldsymbol{\eta}]\left(|\hat\eta^\beta|_0-|\eta^\beta|_0\right).
\end{equation}
Squaring and integrating,  
\begin{equation*}
\begin{split}
\int_M\omega_0\Bigl(\sum_\alpha  \nu'^{\alpha}|
\eta'^{\alpha}|_0\Bigr)^2 
&\le
\int_M\omega_0  \Bigl( \sum_\alpha\nu^\alpha[\boldsymbol{\eta}]\, |\eta^\alpha|_0    \Bigr)^2   
+ 2\epsilon\nu^\beta[\boldsymbol{\eta}]\int_M\omega_0
\rho[\boldsymbol{\eta}]
\left(|\hat\eta^\beta|_0-|\eta^\beta|_0\right)+O(\epsilon)^2 \\
&< \int_M\omega_0\Bigl( \sum_\alpha\nu^\alpha[\boldsymbol{\eta}]\, |\eta^\alpha|_0    \Bigr)^2 \,,    
\end{split}
\end{equation*}
where in the second line we used \eqref{etaboundviolate} and \eqref{etasaturate} and we set $\epsilon$ to be small enough that the order $\epsilon^2$ term is negligible compared to the order $\epsilon$ term. 
Since $\nu'^\alpha = \nu^\alpha[\boldsymbol{\eta}]$,  
 it follows from the definition of $g_{\boldsymbol{\eta}}$ that $g_{\boldsymbol{\eta}'} [\,\boldsymbol{\nu}']>g_{\boldsymbol{\eta}}[\, \nu[\boldsymbol{\eta}]]$.

For the case $\nu^\beta[\boldsymbol{\eta}]=0$, we instead choose
\begin{equation}
\begin{split}  
\eta'^{\alpha} = \eta^\alpha \quad & \text{for }\alpha\neq\beta\,,\qquad
\eta'^{\beta} = \hat\eta^\beta\,, \\
\nu'^{\alpha} = \nu^\alpha[\boldsymbol{\eta}]\quad &  \text{for }\alpha\neq\beta\,,\qquad
\nu'^{\beta} = \epsilon\,,
\end{split}
\end{equation}
where $\epsilon>0$. We have
\begin{equation}
\sum_\alpha\nu^{\alpha\prime}|\eta^{\alpha\prime}|_0 = 
\sum_\alpha\nu^\alpha[\boldsymbol{\eta}]\, |\eta^\alpha|_0     
+ \epsilon|\hat\eta^\beta|_0\,.
\end{equation}
Expanding the definition of $g_{\boldsymbol{\eta}}$ gives
\begin{equation}
g_{\boldsymbol{\eta}'} [\, \boldsymbol{\nu}'] =
g_{\boldsymbol{\eta}}[\, \boldsymbol{\nu}[\boldsymbol{\eta}]]+2\epsilon\left(\ell_\beta - \int_M\omega_0\rho[\boldsymbol{\eta}] \, |\hat\eta^\beta|_0\right)-O(\epsilon)^2\,.
\end{equation}
Again, given \eqref{etaboundviolate}, for sufficiently small $\epsilon$, $g_{\boldsymbol{\eta}'}[\, \boldsymbol{\nu}']>g_{\boldsymbol{\eta}} [\,\nu[\boldsymbol{\eta}]]$.

\subsubsection{From homology MAP}
\label{sec:directdual}

The second derivation we present is the shortest one (although it contains a step that we will not fully justify). We start with the original homology minimal-area program written in terms of $\rho$ \eqref{firstprogramrho}, which we reproduce here:
\begin{equation}\label{firstprogramrho2}
\begin{split}
& \text{Minimize } \,  
\int_M\omega_0 \,\rho^2 \ \   
\ \ \text{over }\rho\ge0 \text{ (function)}\\
& \hbox{subject to} \ \ \ \ 
\, \ell_\alpha-\int_m\rho \, |\dot x |_{{}_0} \le0\,, \ \ 
  \forall\, m\in C_\alpha\,, \ \ \forall\,   \alpha \in J  \,  \,.  
  \end{split}
\end{equation}
We can formally combine the constraints from all the representatives of a given class $C_\alpha$ into one constraint:
\begin{equation}\label{alphaconstraint}
\ell_\alpha - \inf_{m\in C_\alpha}\int_m\rho\,|\dot x|_0\le0\,.
\end{equation}
Notice that the second term on the left-hand side, being the infimum over a set of concave (in fact linear) functionals of $\rho$, is itself a concave functional of $\rho$, so the left-hand side is indeed a convex functional. It will be useful momentarily to have the minimization here over a convex domain, so we appeal to the equivalence of the min cut and convex min cut programs \eqref{min-cut=convex-min-cut} to write \eqref{alphaconstraint} as
\begin{equation}\label{alphaconstraint2}
\ell_\alpha - \inf_{\eta^\alpha\in\tilde C^\alpha}\int_M\omega_0\,\rho|\eta^\alpha|_0\le0\,.
\end{equation}
 We now have
\begin{equation}\label{firstprogramrho3}
\begin{split}
& \text{Minimize } \,  
\int_M\omega_0 \,\rho^2 \ \   
\ \ \text{over }\rho \text{ (function)}\\
& \hbox{subject to} \ \ \ \ 
\ell_\alpha - \inf_{\eta^\alpha\in\tilde C^\alpha}\int_M\omega_0\,\rho|\eta^\alpha|_0\le0\,, \quad\forall\,   \alpha \in J\,.
  \end{split}
\end{equation}

We now dualize \eqref{firstprogramrho3}  
using Lagrange multipliers $\nu^\alpha\ge0$. As in the previous derivation, for sufficiently large $\rho$ the inequality constraints are strictly satisfied, so Slater's condition and therefore strong duality hold. The Lagrangian is
\begin{equation}\label{Lagrangian}
\int_M\omega_0\rho^2+2\sum_\alpha\nu^\alpha\left(\ell_\alpha - \inf_{\eta^\alpha\in\tilde C^\alpha}\int_M\omega_0\,\rho|\eta^\alpha|_0\right) =
2\sum_\alpha\nu^\alpha\ell_\alpha+\sup_{\boldsymbol{\eta}\in\mathbf{\tilde C}}\int_M\omega_0\left(\rho^2-2\rho\sum_\alpha\nu^\alpha|\eta^\alpha|_0\right),
\end{equation}
where, as in the previous derivation, $\boldsymbol{\eta}$ represents a set $(\eta^\alpha)_{\alpha\in J}$, and $\mathbf{\tilde C}$ represents the corresponding product of classes $\tilde C^\alpha$. We now 
minimize over $\rho$, focusing on the second term:
\begin{eqnarray}\label{directderiv}
\inf_\rho\sup_{\boldsymbol{\eta}\in\mathbf{\tilde C}}\int_M\omega_0\left(\rho^2-2\rho\sum_\alpha\nu^\alpha|\eta^\alpha|_0\right) \nonumber &=& 
\sup_{\boldsymbol{\eta}\in\mathbf{\tilde C}}\inf_\rho\int_M\omega_0\left(\rho^2-2\rho\sum_\alpha\nu^\alpha|\eta^\alpha|_0\right) \nonumber\\
&=& \sup_{\boldsymbol{\eta}\in\mathbf{\tilde C}}
\int_M   -  
\omega_0\left(\sum_\alpha\nu^\alpha|\eta^\alpha|_0\right)^2\,,
\end{eqnarray}
with the minimum over $\rho$ given by \eqref{rhomin}. Adding the first term in the Lagrangian, $2\sum_\alpha\nu^\alpha\ell_\alpha$, and maximizing over the $\nu^\alpha$s gives the dual program \eqref{dualeta}.

In the first equality of \eqref{directderiv}, we switched the minimization over $\rho$ and the maximization over $\boldsymbol{\eta}$. 
This is justified as long as the functional admits a saddle point, i.e.\ a configuration that is simultaneously a minimum with respect to $\rho$ and a maximum with respect to $\boldsymbol{\eta}$; in that case, convexity with respect to $\rho$ and concavity with respect to $\boldsymbol{\eta}$ guarantee that both the minimax and the maximin equal the functional's value at the saddle. We would indeed expect, but will not attempt to prove, that a saddle point exists. Alternatively, one could perhaps justify the exchange of the minimization and maximization by appealing to Sion's minimax theorem \cite{MR0097026}.

\subsubsection{From primal MAP}
\label{sec:dualfromprimal}

We will now derive the (version 2) dual program \eqref{seconddual} in a different way, by applying Lagrangian duality to the program \eqref{secondprogram}, the primal whose dynamical variables are the scale factor $\Omega$ of the metric and a set of calibrations $u^\alpha$. As reviewed in subsection~\ref{prim-dual-opt}, this involves introducing Lagrange multipliers to enforce the constraints and then solving for the original variables, leaving a convex program expressed in terms of the Lagrange multipliers. This justifies the name ``dual minimal-area program'' for \eqref{seconddual}.
We also note that one can dualize instead the program \eqref{secondprogramrho}, which is written in terms of $\rho$ rather than $\Omega$. The derivation is very similar and the resulting dual program is the same.

If the optimum of the dual program is equal to the optimum
of the primal we have \emph{strong duality}.  As reviewed
earlier, a sufficient condition for strong duality  is \emph{Slater's condition}, which requires the existence of a feasible point at which all the inequality constraints are strictly obeyed (i.e.\ none are saturated). This clearly applies to the program \eqref{secondprogram}, since closed one forms
$u^\alpha$ with correct periods can be found and then
$\Omega$ can be chosen arbitrarily large to make the 
inequalities strictly obeyed.

Before dualizing \eqref{secondprogram}, we 
 wish to write the period $\int_{m_\alpha}u^\alpha$ as an integral over the whole manifold, which we can do using the ``bump form'' $\eta_{m_\alpha}$ defined in \eqref{bumpform}:
\begin{equation}
\int_{m_\alpha}u^\alpha = \int_Mu^\alpha\wedge\eta_{m_\alpha}\,.
\end{equation}
We can in fact choose an arbitrary representative $\eta^\alpha_0$ of the Poincar\'e(-Lefschetz) dual cohomology class $\tilde C^\alpha$, and the same result will be obtained if $u$ is closed. 
 In terms of $\eta_0^\alpha$, the program \eqref{secondprogram} becomes
\begin{equation}\label{fourthprogram}
\begin{split}
&\text{Minimize }\, \ \int_M\omega_0\Omega
\ \ \ \text{ over $\Omega$ (function), $u^\alpha$ (one-forms)} \\
& \text{subject to:} \ \qquad  |u^\alpha|_0^2-\Omega\le0\,, \\
 & \, \hskip100pt   du^\alpha=0\,, \\
 & \hskip42pt \ell_\alpha - \int_Mu^\alpha\wedge \eta_0^\alpha=\, 0 \,,  
\ \forall \alpha \in J \,.  
\end{split}
\end{equation}

We now introduce three Lagrange multipliers to enforce the three constraints in \eqref{fourthprogram}: 
a function $\lambda^\alpha$ required to obey $\lambda^\alpha\ge0$ since it enforces the inequality constraint,
and a function $\varphi^\alpha$ and a constant $\nu^\alpha$ to enforce the equality
constraints:
\begin{equation}
\begin{split} 
\lambda^\alpha\ge0  \ & \ \ \text{for}  
\ \ |u^\alpha|^2_0 - \Omega\,  \le \,  0 \,,\\
\varphi^\alpha  \ & \ \ \text{for}  \ \ du^\alpha=0 \,, \\ 
\nu^\alpha  \ & \ \ \text{for}  
\ \ \ell_\alpha - \int_M u^\alpha\wedge\eta_0^\alpha = 0 \,,  \\[0.5ex] 
\end{split}
\end{equation}
Adding the Lagrange multiplier terms to the objective, 
and including some factors of 2 and signs for later convenience, we obtain the following Lagrangian functional:\begin{multline}
L  = \int_M\omega_0
\,\Bigl[\Omega+\sum_\alpha\lambda^\alpha(|u^\alpha|_0^2-\Omega)\Bigr]
-2\sum_\alpha\int_M\varphi^\alpha du^\alpha+2\sum_\alpha\nu^\alpha
\Bigl(\ell_\alpha-\int_M u^\alpha\wedge\eta_0^\alpha\Bigr)\,. 
\end{multline}
To find the dual objective, 
we now minimize $L$ with respect to the variables 
$\Omega$ and $u^\alpha$ of the primal. 
We start by integrating by parts and reorganizing terms:
\begin{equation} 
\label{btflvm} 
\begin{split} 
L = &  \  2\sum_\alpha\nu^\alpha \ell_\alpha 
+\int_M \omega_0\,
\Omega\Bigl(1-\sum_\alpha\lambda^\alpha\Bigr)\\[1.0ex]
& \  +\sum_\alpha \Bigl(\, 
 \int_M \omega_0
 \lambda^\alpha|u^\alpha|_0^2 -2
\int_M u^\alpha\wedge(\nu^\alpha \eta_0^\alpha+d\varphi^\alpha)
-2 \int_{\partial M}\varphi^\alpha u^\alpha \Bigr) \, .  
\end{split}
\end{equation}
With the derivatives off of $u^\alpha$, we can now minimize the functional pointwise. All the dependence on $u^\alpha$ is on the second line of \eqref{btflvm}. Using
(\ref{form-wedge-form}), the terms in the bulk integrand involving $u^\alpha$ can be written as
\begin{equation}
\label{Ludependent}
  \lambda^\alpha  \langle u^\alpha, u^\alpha\rangle_0 
  +    2 \langle  *   
(\nu^\alpha \eta_0^\alpha+d\varphi^\alpha ) \,, u^\alpha\rangle_0 
\end{equation}
where $*$ is the Hodge star with respect to any metric in the Weyl
class of  $g^0$.    
This has a minimum with respect to $u^\alpha$ as it 
is the sum of a quadratic
function of $u^\alpha$ with positive coefficient and a linear function of $u^\alpha$.
At the minimum,
\begin{equation}\label{usolution}
u^\alpha = \, - 
\frac1{\lambda^\alpha}
* (\nu^\alpha \eta_0^\alpha+d\varphi^\alpha)\,.
\end{equation}
The boundary integrand in \eqref{btflvm} is $\varphi^\alpha u^\alpha$; this has a minimum only if
\begin{equation}\label{phibc}
\varphi^\alpha|_{\partial M}=0\,,
\end{equation}
in which case it vanishes. 
This boundary condition is a constraint on the dual variable $\varphi^\alpha$.
With these results, the Lagrangian is now
\begin{equation} 
\label{btflvm9} 
\inf_{u^\alpha}L  =    2\sum_\alpha\nu^\alpha \ell_\alpha 
+\int_M\omega_0\,
\Omega\Bigl(1-\sum_\alpha\lambda^\alpha\Bigr)-\int_M \omega_0
\,\sum_\alpha\frac1{\lambda^\alpha}
\bigl|\nu^\alpha \eta_0^\alpha+d\varphi^\alpha\bigr|_0^2\,. 
\end{equation}
We still have to minimize with respect to $\Omega$, which appears linearly in $L$.  If we think of $\Omega$
as an unconstrained variable, the answer is clear:  if the factor multiplying $\Omega$
is non-zero the minimum does not exist. So the existence of a minimum requires
that this factor vanishes, leaving $\Omega$ undetermined:  
\begin{equation}
\sum_\alpha\lambda^\alpha=1\,.
\end{equation}
Using this we finally obtain the dual objective:
\begin{equation} 
\label{btflvm99} 
   2\sum_\alpha\nu^\alpha \ell_\alpha 
-\int_M\omega_0\,\sum_\alpha\frac1{\lambda^\alpha}
\bigl|\nu^\alpha \eta_0^\alpha+d\varphi^\alpha\bigr|_0^2 \,. 
\end{equation}
The dual program is therefore
\begin{equation}
\begin{split} 
\label{firstdual}
&\text{Maximize }\ 
\left[ \, 2\sum_\alpha \nu^\alpha \ell_\alpha
-\int_M\omega_0
\,\sum_\alpha\frac1{\lambda^\alpha}
\bigl|\nu^\alpha \eta_0^\alpha+d\varphi^\alpha\bigr|_0^2\ \right] \\
&\hbox{over}\quad  \lambda^\alpha\,, \, \varphi^\alpha\text{ (functions)},\, \nu^\alpha\text{ (constants)}\\
& \text{subject to:}  \qquad\quad \lambda^\alpha\ge \, 0\,,  \ \ 
 \\
& \hskip33pt  -1+\sum_\alpha\lambda^\alpha=0\,,  \\[-0.7ex]
&  \hskip65pt  \varphi^\alpha|_{\partial M}=0\,, \ \ \forall \alpha \in J\,.  
\end{split}
\end{equation}

Since there are no derivatives acting on $\lambda^\alpha$, we might as well go ahead and solve for it.   To maximize over $\lambda^\alpha \geq 0$, we only need to consider the second term in the objective since $\lambda^\alpha$ does not appear in the first.  
Since the second term is the sum of non-positive terms  we must {\em minimize} the second term with its sign flipped.
Since the integrand is manifestly positive, this amounts to solving, pointwise and for fixed $\alpha$, the following mini-program:
\begin{equation}
\begin{split}
&\text{Minimize }  
\sum_\alpha\frac{c_\alpha^2}{\lambda^\alpha} \ \
\text{ over } \ \ \lambda^\alpha \\
& \hbox{subject to:} \ \ \ \ \ \ \lambda^\alpha\ge0\,, \\
&  \hskip20pt  \  -1+\sum_\alpha\lambda^\alpha=0 \,.
\end{split}
\end{equation}
Here the $c_\alpha$'s are defined by
\begin{equation}\label{cdef}
c_\alpha := \bigl|\nu^\alpha \eta_0^\alpha+d\varphi^\alpha\bigr|_0\,.
\end{equation}
The $c_\alpha$'s and the $\lambda^\alpha$'s are functions on the surface, but for the mini-program which is defined pointwise, they are just some non-negative constants.   We can quickly solve this
program but a slightly more general version is useful:
\begin{equation}
\begin{split}
&\text{Minimize }  
\sum_\alpha\frac{c_\alpha^2}{\lambda^\alpha} \ \
\text{ over } \ \ \lambda^\alpha \\
& \hbox{subject to:} \ \ \ \ \ \ \lambda^\alpha\ge0\,, \\
&  \hskip15pt  \ -1+ \sum_\alpha\lambda^\alpha\,  \leq \, 0 \,.
\end{split}
\end{equation}
In here we replaced  the equality setting the sum of $\lambda^\alpha$'s equal to one
by an inequality.   In fact we will show that at the minimum the equality
holds. This modified program would
arise had we treated $\Omega$ as a variable
satisfying the implicit constraint $\Omega \geq 0$.  In this case the earlier minimization
over $\Omega$ in (\ref{btflvm}) is only possible when the coefficient multiplying 
$\Omega$ is positive or zero, i.e.
\begin{equation}
\sum_\alpha \lambda^\alpha \leq 1 \,. 
\end{equation}
Note that if this inequality is not saturated then 
we must have $\Omega =0$ at the minimum.  This
indeed suggests that the inequality must be saturated.  
It is now a short exercise to show that the minimum
of the objective occurs for
\begin{equation}\label{lambdaexercise}
\lambda^{\alpha*} = \frac{c_\alpha}{\sum_\beta c_\beta}  \,.
\end{equation}
For this value of $\lambda$ the objective is 
\begin{equation}
\sum_\alpha\frac{c_\alpha^2}{
\lambda^{\alpha*}} = \Bigl(\sum_\alpha c_\alpha\Bigr)^2 = \Bigl( \sum_\alpha
\bigl|\nu^\alpha \eta_0^\alpha+d\varphi^\alpha\bigr|_0\Bigr)^2 \,.
\end{equation}
Substituting this optimum back into the program \eqref{firstdual} gives \eqref{seconddual}.

\section{Simple examples}\label{sec:examples}

In order to gain some intuition for the primal and dual programs, in this section we apply them to two exactly solvable cases where the minimal-area metric is known, namely the cylinder (or annulus) and the torus.

\subsection{Cylinder}
\label{subsec:cyl}

An annulus, or  finite cylinder or ring domain, $R$
 is a planar connected Riemann surface with two homotopic disjoint
 boundary components, each of which is a circle.  
 A natural minimal-area problem asks for the
 conformal metric of least area such that all curves
 homotopic to the boundaries are longer than or equal to $\ell_s$.
 The solution of this problem is well-known and will be
 briefly explained below.  
 We will also solve the problem 
 using the primal program as well
as using the dual program.
 
 \begin{figure}[!ht]
\leavevmode
\begin{center}
\epsfysize=6.0cm
\epsfbox{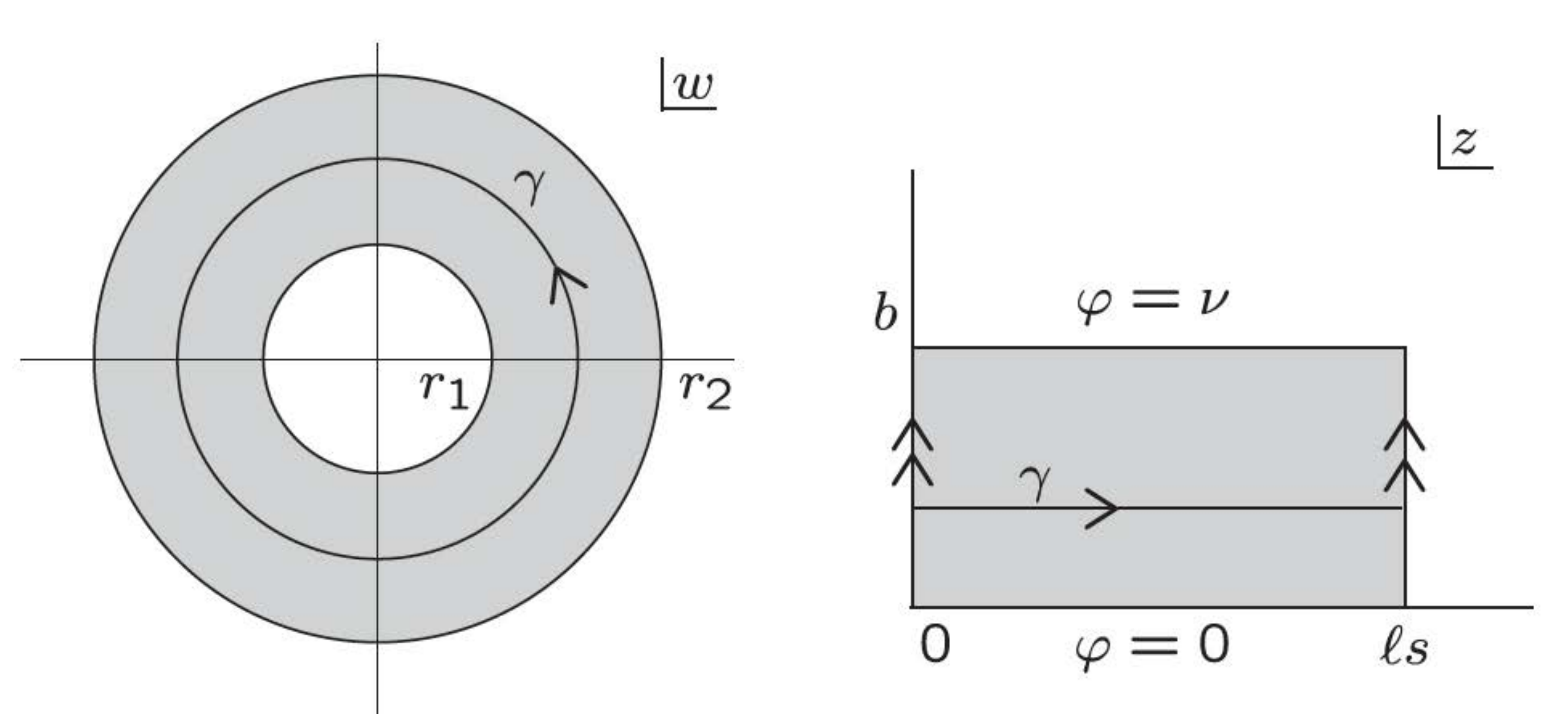}
\end{center}
\caption{\small  An annulus $r_1 \leq |w| \leq r_2$, with a core curve $\gamma$.
This annulus is mapped to a rectangle in the $z$ plane, with its vertical edges
identified.  The map $z =  {\ell_s\over 2\pi i}  \, \ln \,  {w\over r_2}$.}
\label{cYLA}
\end{figure}
 
 It is a familiar result 
 of conformal mappings that any annulus can be
 presented as a canonical annulus   $r_1 \leq |w| \leq r_2$
 on the complex $w$ plane, for
 some fixed value of  the ratio $r_2/r_1$, as shown in Figure~\ref{cYLA}.  This presentation
 is unique up to a constant scaling of $w$.
 The modulus $M$ of the annulus is defined by
 \be
 \label{M-annulus}
M \ \equiv  \ {1\over 2\pi}  \ln {r_2\over r_1} \,.
\ee
The $w$-plane annulus can be mapped to a $z$-plane rectangle with 
its vertical edges identified:
\be
z \ = \, {\ell_s\over 2\pi i}  \, \ln \,  {w\over r_2} \,  \,. 
\ee
With $z = x+ iy$,  the rectangle extends over $x\in [0, \ell_s]$ and
$y \in [0,b]$, where 
$b$ is
\be
b \ = \ {\ell_s \over 2\pi} \ln {r_2\over r_1}  \,. 
\ee
$b$ is 
the height of the cylinder form by the
identification of the vertical edges of the rectangle. 
In the rectangle picture the 
modulus $M$ is defined as the ratio of the height $b$ 
and  the circumference 
$\ell_s$:
\be
M \ = \ {b\over \ell_s} \,,
\ee
which on account of the expression for $b$ is manifestly equal to $M$
as given in (\ref{M-annulus}).  It is simplest to think of the annulus
as 
the identified rectangular region in the $z$ plane and we will do that in the following.

The minimal area problem would then state
that all closed curves beginning on the left vertical edge and ending 
on the right vertical edge should be longer than
or equal to $\ell_s$.  Those are the curves homologous to the
core curve $\gamma$ of the annulus, shown in Figure~\ref{cYLA}.  Let $A$ be the minimal area metric.  The metric
$ds= \rho |dz|$ with $\rho = 1$ is admissible and has area $\ell_s b$.
Therefore, 
\be
\label{eiu3}
A \leq  \ell_s b \,.
\ee
On the other hand, the constraint that any closed curve with fixed $y$ be long enough is
\be
\int_0^{\ell_s} dx  \rho (x, y)  \geq \ell_s\,.
\ee
Integrating this equation over $y$ to get a full two-dimensional
integral over the annulus $R$, 
\be
\label{dfer}
\int_R  dx dy \,  \rho  \ \geq  \ell_s b \,.
\ee
By Schwarz's inequality with $\rho = \rho \cdot 1$, 
the left-hand side above, which is manifestly positive, satisfies
\be
\int_R  dx dy\,   \rho  \,   \leq \  \sqrt{\int_R dx dy \rho^2 \int_R dx dy}  \ = \ 
\sqrt{ A(\rho)  \ell_s b} \,,
\ee
where $A(\rho)$ is the area of the metric $\rho$.  Squaring this relation
we get
\be
A(\rho)  \geq  {1\over \ell_s b} \Bigl( \int_R  dx dy  \rho\Bigr)^2 \geq  \ell_s b\,,
\ee
making use of (\ref{dfer}).  This implies that
\be
A  \geq \ell_s b \,.
\ee
Together with (\ref{eiu3}) this proves that $\rho=1$ is the extremal
metric and the extremal area is $\ell_s b$:
\be
\label{area-modulus}
A  = \, \ell_s b \, = \, \ell_s^2 \, M  \, = \, {b^2\over M}  \,. 
\ee

Let us now consider the solution via the primal program
in its version 2 (\ref{thirdprogram}).  
We have a single homology $C$ 
represented by  closed curves, like $\gamma$, stretching from
the left to the right boundaries of the identified rectangle.  Let us call the
corresponding calibration $u$.  This calibration takes the form
\be
u \ = \ dx  + d \phi\,.
\ee
Here the piece $dx$ is required for the calibration to have the right
period:  $\int_C u = \int_C dx = \ell_s$.   The function $\phi$ on the
surface generates the trivial part of the calibration.  We do not include
a $dy$ component to the calibration because the surface is symmetric
under reflections about a horizontal line $y = b/2$ and $dy$ would not
be invariant under such reflection while $u$ must be because it contains
$dx$.   The constraint $\Omega \geq |u|_0^2$ with fiducial metric $\rho_0=1$
gives 
\be
\Omega \geq  (1 + \partial_x \phi)^2  + (\partial_y \phi)^2  \,.
\ee
We can impose the symmetry $(x, y) \to (x + c , y)$ for any real constant
$c$, this is the rotational symmetry of the annulus.  Applied to our calibration
this requires that $\phi$ be constant along $x$.  The above condition
then gives 
\be
\Omega \geq  1  + (\partial_y \phi)^2  \,.
\ee
Since we are trying to minimize $\Omega$ pointwise, we find $\partial_y \phi=0$
and conclude that $\Omega =1$.  The minimum of the primal is therefore
$\int_R \Omega dx dy = \int_R dx dy = \ell_s b$, as found before.

Now consider the maximization of the dual objective in (\ref{thirddual}),  
called here  ${\cal F}$,  
applied to the  annulus $R$:
\be
{\cal F} \ = \ 2\nu \ell_s  - \int_R \omega_0\,
|d\varphi|_0^2 \,.
\ee
Here we take 
the fiducial metric 
to be the constant unit metric on the identified rectangle in the $z$ plane, so $\omega_0=dx\,dy$. We want to show that 
maximizing ${\cal F}$ over $\varphi$ and $\nu$ leads
to an optimum where ${\cal F}$ equals the previously determined minimal area. 

We fix the value $\varphi=0$ at the outer radius of the annulus, namely the
horizontal segment $x \in [0, \ell_s], y= 0$ (see Figure~\ref{cYLA}).  This is sensible as the dual requires
the value of $\varphi$ at a boundary to be zero.
 The value $\varphi= \nu$ is
fixed at the inner circle of the annulus or the horizontal 
segment  $x \in [0, \ell_s], y = b$.  
As explained before (see (\ref{boundaryrep})),   
this takes into
account the discontinuity requirement on~$\varphi$. 
Considering the rotational symmetry of the boundary conditions and fiducial metric, we could impose a rotational symmetry on $\varphi$; however, it turns out to be just as easy to solve for $\varphi$ without imposing this symmetry. Consider now the variation of $\varphi$ with these boundary conditions, 
taking $\nu$ to be fixed.  We then have
\be
\begin{split}
\delta {\cal F} \ = \ & \ 
- \delta \int_R \, d^2 x \,  \nabla \varphi\cdot \nabla \varphi 
\ = \  -  2 \int_R \, d^2 x \,  \nabla \delta \varphi\cdot \nabla \varphi \\
\ = \ & \ -  2 \int_{\partial R} \delta \varphi \nabla \varphi \cdot \hat n 
+   \ 2 \int_{R} \delta \varphi \nabla^2 \varphi \,.
  \end{split}
\ee
Since $\delta \varphi$ vanishes at the boundary, the equation that fixes
$\varphi$ is $\nabla^2 \varphi=0$.   The solution satisfying the boundary
conditions is thus unique and takes the form
\be
\varphi \ = \   \nu\,  {y\over b} 
\quad \to \quad  \nabla \varphi =  {\nu\over b} \, \hat y 
\quad \to \quad  |\nabla \varphi|^2 \ = \ {\nu^2 \over b^2}  \ = \ {\nu^2 \over
 \ell_s^2 M^2} \,,\ee
recalling that $b = M\ell_s$.
We now evaluate the objective, finding
\be
{\cal F} \ = \ 2\nu \ell_s  - {\nu^2 \over \ell_s^2 M^2} \int_R dx dy 
\ = \ \ 2\nu \ell_s  - {\nu^2 \over \ell_s^2 M^2} \ell_s b  \ = \ 2 \nu \ell_s 
- {\nu^2 \over M} \,.
\ee
We must now maximize over $\nu$.   The critical point is $\nu = \ell_s M$
and this gives, as desired
\be
{\cal F} \ = \ \ell_s^2 M  = A\,. 
\ee

\subsection{Torus}\label{sec:torus}

\begin{figure}[!ht]
\leavevmode
\begin{center}
\epsfysize=7.0cm
\epsfbox{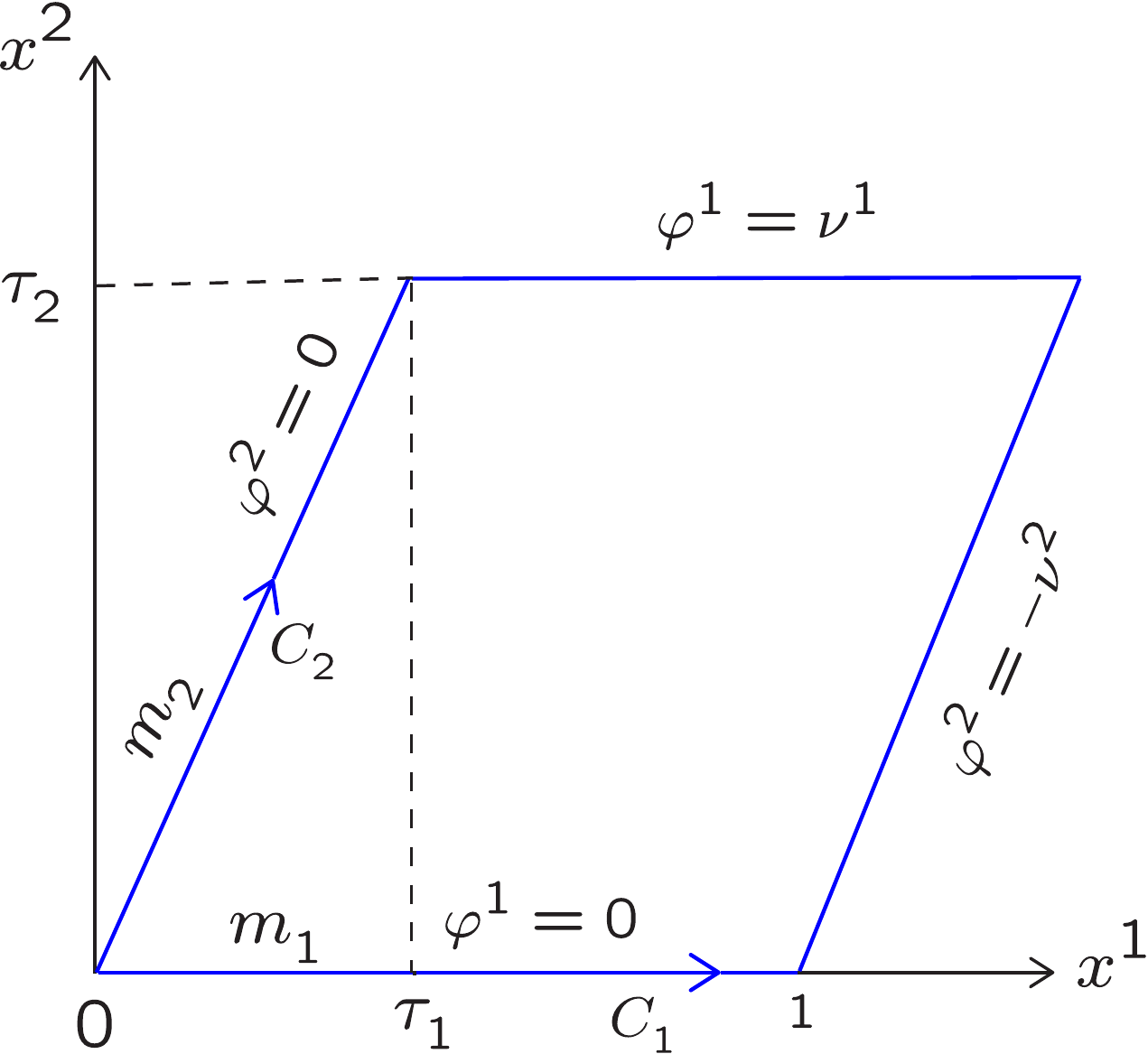}
\end{center}
\caption{\small  Torus with modular parameter $\tau = \tau_1 + i \tau_2$.  Indicated are the homology cycle $C_1$, along the
bottom side, and $C_2$ along the left side the parallelogram. 
In the presentation \eqref{varphis} of the dual configuration, the function $\varphi^1$ is discontinuous on the representative
$m_1$ (bottom or top side)
and the function $\varphi^2$ is discontinuous on the 
representative $m_2$ (left or right side).}
\label{ffsg1}
\end{figure}

We now consider the torus with length constraints on two intersecting cycles. We present the torus in the usual form, with modular parameter $\tau=\tau_1+i\tau_2$, coordinates $x^1,x^2$ with periodicities $(x^1,x^2)\sim (x^1+1,x^2)\sim(x^1+\tau_1,x_2+\tau_2)$, and constant 
fiducial metric $g^0_{\mu\nu} = \delta_{\mu\nu}$.   The torus
is shown in Figure~\ref{ffsg1}. 
Letting the cycles $C_1$ and $C_2$ be the bottom and left sides of the parallelogram-shaped fundamental domain of the torus, we require that all representatives of these two cycles have length at least 1. Note that this is not a modular-invariant condition, and does not imply that representatives of \emph{all} non-trivial cycles have length at least 1.

We start with the original form \eqref{firstprogram} of the 
minimal-area problem. The torus with the fiducial metric has a 2-parameter family of continuous isometries. As explained in subsection \ref{sec:convex}, we can impose this symmetry on the configuration, in other words we can set the Weyl factor $\Omega$ to be constant. This also follows from the fact that the minimal-area metric $\Omega^* g^0_{\mu\nu}$ is unique, as explained in section \ref{sec:isosystolic}.

The closed geodesics in the class $C_1$ are the curves of constant $x^2$, which have length $\sqrt\Omega$. The ones in the class $C_2$ are the curves of constant $\tau_1x^2-\tau_2x^1$, which have length $\sqrt{\Omega(\tau_1^2+\tau_2^2)}$.   The length conditions are therefore:
\be
\Omega \geq 1 \,, \quad \hbox{and} \quad \Omega \geq  {1\over 
\tau_1^2+\tau_2^2} \,.
\ee
The coordinate area is $\tau_2$, so the proper area is $\tau_2\Omega$. The minimal area $A_*$ arises from the minimum possible $\Omega$ satisfying the above constraints.  We therefore have
\begin{equation}\label{answer}
A_* \ = \ \tau_2 \max\left\{ 1 ,\frac{1}{\tau_1^2+\tau_2^2}\right\}\,.
\end{equation}
If the first argument of the max function is the largest, the curves saturating the length condition are in the class $C_1$; if the second argument of the max function is the largest, they are in the class $C_2$.

We now consider the program \eqref{thirdprogram} where we work with calibrations $u^1$ and $u^2$ written in terms of basis one-forms. Again, we can assume that these are invariant under the continuous isometries of the fiducial metric on the torus. A basis of closed one-forms is given by $dx^1$ and $dx^2$, which have the following periods:
\begin{equation}
\int_{C_1}dx^1 =1\,,\qquad
\int_{C_1}dx^2 =0\,,\qquad
\int_{C_2}dx^1 =\tau_1\,,\qquad
\int_{C_2}dx^2 =\tau_2\,.
\end{equation}
We now write $u^1$ and $u^2$ as general linear combinations
of $dx^1$ and $dx^2$ with unknown coefficients; invariance under the torus isometries implies the
scalars $\phi^1$ and $\phi^2$ are constant and can be dropped.  Applying
the constraints $\int_{C_1}u^1= \int_{C_2}u^2=1$, we find
\begin{equation}
u^1 = dx^1+c_1 dx^2\,,\qquad
u^2 = \frac1{\tau_2}dx^2+c_2(dx^1-\frac{\tau_1}{\tau_2}dx^2)\,,
\end{equation}
where $c_1$ and $c_2$ are to be determined. We then have
\begin{equation}
\Omega \geq |u^1|^2_0 = 1+c_1^2\,,\qquad
\Omega \geq |u^2|_0^2 = c_2^2+\frac{(1-c_2\tau_1)^2}{\tau_2^2}
\end{equation}
The minimum of $|u^1|_0^2$ over $c_1$ is clearly 1, while the minimum of $|u^2|_0^2$ over $c_2$ is easily computed to equal $1/(\tau_1^2+\tau_2^2)$. The minimum allowed value of $\Omega$ is the larger of these two, so the result agrees with \eqref{answer}. The tangent to the saturating curves is along the vector $\hat u$ of the calibration that saturates the inequality with the scale factor.

Finally, we solve the dual program. In the form \eqref{thirddual}, the program requires choosing representatives $m_\alpha$, which necessarily break the torus's continuous isometries. Therefore we will work instead with the dual in the form \eqref{dualeta}, where the variables are the one-forms $\eta^\alpha$. The requirement $\eta^\alpha\in\tilde C^\alpha$, which implies $\int_Mu\wedge\eta^\alpha=\int_{C_\alpha}u$ for any closed one-form $u$, together with invariance under the isometries, completely fixes the $\eta^\alpha$:
\begin{equation}
\eta^1 = \frac1{\tau_2}dx^2\,,\qquad
\eta^2 = -dx^1+\frac{\tau_1}{\tau_2}dx^2\,.
\end{equation}
Having obtained the $\eta^\alpha$, we can write them in the form of the program \eqref{thirddual}.  That is, we can write 
$\nu^\alpha\eta^\alpha 
= \nu^\alpha\eta_{m_\alpha}+d\varphi^\alpha$ 
for some representatives $m_\alpha$ 
and functions $\varphi^\alpha$. 
Choosing $m_1$ along the bottom (or top) 
edge of the parallelogram and $m_2$ along 
the left (or right) edge, the required functions $\varphi^\alpha$ are 
\begin{equation}\label{varphis}
\varphi^1 = \nu^1\frac{x^2}{\tau_2}\,,\qquad
\varphi^2 = \nu^2\left(-x^1+\frac{\tau_1}{\tau_2}x^2\right).
\end{equation}
These are functions on the fundamental domain of the torus, shown
in Figure~\ref{ffsg1}. 
On the full torus, they are discontinuous.

We now have 
\begin{equation}
\nu^1|\eta^1|_0  = {\nu^1\over \tau_2}\,, \quad
\nu^2|\eta^2|_0 =   {\nu^2\over \tau_2}  \sqrt{\tau_1^2 + \tau_2^2}\,,
\end{equation}
and, in either form of the dual program,  the following value for the objective:
\begin{equation}\label{torusdual}
\hbox{Objective} \ = \ 2(\nu^1+\nu^2)-\frac1{\tau_2}\left(\nu^1+\nu^2\sqrt{\tau_1^2+\tau_2^2}\right)^2\,.  
\end{equation}
We now need to maximize over $\nu^1$ and $\nu^2$. There are three possible cases for the maximum: 
\begin{enumerate}
\item $\nu^1>0 $ and $\nu^2>0$, 
\item $\nu^1> 0$ and $\nu^2=0$,  
\item$\nu^1=0$ and $\nu^2>0$. 
\end{enumerate}
The first case is ruled out by calculating the gradient of the
objective and showing that it doesn't vanish anywhere (except if $\tau_1^2+\tau_2^2=1$, in which case there is a line of critical points at $\nu^1+\nu^2=\tau_2$). 
The maximum on the $\nu^1>0$, $\nu^2=0$ half-line is at $\nu^1=\tau_2$ and has value $\tau_2$. The maximum on the $\nu^1=0$, $\nu^2>0$ half-line is at $\nu^2=\tau_2/(\tau_1^2+\tau_2^2)$ and has value $\tau_2/(\tau_1^2+\tau_2^2)$. The maximum of \eqref{torusdual} thus agrees with \eqref{answer},
consistent with strong duality. As we will see in subsection \ref{sec:saturating}, 
in the presentation \eqref{varphis}  the level sets for $\varphi^\alpha$ with non-zero $\nu^\alpha$
are saturating geodesics in the class~$C_\alpha$.

\sectiono{Properties of the minimal-area metric}\label{sec:optimal}

We now have two convex programs for the (homological)
minimal-area problem: a primal (\ref{secondprogram}), in which the variables
are the Weyl factor $\Omega$ of the metric and a 
set of 
one-forms $u^\alpha$,
and a dual \eqref{dualeta} or
 (\ref{thirddual}) where the variables are a set of one-forms $\eta^\alpha$ or functions
$\varphi^\alpha$ and a set of constants $\nu^\alpha$.  In this section, we will explore properties of the solutions of these programs, their relation to each other, and what they tell us about the geometry of the minimal-area metric. In subsection \ref{sec:saturating}, we explain how to extract the saturating geodesics in each homology class from primal and dual solutions. In subsection \ref{sec:primaldual}, we derive simple formulas relating the dual solution to the primal solution and to its total area, and in subsection \ref{sec:normal} we use them to study properties of bands of geodesics. In subsection \ref{sec:prop-sol-dual}, we get more intuition for the dual by understanding how it implements non-trivial properties that we know are obeyed by the primal solution. Finally, in \ref{sec:qd-and-dual}, we explain how the dual program can be understood as a generalization of a formula by Jenkins and Strebel for the minimal-area metric for the homotopy MAP problem \eqref{homotopy-MAP} where the constrained homotopy classes are represented by non-intersecting curves.

In subsections \ref{sec:saturating}--\ref{sec:normal} we consider only optimal field configurations, so to avoid cluttering the notation we drop the superscript $*$; thus $\Omega$ means $\Omega^*$, etc.

\subsection{Saturating geodesics}\label{sec:saturating}

Consider a representative of $C_\alpha$ that has length $\ell_\alpha$ and therefore saturates the length condition. Such a curve must
be locally length-minimizing, else there exist representatives that violate the length condition. Except where it touches a singularity or boundary of $M$, 
the curve must therefore be a closed geodesic (or a collection of closed
geodesics, if the curve has multiple components).
With a slight abuse of language we will 
call  a $C_\alpha$ representative of length $\ell_\alpha$ an  $\alpha$-\emph{geodesic}. In the original minimal-area problem, in which $\ell_\alpha=\ell_s$ for all $\alpha$, all $\alpha$-geodesics are in fact systolic geodesics.

As we will now show, the location of the $\alpha$-geodesics can be readily extracted from any solution to either the primal or the dual program.

We begin with the primal. An $\alpha$-geodesic is 
calibrated, in the metric $\Omega g^0$,  by the calibration form $u^\alpha$:
 the constraint $|u^\alpha|_0^2\le\Omega$ is saturated on that curve and furthermore the vector $\hat u^\alpha$ is tangent to the curve  (see (\ref{systole-u})). The converse is also true: Any closed integral curve of $\hat u^\alpha$ on which $|u^\alpha|_0^2=\Omega$ everywhere is an $\alpha$-geodesic. This allows one to easily identify the $\alpha$-geodesics from a solution to the primal program (\ref{secondprogram}). 
\begin{equation}\label{alphageoprimal}
\boxed{\phantom{\Biggl(} 
\text{
The $\alpha$-geodesics are the closed integral curves of $u^\alpha$ on which $|u^\alpha|^2_0=\Omega$ everywhere.
 } \ } 
\end{equation} 
We should point out that it is possible to have $|u^\alpha|_0^2=\Omega$ at a 
{\em point} that is not on an $\alpha$-geodesic.  To confirm the existence
of an $\alpha$ geodesic, the full integral curve of $u^\alpha$ must be examined. 

The $\alpha$-geodesics can also easily be extracted from the solution to the dual program, in any of the forms given in subsection \ref{sec:dualstate}. Specifically, we will show below that, given a solution $(\nu^\alpha,\eta^\alpha)$ to \eqref{dualeta}, if $\nu^\alpha>0$ for a given $\alpha$ then the level sets $m_{\eta^\alpha}(t)$ ($t\in\mathbb{R}/\mathbb{Z}$) defined in \eqref{metadef} are $\alpha$-geodesics. By construction, these are the curves orthogonal to $\eta^\alpha$ in the region of $M$ where $\eta^\alpha\neq0$. We can also write this in terms of the variable $\varphi^\alpha$ appearing in \eqref{thirddual}. 
Away from the fiducial representative $m_\alpha$ on which $\varphi^\alpha$ jumps, $\eta^\alpha=d\varphi^\alpha/\nu^\alpha$, so the level sets $m_{\eta^\alpha}(t)$ are the curves of constant $\varphi^\alpha$ in the region where $d\varphi^\alpha\neq0$:
\begin{equation}\label{alphageodual}
\boxed{\phantom{\Biggl(} 
\text{Wherever $d\varphi^\alpha\neq0$, the curves of constant $\varphi^\alpha$ are $\alpha$-geodesics.} \ } 
\end{equation} 
The converse to this statement does not quite always hold; it is possible for $d\varphi^\alpha$ to vanish at a point through which an $\alpha$-geodesic passes.\footnote{For example, if the $\alpha$-geodesics do not foliate the entire surface, then the boundary of the region they foliate is itself an $\alpha$-geodesic. 
But if $d\varphi^\alpha$ is continuous, then it vanishes on that boundary. 
This situation will be seen on the 
Swiss-cross/torus-with-a-boundary surface studied in \cite{headrick-zwiebach2}. A more extreme example is provided by the square torus $\tau_1=0, \tau_2= 1$, using the analysis of the dual problem in
section~\ref{sec:torus}.   In this case the dual objective is maximized
for $\nu^1+ \nu^2 = \tau_2= 1$.  We can choose to take $\nu^1=1$ and 
$\nu^2=0$, in which case $\varphi^2$ is constant and $d\varphi^2=0$.
Despite this vanishing gradient, the torus has vertical $2$-geodesics,
in addition to the horizontal $1$-geodesics.} However, we expect the converse to hold generically.

The statement that the level sets $m_{\eta^\alpha}(t)$ are $\alpha$-geodesics can be deduced from any of the three derivations of the dual program in subsection \ref{sec:dualderive}. For concreteness consider the derivation in subsection \ref{sec:maximin}. Let $(\nu^\alpha,\eta^\alpha)$ be a solution to the dual MAP \eqref{dualeta}. Apply complementary slackness to the duality relating \eqref{rhominimize} and \eqref{numaximize}. If, for some $\alpha$, $\nu^\alpha>0$, then the corresponding constraint in \eqref{rhominimize} is saturated, i.e.\
\begin{equation}\label{etasaturate2}
\int_M\omega_0\,\rho|\eta^\alpha|_0=\ell_\alpha\,,
\end{equation}
where $\rho$ is given by \eqref{rhomin}:
\begin{equation}\label{rhofrometa}
\rho = \sum_\alpha\nu^\alpha|\eta^\alpha|_0\,.
\end{equation}
As argued in the derivation, this is the solution to the homology MAP \eqref{firstprogramrho}. The left-hand side of \eqref{etasaturate2} equals $\int_M\omega|\eta^\alpha|$, which by the coarea formula \eqref{coarea} is the average length of the level sets $m_{\eta^\alpha}(t)$, so
\begin{equation}
\int_0^1dt\,\text{length}(m_{\eta^\alpha}(t)) = \ell_\alpha\,.
\end{equation}
Since $\rho$ is feasible, every representative of $C_\alpha$ has length at least $\ell_\alpha$. The only way the average can equal $\ell_\alpha$ is then if every level set has length equal to $\ell_\alpha$. So the level sets are indeed $\alpha$-geodesics. 

In the language of the Dual MAP v3, for a solution $(\nu^\alpha, \varphi^\alpha)$  we have:
\be
\rho = \sum_\alpha | d\varphi^\alpha|_0\,.
\ee

\subsection{The metric and the primal and dual solutions}
\label{sec:primaldual}

In the previous subsection, we noted that the solutions to the primal and dual MAPs tell us both the minimal-area metric and the location of the saturating geodesics. Here we will explore further these solutions and their relationship to each other.

Denote by $A$ the area of the minimal-area metric, which is the optimal value of the homology, primal, and dual MAPs. Given the solution $\rho$ to the homology MAP and $(\nu^\alpha,\eta^\alpha)$ to the dual MAP, we have
\begin{eqnarray}
\int_M\omega_0\rho^2 = A= 2\sum_\alpha\nu^\alpha\ell_\alpha -\int_M\omega_0\left(\sum_\alpha\nu^\alpha|\eta^\alpha|_0\right)^2\,.
\end{eqnarray}
However, from \eqref{rhofrometa}, the second term on the right-hand side is minus the left-hand side. This implies the following simple relation between the $\nu^\alpha$s and the area:
\begin{equation}
\label{area-heights}
\boxed{\phantom{\Biggl(} \ 
A = \sum_\alpha\nu^\alpha \ell_\alpha\,,
\ \ } 
\end{equation}
This result gives an intuitive interpretation for the value of the parameters $\nu^\alpha$
in the solution.  The total area for the extremal metric equals 
 the sum of areas of {\em flat rectangles} of height $\nu^\alpha$ 
 and length~$\ell_\alpha$.  We get a contribution from each
 band of $\alpha$-geodesics.  While these bands can cross and the
extremal metric is neither flat nor simple, the area is indeed
given by a simple formula, as if the surface were built with
flat rectangles.  We will rederive the result \eqref{area-heights} from a local point of view in the next subsection.

The relation \eqref{rhofrometa} is very useful. Dividing both sides by $\rho$, and noting that $|\cdot| = \rho^{-1}|\cdot|_0$, yields a simple ``sum rule'':
\begin{equation}\label{sumrule}
\sum_\alpha\nu^\alpha|\eta^\alpha| = 1\,.
\end{equation}
In terms of the variables of \eqref{thirddual},
\begin{equation}
\eta^\alpha = \eta^\alpha_{m_\alpha} + \frac1{\nu^\alpha}d\varphi^\alpha\,,
\end{equation}
with $m_\alpha$ the fiducial representative of $C_\alpha$ on which $\varphi^\alpha$ is required
to jump by~$ -\nu^\alpha$.  Away from the $m_\alpha$'s, \eqref{sumrule} thus becomes
\begin{equation}\label{solution2}
\boxed{\phantom{\Biggl(} \ 
\sum_\alpha|d\varphi^\alpha| = 1 \quad \text{at every point on $M$}\,. \ }
\end{equation}
This ``sum rule'' implies that at every point $d\varphi^\alpha$ must be non-zero for at least one $\alpha$, recovering the fact that at least one saturating curve passes through every point in $M$. 
For a given $\alpha$, the $\alpha$-geodesics foliate the region through which they pass. The fact that a region is foliated by such curves does not itself constrain the geometry, but \eqref{solution2} provides extra information. 

Finally, we use the derivation of the dual MAP in subsection \ref{sec:dualfromprimal} to relate the solutions of the primal and dual MAPs. Recall that \eqref{dualmin} says that the primal solution minimizes the Lagrangian function with the dual variables set to the dual solution. Using this fact, and combining \eqref{usolution}, \eqref{cdef}, and \eqref{lambdaexercise}, we find:
\begin{equation}
u^\alpha = \, -\,  
\frac{*(\nu^\alpha\eta^\alpha_0+d\varphi^\alpha)}{|\nu^\alpha\eta^\alpha_0+d\varphi^\alpha|_0}\sum_\beta|\nu^\beta\eta^\beta_0+d\varphi^\beta|_0  \,. 
\end{equation}
Using \eqref{rhofrometa}, we recognize the sum on the right-hand side as $\rho$, which we can use to replace the fiducial norm $|\cdot|_0$ in the denominator by $|\cdot|$:
\begin{equation}
u^\alpha = \, - \,  
\frac{*(\nu^\alpha\eta^\alpha_0+d\varphi^\alpha)}{|\nu^\alpha\eta^\alpha_0+d\varphi^\alpha|}\,.
\end{equation}
If we choose the fiducial one-form $\eta^\alpha_0$ to be the bump form $\eta_{m_\alpha}$, then away from $m_\alpha$ we have simply
\begin{equation}\label{primdual-simplified}
u^\alpha = \, - \, 
\frac{*d\varphi^\alpha}{|d\varphi^\alpha|}\,.
\end{equation}
This is consistent with \eqref{alphageoprimal} and \eqref{alphageodual}.   Equation \eqref{primdual-simplified} does not determine $u^\alpha$ where $d\varphi^\alpha=0$, that is, away from the $\alpha$-geodesics. We might have anticipated this since $u^\alpha$ is highly underdetermined away from these curves.   
The condition $du^\alpha=0$ is in fact 
the Euler-Lagrange equation for $\varphi^\alpha$ following from the dual objective in \eqref{thirddual}.

\subsection{Normal coordinates for bands of geodesics}\label{sec:normal}

\begin{figure}[!ht]
\leavevmode
\begin{center}
\epsfysize=6cm
\epsfbox{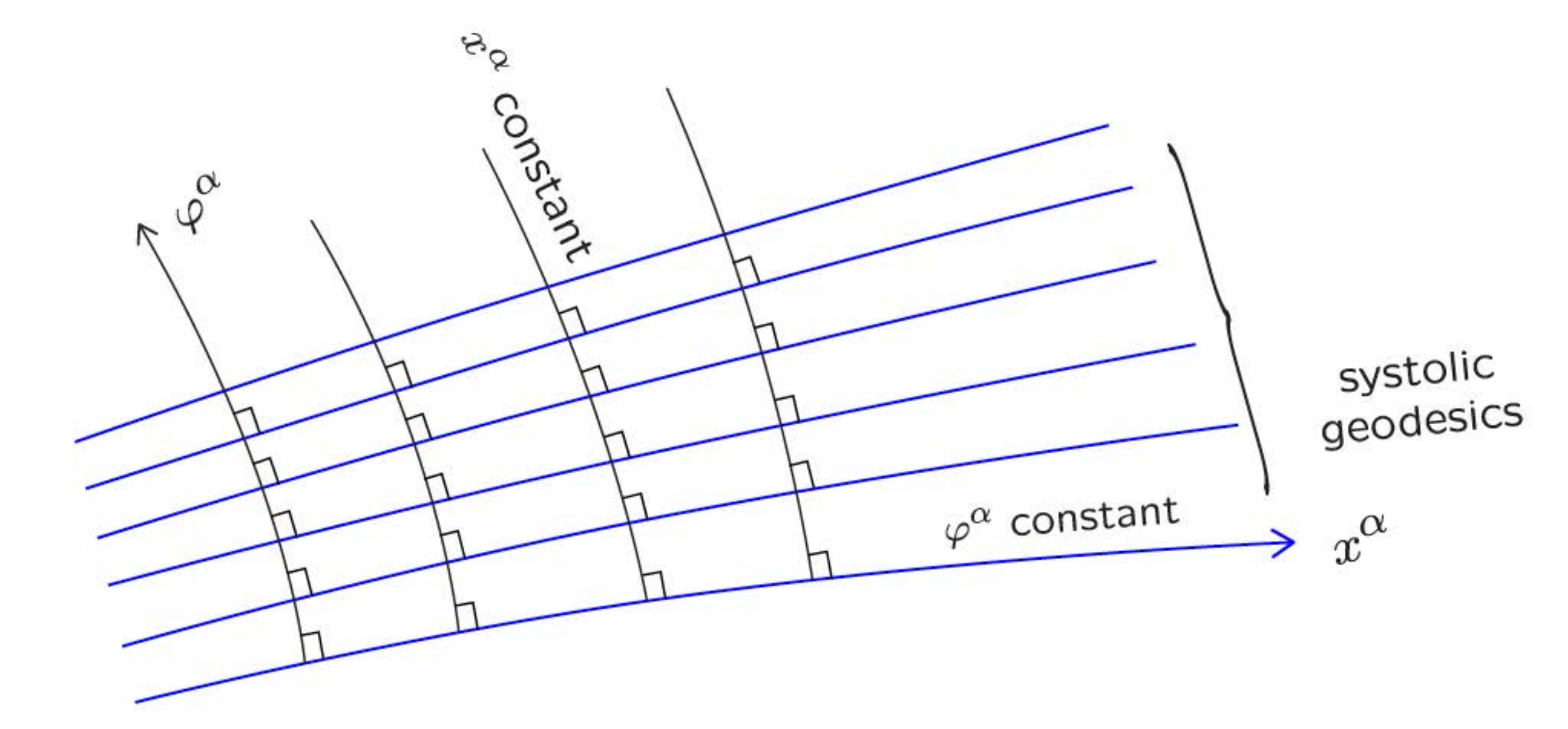}
\end{center}
\caption{\small  Normal coordinates associated with a collection
of $\alpha$-geodesics (in blue).  The coordinate $x^\alpha$ 
is a length parameter along the geodesics and the second coordinate $\varphi^\alpha$ runs normal to them.  The geodesics are curves of constant $\varphi^\alpha$ while the curves orthogonal to them are curves of constant~$x^\alpha$.}
\label{GNC}
\end{figure}

Fix a value of $\alpha$. As discussed in section \ref{sec:saturating}, $\varphi^\alpha$ is constant along any $\alpha$-geodesic and (generically) has non-zero gradient normal to it. 
We can therefore use $\varphi^\alpha$ as a local coordinate for the region foliated by a band of $\alpha$-geodesics. In fact, we can construct a Gaussian normal coordinate system using such a band. 
Consider the tangent vector $\hat u^\alpha$ to the geodesics
and the orthogonal vector field $\widehat {d\varphi}{}^\alpha$. Now use the orthogonal vector field to build integral curves orthogonal to the geodesics.  By
Gauss's lemma the distance along the geodesics
 between two orthogonal
integral curves is a constant.\footnote{As explained in~\cite{hicks} p.137:  If segments of equal length are laid off along geodesics orthogonal
to a univalent curve, their endpoints determine an orthogonal trajectory to the family of geodesics.  For our case, the univalent curve is any integral curve of $\widehat{d\varphi}{}^\alpha$.}  Thus there is a function $x^\alpha$ 
that provides a parameterization of the $\alpha$-geodesics
by length and is constant along the orthogonal integral curves.  
The function $x^\alpha$ will be our first coordinate.
The calibration $u^\alpha$ can be identified locally
with the exterior 
derivative of the function~$x^\alpha$:
\begin{equation}
 u^\alpha  \ = \ d x^\alpha \,.
\end{equation}
Since the period of $u^\alpha$ is $\ell_\alpha$, $x^\alpha$ ranges from 0 to $\ell_\alpha$. 
The second coordinate is $\varphi^\alpha$
which, as required, is constant along the systolic geodesics (see Figure~\ref{GNC}).  The form of the cotangent 
space metric $g^{-1}$  is constrained
 by the conditions $| d x^\alpha| =1$ (since $|u^\alpha|=1$)
and the orthogonality 
$\langle d x^\alpha , d\varphi^\alpha \rangle = 0$.   This determines
$g^{-1}$ up to one unknown function $h_\alpha$:
\begin{equation}
g^{-1} \ = \ \begin{pmatrix} 1 & 0 \\ 0 &\  h_\alpha^2 (x^\alpha,\varphi^\alpha)  \end{pmatrix}\, .
\end{equation}  

The interpretation of $h_\alpha$ arises from the length condition
\be
\label{h-explained}
 |d\varphi^\alpha| =  |h_\alpha| \,.
\ee
Given the above form of $g^{-1}$ the metric $g$ follows immediately and gives
\begin{equation}
\label{Gaussian}
\boxed{\phantom{\Biggl(} 
ds^2 = (dx^\alpha)^2+{1\over h_\alpha^2} (d\varphi^\alpha)^2\,. \ } 
\end{equation}
For this metric the Gaussian curvature $K$ can be calculated
using equation (\ref{gaussian-simple}) and the result is:
\begin{equation}
\label{gc-metric} 
K  \ = \  - {h_\alpha}  
\, {\partial^2 \ \over (\partial {x^\alpha})^2} \,  {1\over h_\alpha}\,.
\end{equation}
As a consistency check, the relation 
 (\ref{primdual-simplified})  between $u^\alpha$ and  
 $\varphi^\alpha$, 
 \begin{equation}\label{primdual-simplifiede}
u^\alpha =  \, - \,  \frac{*d\varphi^\alpha}{|d\varphi^\alpha|} \,,   
\end{equation}
is  satisfied in our construction.  Indeed, a short calculation
using (\ref{hodge-dual}) gives  
$*d\varphi^\alpha = - |h_\alpha| dx^\alpha$   and, given that $|d\varphi^\alpha| = |h_\alpha|$, the equality holds. 

The expression (\ref{Gaussian}) 
holds at a point $P$ for {\em each} class $C_\alpha$ 
of systolic geodesics that goes through $P$, and for each class one has 
$|d\varphi^\alpha | = |h_\alpha|$.  Using these coordinates, our earlier
constraint \eqref{solution2} on the sum of $d\varphi$ norms constrains the metric components in a simple way:
\begin{equation}
\label{solution3}
\boxed{\phantom{\Biggl(}   
\sum_{\alpha:\, d\varphi^\alpha\neq0} | h_\alpha| \,  = \, 1\,.   \ \ }  
\end{equation}
This is a nontrivial constraint on the extremal metric.

The function $h_\alpha$ captures the {\em density} $\rho_\alpha$ of  $\alpha$-geodesics.  Consider two geodesics corresponding to $\varphi^\alpha=0$ and $\varphi^\alpha = \epsilon$,
with $\epsilon$ infinitesimal.   The distance between them is $\epsilon/|h_\alpha| $.  We can use an arbitrarily small $\epsilon$ as
a fixed $\varphi^\alpha$ interval that allows us to pick geodesics from
the continuum and enumerate them.  
In particular with $\varphi^\alpha\in [0,\nu^\alpha]$
we get a total of $\nu^\alpha/\epsilon$ $\alpha$-geodesics. 
It follows  that the
density $\rho_\alpha$ of  $\alpha$-geodesics, defined as the number of 
geodesics per unit transverse length, is given~by 
\begin{equation}
\rho_\alpha =  \hbox{density of $\alpha$-geodesics}  \ = \ {1\over  \epsilon /|h_\alpha|} 
\ = \ {1\over \epsilon } \, |h_\alpha| \,.\end{equation} 
The constraint (\ref{solution3}) now translates into 
\begin{equation}\label{density}
\sum_\alpha  \rho_\alpha \ = \ {1\over \epsilon}  \,, 
\end{equation}
meaning that at any point on the surface 
the sum of geodesic densities over all the classes that go through that point
is the same.  

\begin{figure}[!ht]
\leavevmode
\begin{center}
\epsfysize=7.0cm
\epsfbox{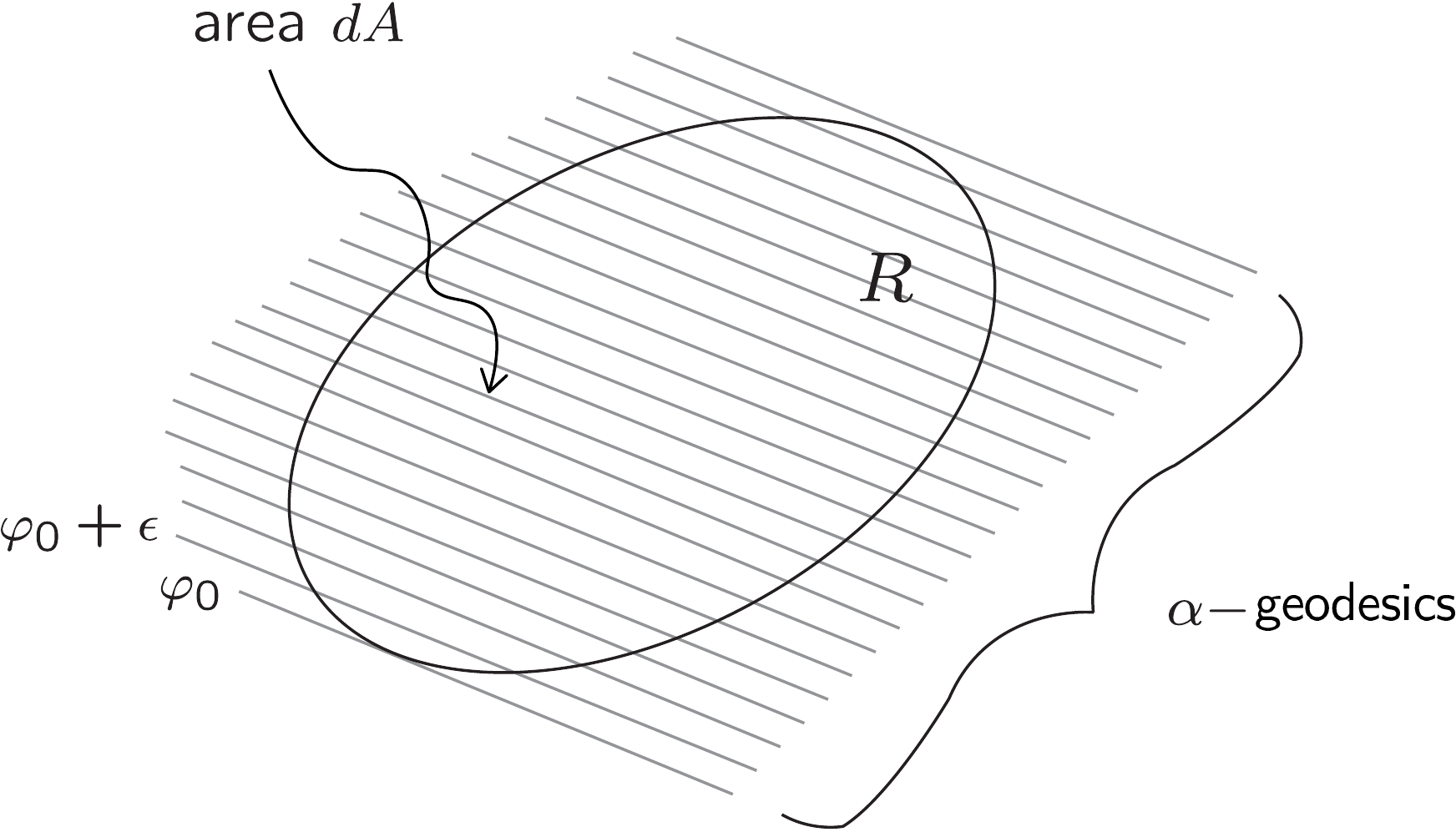}
\end{center}
\caption{\small  A set of $\alpha$-geodesics going through a tiny region $R$ of area $dA$. The geodesics are lines of constant $\varphi^\alpha$ values that
differ by $\epsilon$.  The area $dA$ can be expressed in terms of the length
of all $\alpha$-geodesic segments going through the region.}
\label{fig:lengths}
\end{figure}

We now explain that Eq.\ \eqref{density} stating that the sum of densities
$\rho_\alpha$ is a constant over the surface is  the  local version of the formula \eqref{area-heights} giving the area of the extremal metric as a sum of products   $\nu^\alpha \ell_\alpha$. Indeed, we can relate the two formulas by computing the total length of all the (denumerable) $\alpha$-geodesics on the surface in two different ways. Consider first a small region $R$ of area $dA$, small enough that all metric components are accurately constant over $R$.  By picking $\epsilon$ sufficiently small we can have a large number of $\alpha$-geodesics going through
$R$.  The area $dA$ is given by the {\em total} length $dL_\alpha (R)$ of the $\alpha$-geodesic segments within the region $R$, times the separation $\epsilon/|h_\alpha|$ between the geodesics (see figure \ref{fig:lengths}):
 \be
 \label{dA-geod}
dA =  dL_\alpha (R)  {\epsilon\over |h_\alpha|} \quad \Rightarrow \quad
dL_\alpha (R)  \ = \ \rho_\alpha \,  dA\,.    
\ee
 Let $dL(R)\equiv \sum_\alpha dL_\alpha (R)$  
 denote the total length
 of all geodesics in the small region $R$.   Summing over $\alpha$ in the above 
 equation and using \eqref{density} we find
 \be
 \label{dA-geode}
dL (R)  \ = \ {1\over \epsilon}\,  dA\,.   
\ee
 The total length of the geodesics contained within a region 
 depends only on its area! 
 Integrating now over the whole surface $M$, we find that the total length $L(M)$ of all the geodesics is the total area of the surface divided by $\epsilon$:
 \be
 L(M)  =  {1\over \epsilon} A \,. 
 \ee
  On the other hand, the number of $\alpha$-geodesics is $\nu^\alpha/\epsilon$, since $\varphi^\alpha$ changes monotonically by $\nu^\alpha$ over the full set of $\alpha$-geodesics. Each $\alpha$-geodesic has length $\ell_\alpha$, so their total length is $\ell_\alpha\nu^\alpha/\epsilon$. Summing over $\alpha$:
  \be
  L(M) = \sum_\alpha  \ell_\alpha\nu^\alpha/\epsilon  = {1\over \epsilon} \sum_\alpha \ell_\alpha\nu^\alpha  \,.  
  \ee
Equating the right-hand sides of the two equations above
we recover the area formula \eqref{area-heights}.

Consider now the case when there is only a single set of systolic 
geodesics going through the points in a given region of the surface $M$. Then the constraint \eqref{solution3}, having just one term,  fixes $h_\alpha=1$.  The metric is thus completely determined, and in fact is flat:
\begin{equation}\label{singleband}
ds^2 = (dx^\alpha)^2+(d\varphi^\alpha)^2\,.
\end{equation}
This is a simple derivation of a previously known result that had rather
intricate proofs~\cite{Ranganathan:1991qd,Wolf:1992bk}.
In a region containing more than one set of systoles, the sum rule \eqref{solution3} still provides a strong constraint on the geometry, but it is less obvious what the general solution is.

Suppose we have exactly 
two types of systolic geodesics, arising from calibrations
$u^\alpha$ and $u^\beta$ ($\beta\neq\alpha$), going through the points in 
a given region of the surface.  We then have
\begin{equation}
ds^2  \ =  \  (dx^\alpha)^2+{1\over h_\alpha^2} (d\varphi^\alpha)^2
\,, \qquad \hbox{and} \qquad 
ds^2 \ =  \  (dx^\beta)^2+{1\over h_\beta^2} (d\varphi^\beta)^2 \,. 
\end{equation}
We also have sum rule constraint:
\be
\label{2band}
|h_\alpha| + | h_\beta | \, = \, 1 \,. 
\ee
Another coordinate system gives additional insight. 
Using coordinates $x^\alpha$ and $x^\beta$ and recalling that
$|dx^\alpha | = |dx^\beta| =1$ the inverse metric takes the form
\begin{equation}
g^{-1} = \begin{pmatrix} 1 & f \\ f & 1 \end{pmatrix} 
\,,  \quad \  f \ = \ \langle dx^\alpha, dx^\beta\rangle \ = \ 
\langle \hat u^\alpha , \hat u^\beta \rangle   \,.
\end{equation}
Here we learn something interesting.  If the angle between the $\alpha$ and 
$\beta$ systoles is a constant, $g^{-1}$ is a constant matrix and
 the metric is flat.   In $(x^\alpha, x^\beta)$ coordinates the metric reads 
\begin{equation}
ds^2 =  {1\over 1 - f^2}  \bigl[\,   (dx^\alpha)^2 - 2 f \, 
dx^\alpha dx^\beta  + (dx^\beta)^2\,  \bigr]\,. 
\end{equation}
In this oblique coordinate system, the vector ${\partial \over \partial x^\alpha}$ is
orthogonal to the $\beta$ systoles and the vector ${\partial \over \partial x^\beta}$ is
orthogonal to the $\alpha$ systoles.
We have not been able to `solve' the constraint (\ref{2band})
to find a parameterization of the allowed metrics in a 
two-band region.   Our results in~\cite{headrick-zwiebach2}  indicate that both
positive and negative Gaussian curvature are possible.

We conclude this section with some remarks on
{\em geodesic bands}, namely bands
formed by a continuous collection of $\alpha$-geodesics for some $\alpha$.  Since
these geodesics are closed curves of constant $\varphi^\alpha$,  
a geodesic band can be described as
 collection of homotopic  geodesics in some range
$\varphi^\alpha \in [a, b]$.  A geodesic band is thus 
an annulus whose two boundaries are geodesics.  An annulus
has Euler number $\chi= 2 -b= 0$, since the number $b$ 
of boundaries is equal to two.  Since the boundary curves 
of the geodesic band are geodesics, the Gauss-Bonnet formula does not receive contributions from the boundary, and we conclude that 

\medskip 

\centerline{
 {\em The integral 
of the Gaussian curvature over any geodesic
band vanishes.  }}

\noindent
We can readily show that our earlier results confirm 
this explicitly.   Over a geodesic band ${\cal B}$ 
with $\varphi^\alpha \in [a,b]$
the metric takes the form (\ref{Gaussian}) with Gaussian curvature (\ref{gc-metric}).  We therefore have
\begin{equation}
\int_{\cal B}  K \sqrt{g} \, d^2 x  \ = \ \int_a^b d\varphi^\alpha\int_0^{\ell_s} dx^\alpha 
\, {1\over h_\alpha } \, K \,  \ = \ - \int_a^b d\varphi^\alpha
\int_0^{\ell_s} dx^\alpha 
 {\partial^2  \ \over \partial {x^\alpha}^2} {1\over h_\alpha} \ = \ 0 \,,  
\end{equation}
since the geodesics are closed curves and $h_\alpha$ is single-valued.

For a band that does not intersect any other bands, the metric is flat (see \eqref{singleband}), so the above claim is trivially satisfied. For example, on a genus $g$ Riemann surface where the minimal-area metric arises from a quadratic differential, there is
a single systolic geodesic almost everywhere on the surface, so the metric is flat almost everywhere. 
The negative
curvature needed from the topology is given by delta functions at the zeroes of the quadratic differential.  In general, however, the metric will not arise from a quadratic differential and  we expect regions
with one systolic band, regions with two systolic bands, and so on and so forth.  There may be curvature over regions with two or more bands, but then on each band there
must be both positive and negative curvature so that integrated
curvature vanishes. 

Focusing on a single band $R_\alpha$ of geodesics in the class $C_\alpha$ we
can use the metric (\ref{Gaussian}) to define the height $b^\alpha(x)$ 
of the band as a function of the position $x^\alpha$ along the band:
\be
b^\alpha (x)  \equiv \int_0^{\nu^\alpha}  {d\varphi^\alpha \over |h_\alpha (x , \varphi) |}\,. 
\ee
The height $b_\alpha (x)$ is
the {\em distance} between the boundary geodesics of the band at $x^\alpha$
measured along the segment $I(x^\alpha) = (x^\alpha, \varphi^\alpha\in [0, \nu^\alpha])$ 
perpendicular  to the geodesics at $x^\alpha$.
Assume now there is a value $x_0^\alpha$ of $x^\alpha$ for 
which $I(x_0^\alpha) \subset U_1$, that is it lies the region
where $R_\alpha$
is the {\em only} systolic band. Then $|h_\alpha | = 1$ along this segment~and 
\be
\label{identify-nu-height}
b_\alpha (x_0^\alpha ) =  \nu^\alpha \,.
\ee
This gives simple characterization of $\nu^\alpha$: it is the height
of the band at a region where it is the only band on the surface.  It is also clear that due to the sum rule (\ref{solution3}) in general we have
\be
\label{identify-nu-height}
b_\alpha (x ) \geq   \nu_\alpha \,.
\ee
If we call $A_\alpha$ the area of the band $R_\alpha$ we have 
\be
A_\alpha =  \int_0^{\ell_{\alpha}} \hskip-5pt \int_0^{\nu^\alpha}  {dx^\alpha d\varphi^\alpha\over |h_\alpha|}  \geq  \nu^\alpha \ell_\alpha \,. 
\ee
This provides a lower bound for $A_\alpha$.  
Let $M_\alpha$ denote the modulus of the band, or ring domain, $R_\alpha$.
As noted in~(\ref{area-modulus}), the minimal area of a ring domain whose
core curves are longer than or equal to $\ell$ is $\ell^2 M$, where $M$ is
the modulus of the ring domain. 
This gives us a second lower bound for $A_\alpha$:
\be
A_\alpha  \geq  \, \ell_\alpha^2  M_\alpha  \,. 
\ee
It would be interesting to see how the two lower 
bounds above compare. In other words: How does 
$M_\alpha$ compare with the ratio $\nu^\alpha/\ell_\alpha$?

\subsection{Properties of the solution from the dual}
\label{sec:prop-sol-dual}

Our aim in this subsection is to get more intuition for the dual program by showing how it enforces certain properties of the solution. These properties, having to do with interactions between bands of level sets, can also be understood from the primal.

Unlike in subsections \ref{sec:saturating}--\ref{sec:normal}, here a configuration $(\nu^\alpha,\varphi^\alpha)$ is not necessarily a solution.

\subsubsection{Segregation of level sets}

As we've seen, on a solution the level sets for the function $\varphi^\alpha$ are the $\alpha$-geodesics. 
The first lemma we prove essentially says that geodesic bands  
do not overlap unless they are forced to do so topologically.

\begin{figure}[!ht]
\leavevmode
\begin{center}
\epsfysize=4cm
\epsfbox{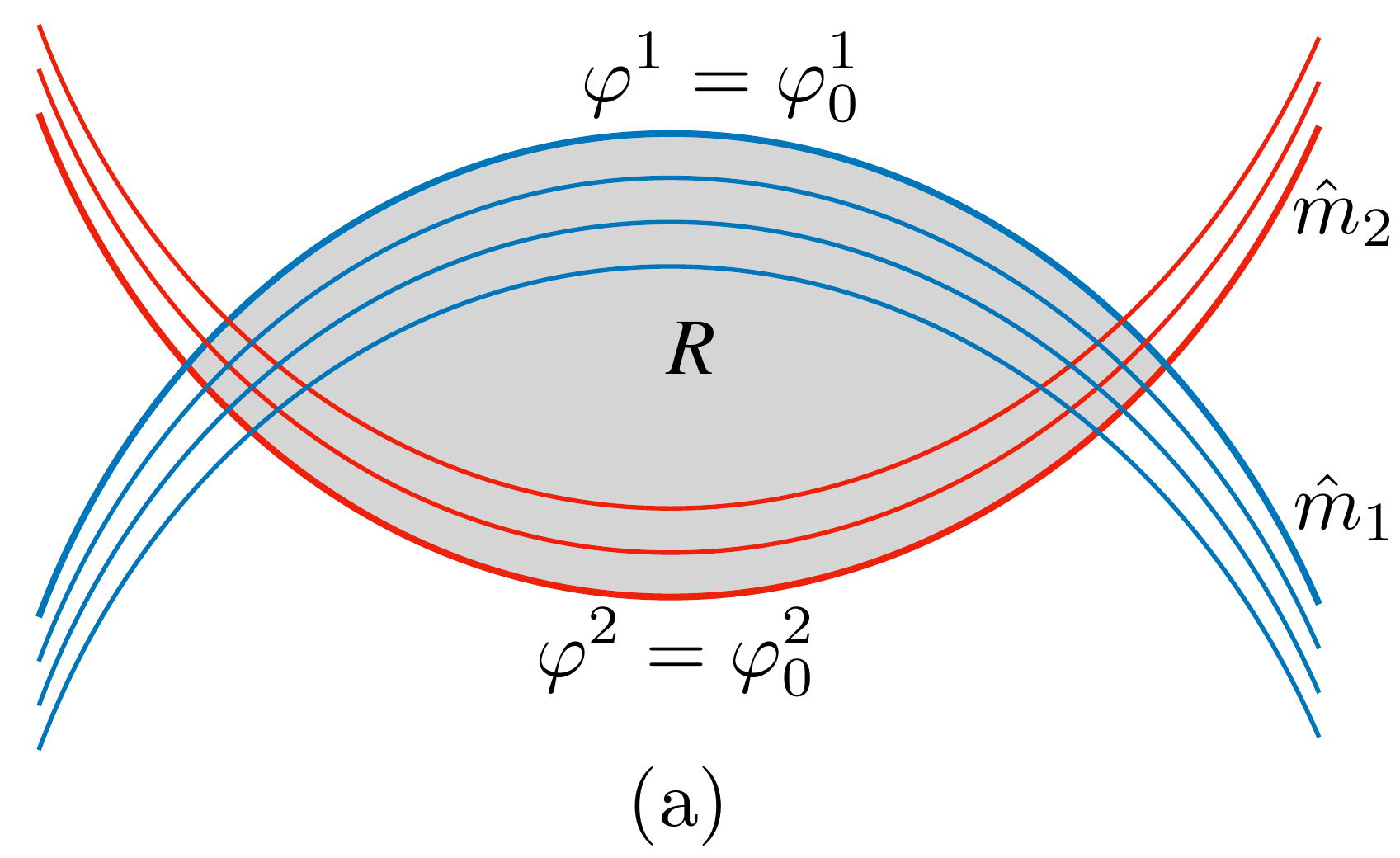}
\hspace{2cm}
\epsfysize=4cm
\epsfbox{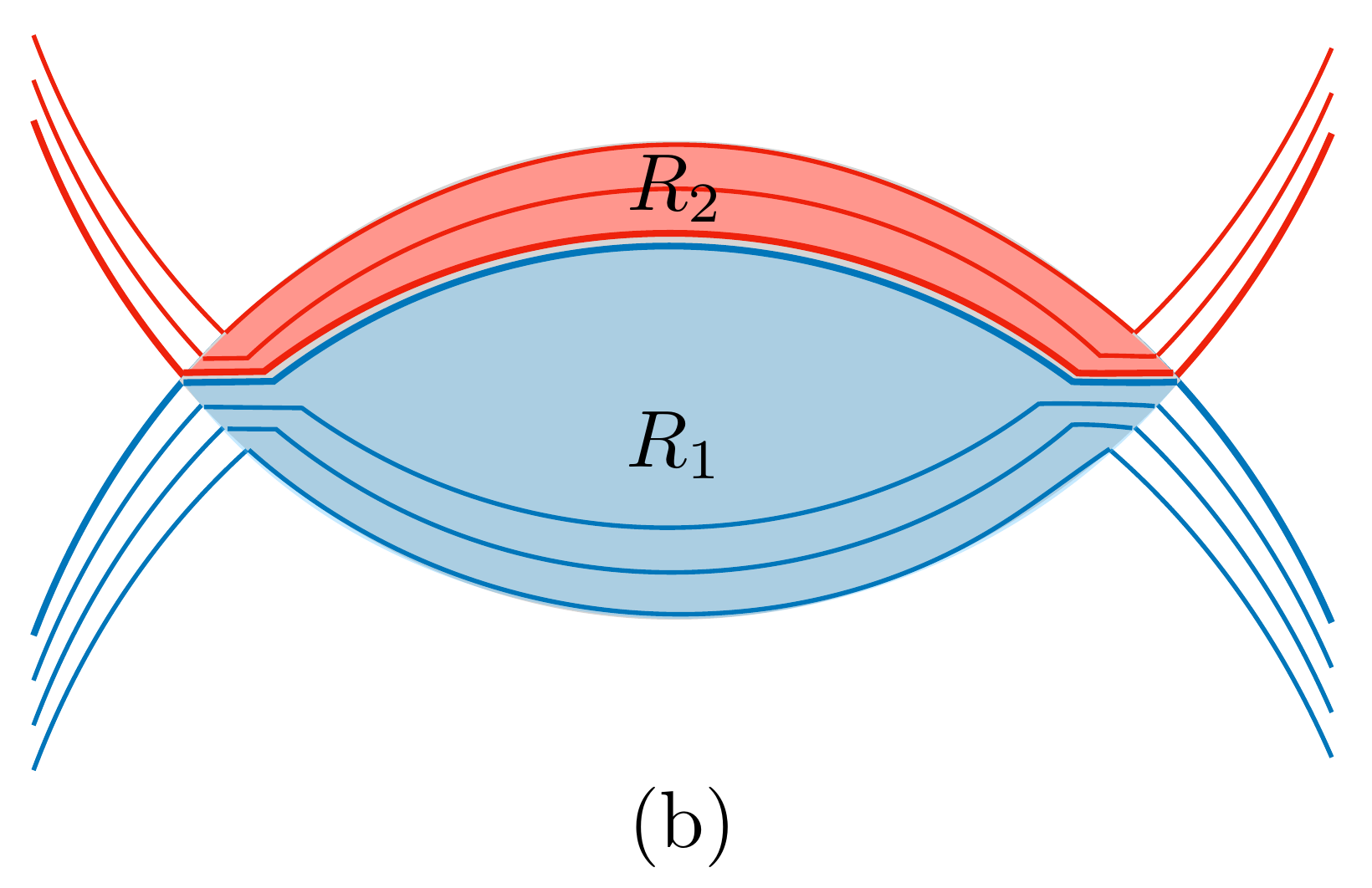}
\end{center}
\caption{\small
Illustration of the segregation lemma. 
 (a) The region $R$ (gray) is bounded by the $\varphi^1$ level set $\hat m_1$ on which $\varphi^1=\varphi^1_0$ and the $\varphi^2$ level set $\hat m_2$ on which $\varphi^2=\varphi^2_0$, and contains overlapping bands of level sets. The blue curves are $\varphi^1$ level sets and the red curves are $\varphi^2$ level sets. (b) The level sets of the new configuration $\varphi^{1,2\prime}$ and regions $R_1$ (light blue) and $R_2$ (pink). 
 On $R_2$, $\varphi^{1\prime}(x) =\varphi^1_0$, so there are no $\varphi^{1\prime}$ level sets in $R_2$, and similarly with 1 and 2 switched. Note how the blue and red 
level sets that cross in (a) segregate and do not cross in~(b).} 
\label{fig:segregation}
\end{figure}

Given a configuration in which a band of level sets of $\varphi^1$ and $\varphi^2$ overlap within some region $R$ bounded by a level set of $\varphi^1$ and a level set of $\varphi^2$ (see Figure \ref{fig:segregation}(a)), we will show that there exists a configuration which is the same outside of $R$, but in which they don't overlap in $R$, and in which the dual objective,
\begin{equation}\label{dualobjective}
2\sum_\alpha\nu^\alpha\ell_\alpha-\int_M\omega_0
\Bigl(\sum_\alpha|d\varphi^\alpha|_0\Bigr)^2 \,, 
\end{equation}
is at least as large as in the original configuration (see Figure \ref{fig:segregation}(b)). (This holds even if level sets of other $\varphi^\alpha$s also pass through $R$.) Therefore, in solving the dual program, we can restrict ourselves to configurations without such overlaps, 
in other words  configurations where the level sets segregate.  
(However, it should be noted that imposing this constraint makes the dual program non-convex.) Bands of geodesics can, however, meet along a mutual boundary, which coincides with the last geodesic of each band.

\paragraph{Lemma 1 (Segregation).}  
 We consider the dual program in the form \eqref{thirddual}, with $\alpha$ taking at least two values, including 1 and 2. Let $(\varphi^\alpha)$, $(\nu^\alpha)$ be a feasible configuration, $R$ a region of $M$ 
 (with no restriction on its topology), and $\varphi^1_0$, $\varphi^2_0$ constants such that: 
\begin{enumerate} 
\item[(1)] $R$ does not intersect the curves $m_1$, $m_2$ on which $\varphi^1$, $\varphi^2$ jump; 
 
\item[(2)] at every point of $\partial R$, 
$\varphi^1=\varphi^1_0$ or $\varphi^2=\varphi^2_0$ (or both). 
\end{enumerate}
 
\noindent
Then there exist regions $R_1,R_2\subseteq R$ and functions $\varphi^{1\prime}$, $\varphi^{2\prime}$ on $R$ such that 
\begin{enumerate} 
\item[(a)] $R_1\cup R_2=R$; 
 
 \item[(b)] $\varphi^{1\prime}=\varphi^1_0$ on $R_2$ and $\varphi^{2\prime}=\varphi^2_0$ on $R_1$; 
 
\item[(c)] $\varphi^{1\prime}=\varphi^1$ and $\varphi^{2\prime}=\varphi^2$ on $\partial R$; 
 
\item[(d)] the dual objective with $\varphi^1$ replaced by $\varphi^{1\prime}$ and $\varphi^2$ replaced by $\varphi^{2\prime}$ within $R$ (and other $\varphi^\alpha$s and all $\nu^\alpha$s unchanged) is at least as large as for the original configuration. 
\end{enumerate}
Note that, since $\varphi^{1,2\prime}$ agrees with $\varphi^{1,2}$ on $\partial R$, the new configuration does not introduce any discontinuities on $\partial R$ and obeys the boundary conditions for $\varphi^\alpha$ 
on $\partial M$, 
and is therefore feasible.

\noindent
\emph{Proof:} It is easy to check that the function $f:\mathbb{R}^2\to\mathbb{R}$ defined by
\begin{equation}
f(y_1,y_2) := \begin{cases}
y_1-\text{sgn}(y_1y_2)y_2\,,& |y_1|>|y_2|\\  
0\,,& |y_1|\le|y_2|
\end{cases}
\end{equation}
is continuous and has gradient either 0 or $(1,\pm1)$ everywhere. Define $R_{1,2}$ as follows,
\be
\begin{split}
R_1 =&\  \left\{x\in R:|\varphi^1(x)-\varphi^1_0|\ge|\varphi^2(x)-\varphi^2_0|\right\} \,, \\
R_2 =& \ \left\{x\in R:
|\varphi^1(x)-\varphi^1_0|\le|\varphi^2(x)-\varphi^2_0|
\right\},  
 \end{split}
\ee
and set
\begin{equation}
\begin{split}
\varphi^{1\prime}(x) =& \  \varphi^1_0 + f(\varphi^1(x)-\varphi^1_0,\varphi^2(x)-\varphi^2_0)\,,\\
\varphi^{2\prime}(x) =& \  \varphi^2_0 + f(\varphi^2(x)-\varphi^2_0,\varphi^1(x)-\varphi^1_0)\,
\end{split}
\end{equation}
Properties (a), (b), (c) are easy to check. We now consider the effect on the objective of replacing $\varphi^{1,2}$ with $\varphi^{1,2\prime}$. On $R_1$, we have
\be 
\varphi^{\prime1} (x) =  \varphi^1(x) \pm (\varphi^2 (x) - \varphi^2_0)  \,, \ \ \varphi^{2\prime} (x) = \ \varphi^2_0 \,, 
\ee
and therefore, 
\begin{equation}
d\varphi^{\prime1} = d\varphi^1\pm d\varphi^2\,,\qquad d\varphi^{2\prime}=0\,,
\end{equation}
so by the triangle inequality
\begin{equation}\label{triangle}
|d\varphi^{\prime1}|_0+|d\varphi^{\prime2}|_0 = |d\varphi^1\pm d\varphi^2|_0 \le |d\varphi^1|_0+|d\varphi^2|_0\,.
\end{equation}
By the same reasoning with 1 and 2 switched, \eqref{triangle} applies also in $R_2$. Therefore the integrand in the second term of the dual objective is not increased:
\begin{equation}
\Biggl(|d\varphi^{1\prime}|_0+|d\varphi^{2\prime}|_0+\sum_{\alpha\neq1,2}|d\varphi^\alpha|_0\Biggr)^2 \le
\left(\sum_\alpha|d\varphi^\alpha|_0\right)^2 \,. 
\end{equation}
Since the $\varphi^\alpha$s outside of $R$ and the $\nu^\alpha$s are unchanged, the dual objective is not decreased. This establishes property (d). $\square$

The objective will in fact \emph{increase} under the substitution $\varphi^{1,2}\to\varphi^{1,2\prime}$, unless everywhere in $R$ the triangle inequality in \eqref{triangle} is saturated, which requires that $d\varphi^1$ and $d\varphi^2$ be parallel (or anti-parallel) wherever they are both non-zero. Since the tangent vector of a level set of $\varphi^{1,2}$ is given by $\epsilon^{\mu\nu}\partial_\nu\varphi^{1,2}$, this in turn requires that any $\varphi^1$ level set and any $\varphi^2$ level set must fully coincide within $R$. (More precisely, connected components within $R$ must fully coincide.) Thus in a situation like the one shown 
in Figure \ref{fig:segregation}(a), in which the level sets of $\varphi^1$ and $\varphi^2$ intersect transversely, the objective can be increased, so such a situation cannot occur in a solution.

The fact that a configuration like the one in Figure \ref{fig:segregation}(a) cannot occur in a solution can also be understood from the primal program. Specifically, we can easily show that a region $R$ cannot be bounded by the union of a segment $\gamma_1$ of a 1-geodesic $\hat m_1$ and a segment $\gamma_2$ of a 2-geodesic $\hat m_2$. Since together they bound a region, $\gamma_1$ and $\gamma_2$ are homologous (relative to their intersection points). Therefore, replacing $\gamma_1$ with $\gamma_2$ on $\hat m_1$ 
and $\gamma_2$ with $\gamma_1$ on $\hat m_2$ 
gives an element $\tilde m_1$ of the class $C_1$, that must therefore have length at least $\ell_1$, and an element  $\tilde m_2$ of the class $C_2$ 
that must therefore have length at least~$\ell_2$. 
 On the other hand, since we traded pieces of curves, the total length is unchanged, 
\begin{equation}
\text{length}(\tilde m_1)+\text{length}(\tilde m_2) =
\text{length}(\hat m_1)+\text{length}(\hat m_2) =\ell_1+\ell_2\,,
\end{equation}
so $\tilde m_1$ and $\tilde m_2$  must have length equal to $\ell_1$ and 
$\ell_2$, respectively.  They must therefore be 1,2-geodesics respectively. But this is impossible since they have corners.

\subsubsection{Cylinder versus pants}

\begin{figure}[!ht]
\leavevmode
\begin{center}
\epsfysize=6cm
\epsfbox{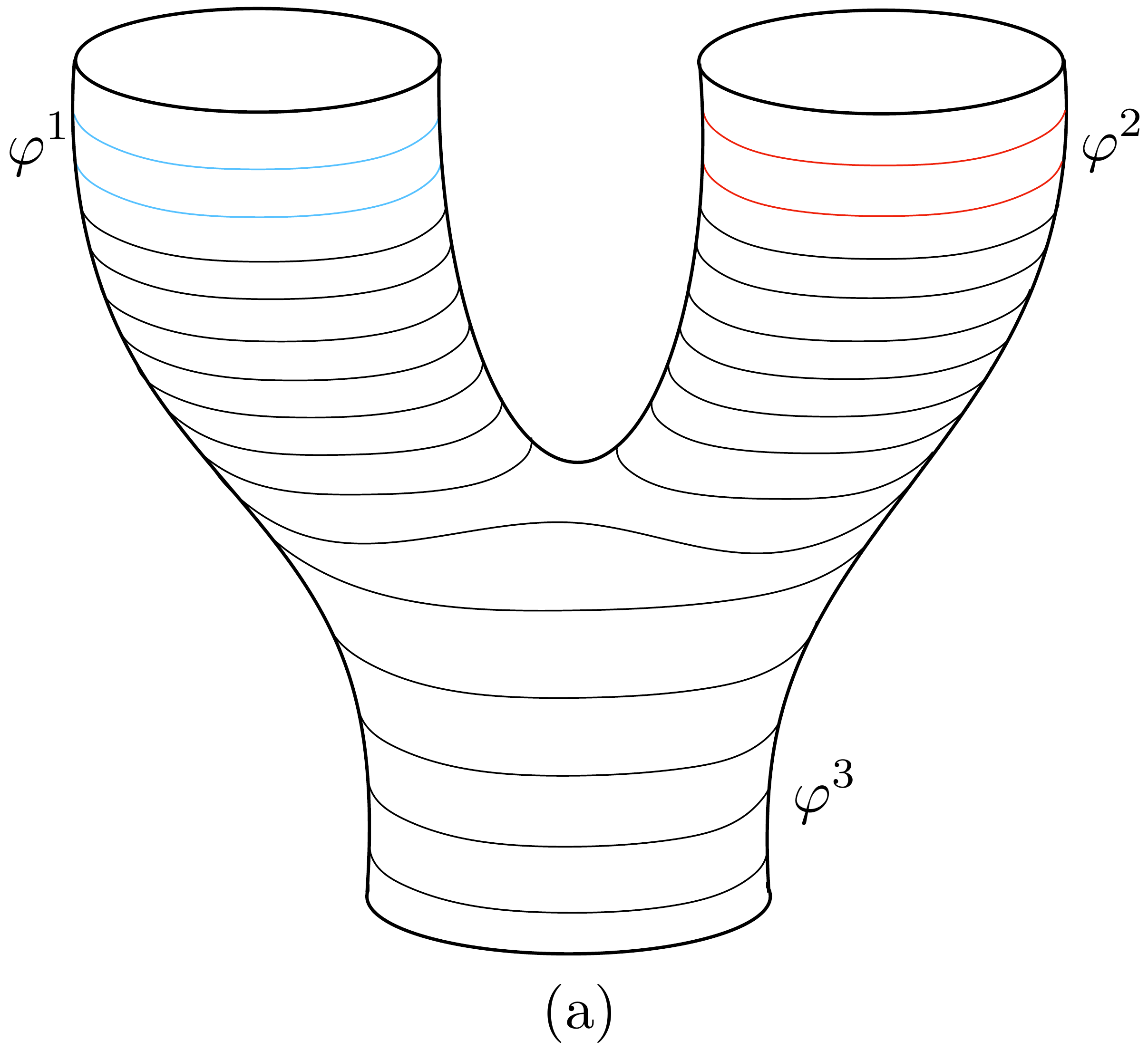}
\hspace{1cm}
\epsfysize=6cm
\epsfbox{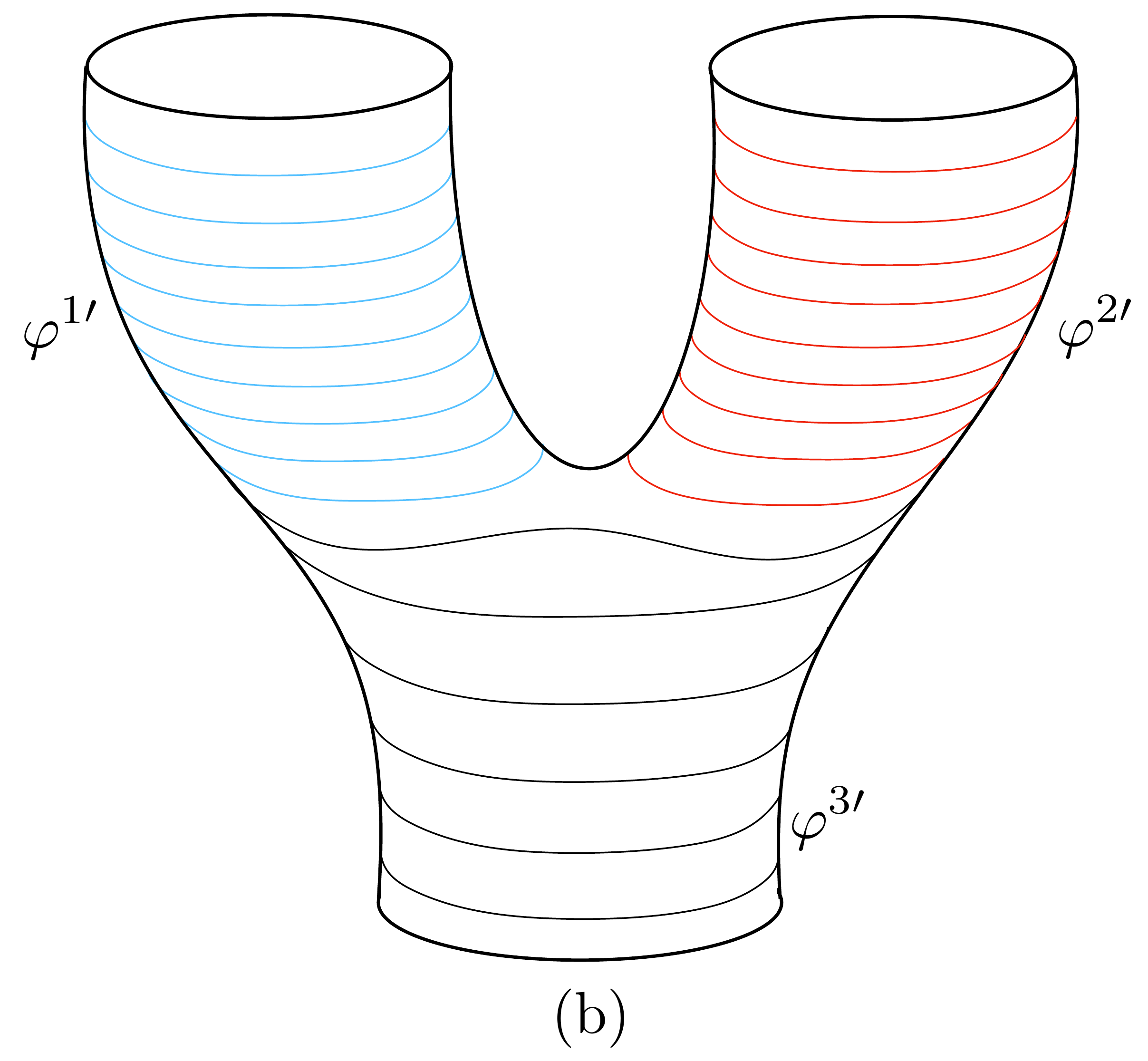}
\end{center}
\caption{\small
(a) Pants diagram with three 
types of 
level sets, corresponding to $\varphi^1, \varphi^2$,
and $\varphi^3$.  Above the crotch the $\varphi^3$ level set
 is the union 
of representatives of $C_1$ and $C_2$. (b) The same diagram, where
the $\varphi^3$ level sets above the crotch now belong to the $\varphi^1$ and $\varphi^2$ level sets.   This rearrangement
of curves leads to a larger objective in the dual program when the length parameters satisfy $\ell_1 + \ell_2 > \ell_3$. 
}
\label{fig:pants}
\end{figure}

Even without overlapping, bands of level sets can still interact, in a sense we will now explain. Consider a situation like that illustrated in Figure \ref{fig:pants}, which shows a ``pair of pants'' surface; this could be part of a larger surface. We consider three homology cycles, $C_{1,2,3}$, represented by the top left, top right, and bottom boundaries respectively. In Figure \ref{fig:pants}(a), we have a band of level sets in each class, and these bands don't overlap. Above the crotch, each $\varphi^3$ level set is the union of a representative of $C_1$ and 
a representative of $C_2$.  In Figure \ref{fig:pants}(b), the configuration has been changed so that those curves are level sets of $\varphi^{1\prime}$ and $\varphi^{2\prime}$, and not of $\varphi^{3\prime}$. Correspondingly, in order for the new configuration to obey the boundary and jump conditions, we change the $\nu^\alpha$ values:
\begin{equation}\label{Deltanu}
\nu^{1\prime} = \nu^1+\Delta\nu\,,\qquad
\nu^{2\prime} = \nu^2+\Delta\nu\,,\qquad
\nu^{3\prime} = \nu^3-\Delta\nu\,,
\end{equation}
where $\Delta\nu$ is the ``number'' of level sets that have been transferred. We have
\begin{equation}
\sum_{\alpha=1}^3|d\varphi^{\alpha\prime}|=
\sum_{\alpha=1}^3|d\varphi^\alpha|\,,
\end{equation}
so the second term in the dual objective \eqref{dualobjective} is unchanged (even if there are level sets of other $\varphi^\alpha$s passing through this region). However, by \eqref{Deltanu}, the first term changes by
\begin{equation}
2(\ell_1+\ell_2-\ell_3)\Delta\nu\,.
\end{equation}
So, if the $\ell_\alpha$ obey the strict triangle inequality
\begin{equation}\label{stricttriangle}
\ell_3<\ell_1+\ell_2\,,
\end{equation}
then the objective increases. In effect, the $\varphi^1$ and $\varphi^2$ bands ``eat up'' the $\varphi^3$ band, until no $\varphi^3$ level sets remain that are the union of an element of $C_1$ and of $C_2$. So, in solving the dual, we can restrict our attention to bands that are cylindrical and not pairs of pants. (However, imposing this constraint makes the dual non-convex.)

If the length constraints obey the inequality \eqref{stricttriangle}, the fact that no 3-geodesic is a union of a $C_1$ representative and a $C_2$ representative also follows directly from the primal.  Indeed, any 3-geodesic must have length $\ell_3$, but by the length conditions
on $C_1$ and $C_2$ representatives, the composite one will be longer, at least
of length $\ell_1+ \ell_2$, showing this is not possible. Our purpose here was to show how the dual program enforces this fact.

\subsection{Quadratic differentials and the dual objective}\label{sec:qd-and-dual}

Jenkins and Strebel proved the following result  
(Theorem 21.10 \cite{strebel}):  Let $M$ be a Riemann surface
with a set of {\em admissible} curves $\gamma_1, \ldots , \gamma_k$, that is, a set 
of nontrivial-homotopy, non-homotopic,  simple closed Jordan curves that have non-intersecting
representatives.  Now let $R_i$ with $i = 1, \ldots, k$, be disjoint
ring domains (annuli)  of homotopy type $\gamma_i$ and let $M_i$ be the
modulus of $R_i$.  To each curve $\gamma_i$ we assign a positive 
number $\ell_i$. Consider now the functional ${\cal F}$ defined by
\be
{\cal F} \ = \   \ell_1^2 M_1 + \ldots   \ell_k^2 M_k \ = \ \sum_{i=1}^k  \ell_i^2 M_i\,. 
\ee
Maximizing ${\cal F}$ over the shape of the
{\em disjoint} ring domains gives
\be
\hbox{Max}_{R_i}  {\cal F}  \ = \  ||\Phi|| \,,
\ee
that is,  the norm of a Jenkins-Strebel quadratic differential
$\Phi(z)dz^2$ on the surface $M$.  The $R_i$ are the characteristic ring domains
of the quadratic differential $\Phi$ which has closed trajectories.  Strebel and
Jenkins also show that the norm of the quadratic differential is  equal
to the minimal area of a metric in $M$ such that 
any curve homotopic to $\gamma_i$ is 
longer than or equal to~$\ell_i$.  Thus the maximum of the
objective ${\cal F}$ coincides with the minimum of the objective in
the homotopy minimal area problem. 

We want to understand how the objective ${\cal F}$ 
arises in the context of the dual program.  Of course it will not
arise in general situations when one constrains classes of curves
that have intersections;   the objective in the dual program is a subtle generalization of ${\cal F}$ when the curves do not form an admissible set.   When they do, the dual program objective can 
be related to ${\cal F}$. 
To discuss the relation consider a setup where the homology
program is equivalent to the homotopy problem above.

In this setup we choose an admissible set of curves $\gamma_1, \cdots, \gamma_k$ all of which are {\em also} nontrivial curves in homology   classes $C_1, \cdots C_k$.  We then impose the length conditions $\ell_1, \cdots, \ell_k$ on the respective classes of curves.  Since all curves homotopic to $\gamma_i$ are also
homologous to $\gamma_i$ the dual program imposes all the
conditions of the homotopy program.  Moreover, 
by suitably choosing the length parameters $\ell_i$  it
is possible to guarantee that, at least in some cases,  
the  homology program does not impose additional conditions
absent in the homotopy program.

\begin{figure}[!ht]
\leavevmode
\begin{center}
\epsfysize=5.5cm
\epsfbox{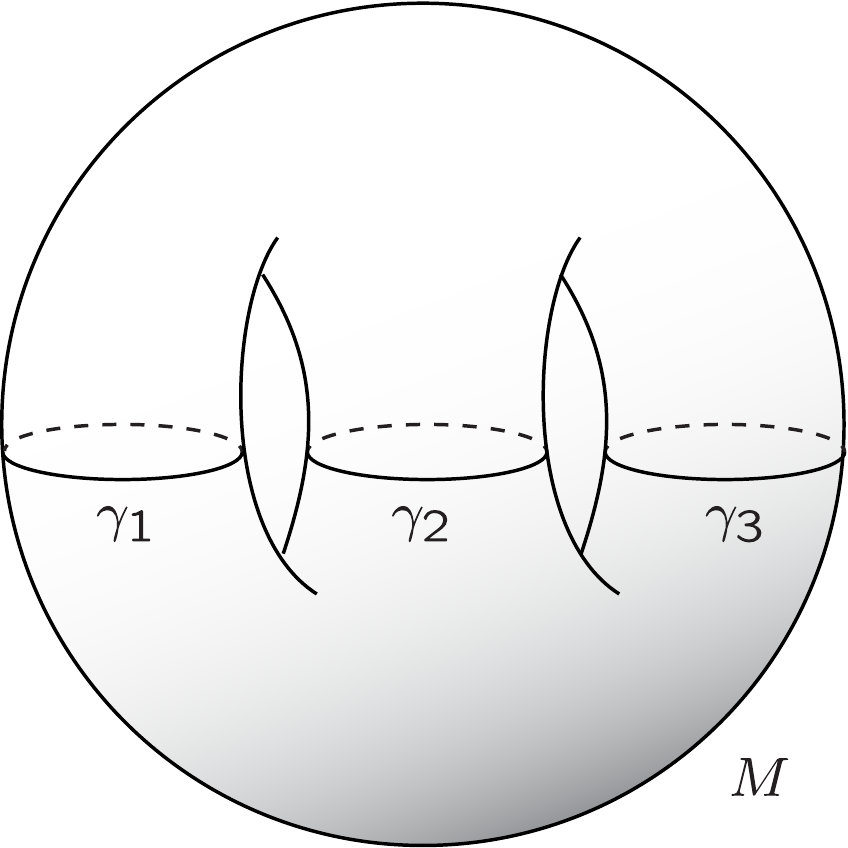}
\end{center}
\caption{\small  A genus two surface with a maximal
admissible set of curves $\gamma_1, \gamma_2, \gamma_3$ of nontrivial
homotopy.  The curves are also of nontrivial homology, and with 
suitable length conditions the dual program and the homotopy minimal area problem are equivalent.}
\label{ffg31}
\end{figure}

This is illustrated in the genus-two surface $M$ shown in Figure~\ref{ffg31}, and equipped with an admissible set of  curves 
$\gamma_1, \gamma_2, \gamma_3$.  The set is maximal:  any curve on a different homotopy class would
intersect at least one of the curves in the set.  The curves are also
nontrivial in homology, belonging to classes denoted as $C_1, C_2$,
and $C_3$.   The curves are related in homology.  Up to signs 
determined by the choice of orientations, we have relations of the form
\be
\gamma_i  \sim   \pm\gamma_j  \pm \gamma_k \,,  \ \ i \not= j \not= k\,,
\ee
for all values of $i, j, k$ in the set $1,2,3$.  As a result, for example, the calibration that guarantees that the length of any curve homologous to $\gamma_1$ exceeds or equals $\ell_1$, will
also imply that for curves $\gamma'_2$ and $\gamma'_3$
homotopic to $\gamma_2$ and $\gamma_3$, respectively, we have 
\be
\hbox{length} (\gamma'_2) + \hbox{length} (\gamma'_3)  \geq  \ell_1 \,.
\ee
This will impose no new condition if $ \ell_2 + \ell_3  \geq \ell_1$.
Considering the other homology relations, we need the length
parameters  to satisfy
\be
\ell_i + \ell_j  \geq   \ell_k\,, \ \  \ \  \forall  \ i \not= j \not= k\,.
\ee
We can argue that these inequalities, which are easily satisfied, guarantee that no new condition is imposed by the homology problem.  Indeed, consider a general representative $m_1$
of $C_1$.  Being a sum of closed curves, $m_1$
may include 
a curve in $C_2$ and/or a curve in $C_3$, as considered above,
but it may contain other curves.  It cannot contain a single curve
in a new homotopy class, because our set is maximal and any new homotopy class must
necessarily intersect one of the representatives in the set.
This, however, cannot happen for a curve representing $C_1$. 

Consider, therefore, any setup as above, where the homotopy and 
the homology programs are equivalent.  Then 
the maximum of the dual objective is the minimum
of the homotopy problem.  As we know from the homotopy
problem, the bands of saturating geodesics at the optimum
form non-intersecting ring domains, or cylinders.
This is consistent with the segregation lemma
considered in the previous subsection as a property of the
dual program. 

It now follows that for any set of {\em fixed} non-overlapping
ring domains $R_1, \ldots R_k$  
of homotopy type  $\gamma_1, \ldots \gamma_k$
and length parameters $\ell_1, \ldots, \ell_k$,  the objective 
of the dual program will have an optimum that equals the sum 
$\sum_i \ell_i^2 M_i$ where $M_i$ is the modulus of $R_i$.
This follows from the calculation in section~\ref{subsec:cyl}, that
proved that the optimum of the dual objective in an annulus of modulus $M$ and length parameter 
$\ell$  is indeed $\ell^2 M$.  Constrained by the condition
of non-overlapping ring domains, further maximization of the dual
objective over the shape of the ring domains clearly becomes maximization of  ${\cal F}$.

\section{Back to the homotopy problem}\label{sec:homotopy}

We now return to the original minimal area problem of closed
string field theory, which constrains all curves of nontrivial
homotopy,  and explain how it can be solved using the
ideas considered in this paper.  As a first step we describe the
natural homological problem associated with the problem in
homotopy.  We then show that the homology problem does not
impose extraneous length conditions.  The main complication is  that there usually are 
nontrivial homotopy closed curves  $\gamma$ that
are trivial in homology and thus their lengths cannot be constrained
by calibrations.  To deal with this we take the original surface $M$
and construct a covering space $\tilde M$ where the curve $\gamma$ becomes
homologically nontrivial.  In principle a double cover suffices
and much of our discussion is couched in this language.  We explain that a calibration $\tilde u$ 
on $\tilde M$, antisymmetric under the exchange of the two sheets, 
together with a metric symmetric under the exchange, correctly
constrains the length of all curves homotopic to $\gamma$ in the original
surface. By introducing such a double cover for each homotopy class
of curves that is trivial in homology, the minimal area homotopy
problem is solved in terms of homology and calibrations.  
The same double cover can also be used for the dual program
by imposing suitable conditions on the functions 
$\varphi^\alpha$ or the one-forms $\eta^\alpha$.

\subsection{The homotopy problem reduced to homology}\label{sec:homhom}

As discussed in the introduction, the minimal
area problem for closed string field 
theory~\cite{Zwiebach:1990ni,Zwiebach:1990nh}
asks for the conformal metric of least area
on a Riemann surface under the condition
that all non-contractible closed curves have length greater than or equal to 1.  
We recall the formulation 
in (\ref{csft-MAP}):
\begin{equation}
\label{csft-MAP-v2}
\begin{split}
& \text{Minimize } \,  
\int_M\omega_0 \,\Omega \ \  
\ \ \text{over }\Omega\ge0 \text{ (function)}\\
& \hbox{subject to} \ \ \ \ 
\, \ell_s-\int_\gamma\sqrt{\Omega} \, |\dot x |_{{}_0} \le0\,, \ \ 
  \forall\,   \gamma \in \Gamma \,. \ \ 
  \end{split} 
\end{equation}
Here  $\Gamma$ is the set of all
homotopically nontrivial simple closed curves $\gamma$ on $M$.
It is known that this problem is not changed if we reduce the
constraint space from $\Gamma$ down to a smaller set 
of nontrivial homotopy classes $S\subset\pi_1 (M)$, where any class in $S$ has a  Jordan closed curve representative.  Given a class
$D \in S$, the class $D^{-1}$, representing oppositely oriented
curves, is not included in $S$.  The minimal-area problem is
now written as
\begin{equation}
\label{csft-MAP-v3}
\begin{split}
& \text{Minimize } \,  
\int_M\omega_0 \,\Omega \ \  
\ \ \text{over }\Omega\ge0 \text{ (function)}\\
& \hbox{subject to} \ \ \ \ 
\, \ell_s-\int_\gamma\sqrt{\Omega} \, |\dot x |_{{}_0} \le0\,, \ \ 
  \forall\,   \gamma \in S \,. \ \ 
  \end{split} 
\end{equation}

Consider the usual map $\Pi$  from homotopy to homology classes, $\Pi :  \pi_1(M) \to  H_1 (M)$.   The map is not
one-to-one and  has a kernel $S_0$.  We can now split $S$ into two non-overlapping subsets 
\begin{equation}
 S  \ = \  S_{1} \cup S_{0}\, , \quad S_0:=\Pi^{-1}(0)  \,. 
\end{equation}
We write the elements of $S_{0}$ as $D_0^k$ (where $k$ is an index):
\begin{equation}
\Pi ( D_0^k ) \ = \ 0 \,,  \ \forall k\,. 
\end{equation}
The representatives of $D_0^k$ are homotopically non-trivial closed curves that cut $M$ into two separate pieces, at least one of which has no other boundary. 
The elements of $S_{1}$, on the other hand, are homotopy classes $D_\alpha^i$
that  are mapped by $\Pi$ to 
a nontrivial homology class~$C_\alpha$:
\begin{equation} 
\Pi (D_\alpha^i )  =  C_\alpha  \not= 0 \,. 
\end{equation}
 The index $i$ on the homotopy class is needed
 because curves in different homotopy
 classes may be homologous.
We call $W$ the image of $S_{1}$ under the map 
$\Pi$
\begin{equation}
W   \ = \ \Pi(S_1) \ = \ 
\bigl\{  C_\alpha \, , \alpha \in J_0  \bigr\}\,,
\end{equation}
where we introduced the index set $J_0$ that labels
all the homology classes in $W$.
By definition, the homology classes in $W$ all have
representatives that are simple closed Jordan curves. 

The natural homological problem associated to the
homotopy problem constrains the homology classes 
$C_\alpha \in W$.  Using calibrations, the program is
just the primal MAP, version 1, given in (\ref{secondprogram}), for the appropriate set of classes and with length constraint 
$\ell_\alpha=\ell_s$ for all classes:
\begin{equation}
\begin{split}
\ \ &\text{Minimize }\ \, \int_M\sqrt{g^0}\,\Omega \quad 
\text{ over }\ \ 
\Omega \, \hbox{(function)},u^\alpha \, (\hbox{one-forms})\,  \ \  \\
&   \hbox{subject to:}  \hskip25pt |u^\alpha|_0^2-\Omega\le0\,,  \\
&  \hskip80pt
\qquad  du^\alpha=0\, ,  \  \\[-0.7ex]
& \hskip70pt  \ell_s-\int_{m_\alpha}\hskip-8pt u^\alpha=0\,, 
\ \forall \alpha \in J_0  \,. 
\end{split}
\end{equation}
This program will have to be supplemented to
solve the original homotopy problem (\ref{csft-MAP-v2}).  The requisite program 
will be written at the end of the section.

This homology program, we claim, gives an optimum in which all curves in each homotopy
class in  $S_{1}$ satisfy the length condition.  Indeed, 
any curve in a class in $S_1$ is itself a representative
of a class in $W$, and its length is properly constrained by
the homology problem.  In fact, each class in $W$ effectively
constraints all the classes in $S_1$ that map to it under $\Pi$. 

We now explain why the homology problem does not impose
extraneous length conditions.   
Any class $C_\alpha \in W$, in addition to having representatives
with a single closed curve that are properly constrained,  also contains representatives
with multiple closed curves, and the sum of their
lengths is also constrained to be greater than or equal to $\ell_s$. 
Consider a $C_\alpha$ 
representative $m_1 + \cdots  + m_k$, with $k\geq 2$,
where each $m_i$ ($i=1, \ldots , k$) is a simple closed curve.  
Since the original class $C_\alpha$
is nontrivial, not all of the $m$'s can be trivial in homology.  At least one 
$m_i$ must be a representative of a nontrivial homology class. 
That $m_i$ must then be a homotopically nontrivial simple
closed curve.  It follows that  $m_i$ belongs to  $S_{1}$ 
therefore has length
greater than or equal to $\ell_s$.  Since lengths add, the representative
$m_1 + \cdots  + m_k$ is automatically guaranteed to satisfy the
length condition.  It does not impose a new constraint. 

Note that the set $W$  contains an infinite number
of homology classes. 
Constraining all of its members therefore requires an infinite number of calibrations $u^\alpha$. In practice,  one would only constrain a finite number of cycles  and  check by hand that the rest of the cycles in $W$ do not contain any representatives with length less than $\ell_s$. If any cycles did, then the program would have to be re-solved including the offending cycles.

Since the curves in $S_1$ are properly constrained by the
homology problem, this leaves the problem of  the curves in $S_{0}$, homotopy  nontrivial curves
that are homologically trivial.
The next subsection will address this issue.

\subsection{Homologically trivial curves}\label{sec:trivial}

One can only use calibrations to constrain the lengths of curves that are homologically non-trivial. The reason is that a calibration $u$  constrains the length of a curve $\gamma$ using the inequality 
\be
\text{length}(\gamma)\ \ge \  \Bigl| \, \int_\gamma u \, \Bigr|\,  .
\ee
For this to work the period $\int_\gamma u$ must be non-zero.
For a homologically trivial curve $\gamma$, however, we have $\gamma = \partial R$ and therefore
\be
\int_\gamma u = \int_R  du  = 0\,,
\ee
because calibrations are closed forms.  We can get around this problem, however, by observing that, on a suitable multiple cover of the surface, a homologically trivial curve lifts to a homologically non-trivial one. Here we will explain how to construct a double cover with this property and thereby constrain the lengths of such curves by calibrations. By a straightforward generalization, it is also possible to construct a higher-multiplicity cover, and in certain cases this may be convenient, for example to preserve symmetries of the surface; we will see an example in \cite{headrick-zwiebach2}.

\subsubsection{Defining a suitable covering space}

\begin{figure}[!ht]
\leavevmode
\begin{center}
\epsfysize=7.5cm
\epsfbox{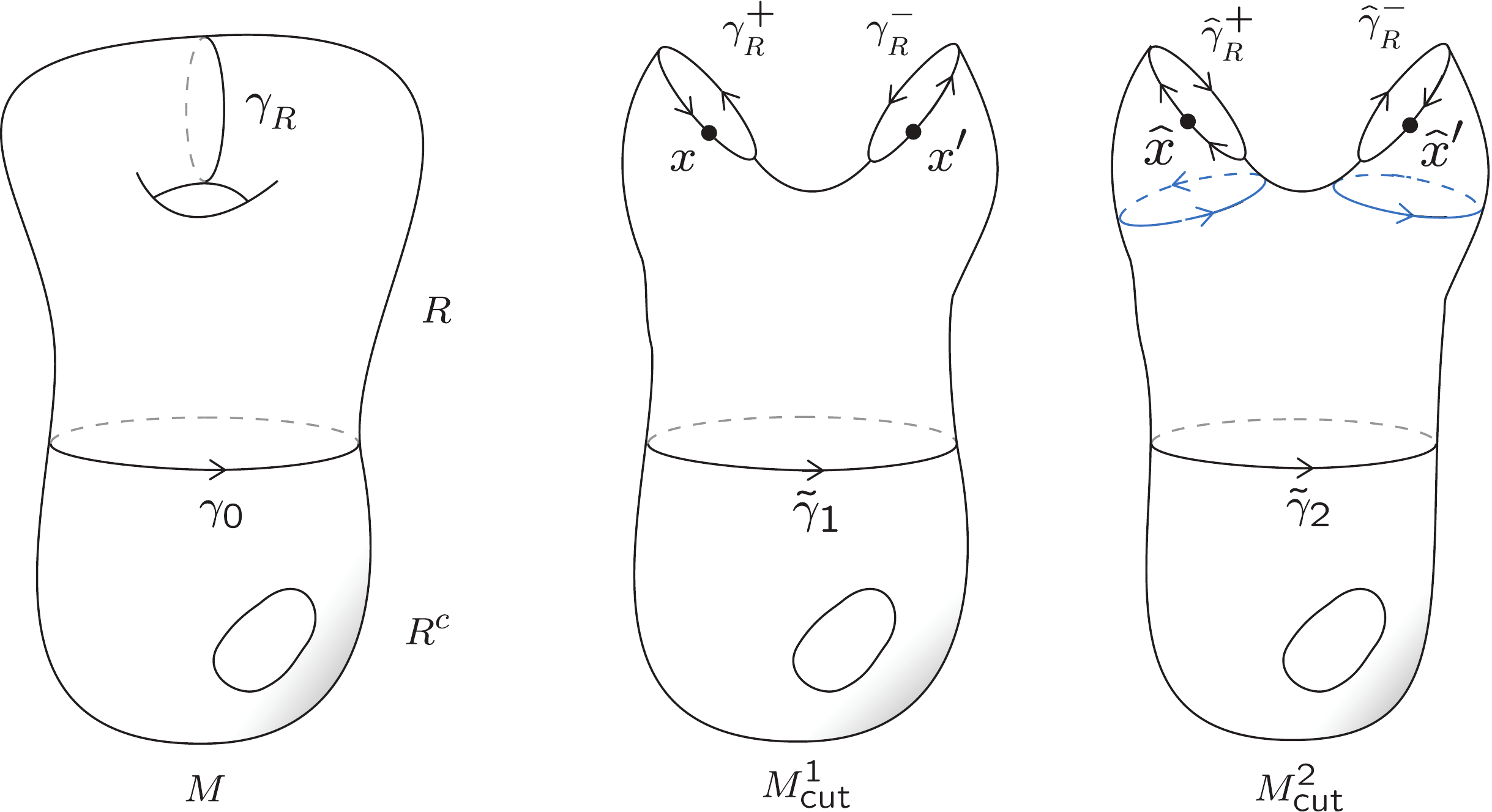}
\end{center}
\caption{\small  Left:  The surface $M$ with a curve $\gamma_0$ in the nontrivial
homotopy class $D$.  The curve $\gamma_0$ is homologically trivial and divides $M$ into $R$ and $R^c$. We choose a curve $\gamma_R$  
along the handle in~$R$.   
Right:  The surface $M$ is cut open along $\gamma_R$ and 
and it is then doubled to form $M_{\text{cut}}^1$ and $M_{\text{cut}}^2$.  
These two are glued to form $\tilde M$, with $\gamma_R^+$ attached to $\hat \gamma_R^-$
and $\gamma_R^-$ attached to $\gamma_R^+$.  The curves $\tilde\gamma_1$ and $\tilde\gamma_2$ are the images of $\gamma_0$ in the double surface $\tilde M$, and the blue curves (homologous to $\tilde\gamma_2$) help
show that $\tilde\gamma_1 = -\tilde\gamma_2$ in $H_1(\tilde M)$.  The curve
$\tilde\gamma_1$ is homologically nontrivial in $\tilde M$. }
\label{ff1}
\end{figure}

Let $D$ be a non-trivial homotopy class in the set $S_{0}$; this means that it has
a representative $\gamma_0$ that is a homotopically non-trivial but homologically
trivial Jordan closed curve on the surface $M$ (see Figure~\ref{ff1}).
Our goal now is to cut $M$ appropriately and glue the cut surface
to another copy of itself to produce a covering surface $\tilde M$ on which $\gamma_0$ lifts to a homologically nontrivial curve $\tilde\gamma_0$.  We do this cut as follows.

Since $\gamma_0$ is homologically trivial, it must be the boundary
of a  region $R$ of $M$:  $\partial R = \gamma_0$. The region $R$ has genus at least one, otherwise $\gamma_0$ would not be homotopically nontrivial. 
Let $\gamma_R$ be a Jordan curve that wraps one
of the  handles of $R$ and does not intersect $\gamma$  (see Figure~\ref{ff1}). The surface
$M$ is then cut along $\gamma_R$ without falling into two pieces. 
On the cut surface $M_{\text{cut}}$, 
denote by $\gamma_R^+$ and $\gamma_R^-$ the 
two  boundaries generated by the cut along $\gamma_R$.  We orient
the curves $\gamma_R^+$ and $\gamma_R^-$ such that their sum is homologous to $\gamma_0$:
\be
\gamma_0 \sim  \gamma_R^+ \, + \,  \gamma_R^-\,. 
\ee

On the other side of $\gamma_0$ we have
the complement $R^c$ of $R$ in $M$.  If the surface $M$ has a boundary component
within $R^c$, as it is the case in Figure~\ref{ff1}, 
then $\gamma_0$ is not homologically trivial within $R^c$ and we
need not do any further cut.  If $M$ has no boundary component, 
$\partial R^c = -\gamma$ and we must also cut $R^c$, which must
have genus at least one. 
Just like we did for $R$,  we select a handle and a Jordan curve 
$\gamma_{R^c}$ that wraps the handle and does not intersect $\gamma$. 
The surface is then cut along $\gamma_{R^c}$.  
We denote by $\gamma_{R^c}^+$ and $\gamma_{R^c}^-$ the 
two new boundaries generated by the cut along $\gamma_{R^c}$.  
 The result is now the
cut surface $M_{\text{cut}}$.

Now consider two copies $M_{\text{cut}}^1$  and $M_{\text{cut}}^2$ of the cut surface $M_{\text{cut}}$.  On $M_{\text{cut}}^2$
we define the analogous curves   $\hat\gamma_R^+$ and $\hat\gamma_R^-$ but with opposite orientations as compared with
$M_{\text{cut}}^1$.

We define $\tilde M$ as the double cover   
obtained 
 as follows:  $\gamma_R^+$ on  $M_{\text{cut}}^1$ is glued to 
 $\hat\gamma_R^-$ on  $M_{\text{cut}}^2$ and 
 $\gamma_R^-$ on  $M_{\text{cut}}^1$ is glued to 
 $\hat \gamma_R^+$ on  $M_{\text{cut}}^2$ (see Figure~\ref{ff1}):
 \be  
 \gamma_R^+ \, \sim \,  \hat\gamma_R^- \,,  \quad
 \hbox{and} \quad  \gamma_R^- \, \sim\,  \hat \gamma_R^+\,.
 \ee
With the gluing done consistent with the 
orientations shown in the figure, the above are relations in
$H_1(\tilde M)$.
An analogous gluing is done for the cuts in the region $R^c$ if they exist.
 The projection map 
 $p$ taking the double cover to the original surface, 
 \begin{equation}
 p: \tilde M \to M\,, 
 \end{equation}
 is obvious from the construction.\footnote{
 We can characterize $\tilde M$ recalling that (see~\cite{singer-thorpe} Ch.\,3, Theorem 4) covering
spaces can be associated to subgroups of the first homotopy group: For $X$ a connected
space and $x\in X$, let $H$ be a subgroup of $\pi_1 (X, x)$.
Then there is a covering space $(\tilde X, p)$ and a point $\tilde x\in\tilde X$ such that $p(\tilde x)=x$ and $\pi_1 (\tilde X, \tilde x)$, when pulled back to $X$, equals $H$.  In our case, $\tilde M$ is associated to the subgroup $H$ of $\pi_1(M,x)$ consisting of  homotopy classes of curves that intersect $\gamma_R$ an even number of times. These are precisely the curves in $M$   that descend from closed curves on $\tilde M$.}

Since $\gamma_0$ does not intersect the cuts, its images $\tilde\gamma_1,\tilde\gamma_2$ on the first and second sheets of $\tilde M$ respectively are both closed curves. 
The objective of the construction is then achieved:  \emph{$\tilde\gamma_1$ and $\tilde\gamma_2$ are nontrivial in $H_1(\tilde M)$}.  
Indeed,  if we cut $\tilde M$ along $\tilde\gamma_1$, it is clear from
the figure that $\tilde M$  will break into two pieces and, in each piece,
  $\tilde\gamma_1$ will be homologous to a boundary component and thus not trivial.  If $R^c$ did not have a boundary, a cut and glue construction similar to that
  for $R$ would have been needed to build $\tilde M$, 
  and a cut along $\tilde\gamma_1$ would have not
  split $\tilde M$ into two surfaces, again showing that $\tilde\gamma_1$ has nontrivial homology  in $\tilde M$.  
  The same line of argument holds for $\tilde\gamma_2$, which is also
  of nontrivial homology.  
  
Noting that on $M_{\text{cut}}^1$ we have
\be
\tilde\gamma_1\sim  \gamma_R^+ \, + \,  \gamma_R^-\,,
\ee
and on $M_{\text{cut}}^2$ we have
\be
\tilde\gamma_2\sim  -\hat\gamma_R^+ \, - \,  \hat\gamma_R^-\,,
\ee
we conclude that
\begin{equation}
\tilde\gamma_1 \sim -\tilde\gamma_2 \,. 
\end{equation}
We define  $\tilde C\in H_1(\tilde M)$ to be the homology class of $\tilde\gamma_1$:
\begin{equation}
\tilde\gamma_1 \ \in \ \tilde C \ \in \ H_1(\tilde M)  \,.  
\end{equation}

\subsubsection{Defining the calibration on $\tilde M$}

We can now apply our methods to the nontrivial cycle $\tilde C$ in $\tilde M$.  By the same reasoning as in subsection \ref{sec:localprimal}, there exists a calibration 
$\tilde u$ on $\tilde M$ such that $\int_{\tilde C}\tilde u=1$ if and only if every representative of $\tilde C_1$ has length at least 1.

Given a metric on $M$, there is a unique way to lift it to one on $\tilde M$ that is invariant under exchange of the two sheets: the metric on $M$ is copied onto each of the
cut surfaces $M_{\text{cut}}^1$  and $M_{\text{cut}}^2$ before they are glued. 
We also impose a symmetry condition on the calibration. Since $\tilde\gamma_1+\tilde\gamma_2\sim 0$ in homology, we have $\int_{\tilde\gamma_2}\tilde u=-\int_{\tilde\gamma_1}\tilde u$. Since $\tilde\gamma_1$ and $\tilde\gamma_2$ are exchanged under exchange of the sheets, we demand that $\tilde u$ is antisymmetric under sheet exchange. In fact, a slight generalization of the max flow-min cut theorem establishes the following statement: given a symmetric metric on $\tilde M$, there exists an antisymmetric calibration $\tilde u$ such that $\int_{\tilde C}\tilde u=1$ if and only if every representative of $\tilde C$ has length at least 1. 

We now show that the antisymmetry of $\tilde u$ under exchange of the two sheets implies that $\tilde u$
on $M^1_{\text{cut}}$ 
is anti-continuous across the cut: it takes  
{\em opposite} values on points immediately across the cut.  Because of the antisymmetry of the calibration, we have 
the following relations among the calibration $\tilde u$ at points on the two
sheets that map to the same point on $M$ 
(see Figure~\ref{ff1}):
\begin{equation}
\tilde u(x)  \ = \ - \, \tilde u(\hat x) \,,  \quad  \tilde u(x')  \ = \ - \tilde u(\hat x') \,.
\end{equation}
On the other hand, because the gluing conditions 
that create $\tilde M$  
must be consistent with the calibration  and we glue $x$ to $\hat x'$ and $x'$ to $\hat x$, we  have 
\begin{equation}
\tilde u(x)  \ = \  \, \tilde u(\hat x') \,,  \quad  \tilde u(x')  \ = \  \tilde u(\hat x) \,.
\end{equation}
From the above equations we immediately get
\begin{equation}
\tilde u(x)  \ = \ - \, \tilde u(x')  \,.  
\end{equation}
Since $x$ and $x'$ are the same point on the curve
$\gamma_R \in M$, we have shown that the antisymmetric calibration $\tilde u$ in 
$\tilde M$ is an anti-continuous calibration on $M^1_{\text{cut}}$ across 
the curve $\gamma_R$.
Thus we can work entirely on $M_{\text{cut}}^1$, supplementing the program \eqref{secondprogram} with the existence of a calibration satisfying, in addition to the usual constraints,
\begin{equation}\label{ubc}
u|_{\gamma_R^+}=-u|_{\gamma_R^-}\,,
\end{equation}
where $\gamma_R^\pm$ are the two  curves created by the cut along $\gamma_R$ (and similarly for $\gamma_{R^c}$). 

In terms of the dual flows, the boundary condition at the cut 
effectively  makes the cut into a source for a flow that eventually reaches the curve $\gamma$, solving the problem that in the
original surface there could be no flux across $\gamma$. 
We also wish to emphasize that the construction of the
double surface $\tilde M$ is not unique.  If $R$, for example, was
of genus two, we could produce the double by cutting either of
two handles.  We would expect that at the optimum 
either cut would yield calibrations that are equal on the band of 
saturating geodesics homotopic
to $\gamma$, but away from these geodesics
the two calibrations would be different, realizing the
ambiguity pointed out below equation~(\ref{thirdprogram}).
 
It is straightforward to see, by dualizing either on $\tilde M$ using the antisymmetry of $\tilde u$ or dualizing on $M$ using the boundary condition \eqref{ubc}, that the corresponding dual scalar $\varphi$ is also anti-continous across the cut(s):
\begin{equation}
\varphi|_{\gamma_R^+}=-\varphi|_{\gamma_R^-}\,.
\end{equation}
This is also consistent with the relation (\ref{primdual-simplified}) between the $u$'s and 
the $\varphi$'s.  
The dual objective will tend to make 
the values of $\varphi$ small at the  sides of the cut.  
Indeed, if we consider a cycle orthogonal to $\gamma_R$
on the handle, the discontinuity at the cut requires an expensive
gradient of $\varphi$ along that cycle.  (Recall that $|d\varphi|$ enters negatively in the dual objective, which is being maximized.) The program will
make the values of $\varphi$ small at the cut, almost as if
the cuts were boundaries, where we know that $\varphi$ must vanish.

For each homotopy class in $S_0$ we must 
include in our programs a calibration $u$ or a scalar $\varphi$ subject to these anti-continuous boundary conditions across the cuts
necessary to produce a double cover rendering the homotopy
class nontrivial in homology.

\subsubsection{Analysis of curves on $\tilde M$}

Of course, our original intent was not to constrain members of the homology class 
$\tilde C\in H_1(\tilde M)$ where $\tilde \gamma_1$ belongs,
 but rather members of the homotopy class $D \in \pi_1(M)$. Given that the metric on $\tilde M$ is obtained by lifting the metric on $M$ symmetrically, any curve on $\tilde M$ descends to a curve on $M$ with the same length. 
Let $C$ be the image under descent of $\tilde C$:
\begin{equation}
C  := \ \bigl\{    
m = p(m') \, ,  m' \in \tilde C     \bigr\} \,.
\end{equation} 
Each element $m$ of $C$ is a curve or a sum of curves on $M$; 
however, $C$ is a priori neither a homotopy nor a homology class. Since $\tilde\gamma_1 \in \tilde C$ and $\tilde\gamma_1$ descends to $\gamma_0$,
we have 
\begin{equation}
\gamma_0 \in C\, .
\end{equation} 
The calibration $\tilde u$ constrains the length of every element in $\tilde C$ and thus 
of every element in $C$.  
First we prove that every curve in the homotopy class $D$
 of $\gamma_0$ is an element of $C$, 
showing we have succeeded in constraining all curves in $D$.  
Then we show that the constraints on any other elements
of $C$ are not new.

\paragraph{Lemma 2.} $C$ contains entire homotopy classes:  
given homotopic curves $\gamma$, $\gamma'$ on $M$, if $\gamma\in C$ then $\gamma'\in C$.

\noindent
\emph{Proof:} Since $\gamma\in C$ there exists an element $\tilde\gamma$ of $\tilde C$ that descends to $\gamma$. 
The homotopy deformation from $\gamma$ to $\gamma'$ 
can be lifted to a homotopy on $\tilde M$ from $\tilde\gamma$ 
to a curve $\tilde\gamma'$ that descends to $\gamma'$. 
This fact follows from the ``covering homotopy theorem'' 
(see~\cite{singer-thorpe}, Chapter 3, Theorem 3).  Homotopic implies homologous, 
so $\tilde\gamma'$ is in $\tilde C$, therefore $\gamma'$ is in $C$. \hfill  $\square$

Since $\gamma_0 \in C$, the lemma shows that, as claimed, the full homotopy class $D\in S_0$ is contained in~$C$.  

We now argue that the constraint on all curves in $C$ does not impose
constraints that go beyond those of the homotopy minimal-area problem.   If a curve in $C$ is a sum of curves in which
at least one is homotopically nontrivial, this is acceptable because we 
aim  to constrain {\em all} homotopically nontrivial curves, even those
not in $D$.    We now show that none of the elements of $C$ are homotopically trivial curves, or sums of homotopically trivial curves.  
This follows from Lemma 1:  if homotopically trivial curves
were contained in $C$, all of them including the constant curve must
be included in $C$.  But a constant curve lifts to a constant curve on $\tilde M$,
and sums of constant curves lift to sums of constant curves on $\tilde M$. 
But neither a constant  curve nor sums of them are elements of $\tilde C$. 

This concludes our proof that the calibration $\tilde u$ on $M$ will properly constrain
the homotopy class $D$ without imposing unacceptable additional constraints.

\subsubsection{Final form of primal program}

We now give the form of the full convex program
for the minimal-area homotopy problem for $M$.
We assume that $S_0$ contains a set of homotopy classes $D_0^k$ 
with $k\in K_0$ (with $K_0$ a set),  with Jordan representatives 
$\gamma_k$ 
that require cutting curves $\gamma_{Rk}$ (with $\pm$ labels
for the resulting curves) and double covers to render them homologically nontrivial.  We also have 
calibrations $u^k$ anti-continuous across the cutting curves $\gamma_{Rk}$.
The program is
\begin{equation}
\begin{split}
\ \ &\text{Minimize }\ \, \int_M\sqrt{g^0}\,\Omega \, \quad 
\text{ over }\ \ 
\Omega\, (\hbox{function}) \, ,u^\alpha\, , u^k \  (\hbox{one-forms})  \ \  \\
&   \hbox{subject to:}  \hskip25pt |u^\alpha|_0^2-\Omega\le0\,,  \\
&  \hskip80pt
\qquad  du^\alpha=0\, ,  \  \\[-0.7ex]
& \hskip77pt  1-\int_{m_\alpha}  
\hskip-8pt u^\alpha=0\,, 
\ \ \forall \alpha \in J_0\,,\\
&   \phantom{\hbox{subject to:}}  \hskip25pt |u^k|_0^2-\Omega\le0\,,  \\
&  \hskip80pt
\qquad  du^k=0\, ,  \  \\[-0.7ex]
&  \hskip27pt
\qquad  u^k|_{\gamma_{Rk}^+}\, +\,\,   u^k|_{\gamma_{Rk}^-} = 0\,, \  \\[-0.7ex]
& \hskip78pt  1-\int_{\gamma_k}\hskip-8pt u^k=\, 0\,, 
\  \ \ \forall k \in K_0  \,. 
\end{split}
\end{equation}
An analogous expression could be written for the dual program
involving the scalars $\varphi$ and the discontinuities $\nu$.

\section*{Acknowledgments}

Barton Zwiebach would like to thank Larry Guth and Yevgeny 
Liokumovich for many instructive discussions on systolic
geometry and on the challenges of proving existence of solutions.   
The work of M.H is supported by the National Science Foundation through Career Award No. PHY-1053842, by the U.S.\ Department of Energy under grant 
DE-SC0009987, and by the Simons Foundation through a Simons Fellowship in Theoretical Physics. The work of B.Z.~is supported by the U.S.\ Department of Energy under grant Contract Number DE-SC0012567. M.H.\ would also like to thank MIT's Center for Theoretical Physics for hospitality and a stimulating research environment during his sabbatical year.

\appendix

\sectiono{Notation and some useful formulae}
\label{sec:notation}

Consider a two-dimensional surface $M$ with a metric 
$g_{\mu\nu}$. For vectors $v_1, v_2$ on the surface  the inner product $\langle \cdot \,, \, \cdot 
\rangle$ and the norm $| \, \cdot \, |$ are defined by
\begin{equation}
\langle v_1, v_2 \rangle \ \equiv \ g_{\mu\nu} v_1^\mu v_2^\nu,  \quad | v|^2 \equiv \langle v ,  v\rangle \,.
\end{equation}
For one-forms $u$, with components $u_\mu$  we take
\begin{equation}
\langle u_1, u_2 \rangle \ \equiv \ g^{\mu\nu} u_{1\mu} u_{2\nu},  \quad
| u|^2 \equiv \langle u,  u\rangle \,. 
\end{equation}
Given a one-form $u$ with components $u_\mu$ we define the 
associated vector $\hat u$  with components
\begin{equation}
\hat u^\mu \ \equiv \ g^{\mu\nu} u_\nu \,.
\end{equation}
A one form $u$ acting on a vector $v$ gives $u(v)$ with
\begin{equation}
u(v) \ = \ u_\mu v^\mu  \ = \ \langle \hat u , v \rangle \,. 
\end{equation}
It follows also that 
\begin{equation}
| u |^2 \ = \  | \hat u |^2 \ = \ | u (\hat u) | \,. 
\end{equation}

\noindent
The Hodge dual $* u$ of a one-form $u$ is a one form with components
\begin{equation}
\label{hodge-dual}
(* u)_\mu  \ \equiv  \,  u_\rho \, \varepsilon^{\rho \nu} \, g_{\nu\mu} \,, \quad 
\varepsilon^{\mu\nu} \equiv \  {\epsilon^{\mu\nu} \over \sqrt{g}} \,, \quad \epsilon^{12} = 1 \,. 
\end{equation}
Additional useful relations are
\be
\epsilon^{\mu\nu} = g g^{\alpha \mu}g^{\beta \nu} \epsilon_{\alpha \beta} \,, \ \ 
\epsilon_{12} = 1 \,.  
\ee
The Hodge dual operation depends only on the 
Weyl class of the metric.  This is clear because 
\begin{equation}
\label{hodge-dual-elaborate}
(* u)_\mu  \ \equiv  \,  u_\rho \, \epsilon^{\rho \nu} \,\,  \tfrac{g_{\nu\mu}}{\sqrt{g}} \,, \end{equation}
and the rightmost ratio is Weyl invariant.  
One can also check that   
\be
(* u)_\mu =  u_\rho \sqrt{g} \, g^{\rho\alpha} \epsilon_{\alpha\mu} \,, \qquad
* dx^\mu  = \sqrt{g} \, g^{\mu\alpha} \epsilon_{\alpha\rho} dx^\rho \,. 
\ee 
For a conformal metric one has
$(*u)_1= - u_2,  \,  (*u)_2 = u_1$.  
In general, one has 
\begin{equation}
\label{hodge-ip}
* ( * u) \ =\   - u \,, \quad \text{and} \quad  \langle * u_1 , * u_2 \rangle \ = \ \langle u_1, u_2 \rangle \,. 
\end{equation}
It follows that the one forms $u$ and $*u$ are orthogonal
\begin{equation}
 \langle u , * u \rangle \ = \ 0 \,, 
 \end{equation}
and so are the vectors associated to $u$ and $* u$.  We define the volume 
form $\omega$   
\begin{equation}
\omega \ \equiv \  \sqrt{g} \,  d^2 x \ \equiv \ \sqrt{g} \, dx^1 \wedge dx^2 \,.
\end{equation}  
We then have 
\begin{equation}
dx^\mu \wedge dx^\nu \ = \ \epsilon^{\mu\nu} dx^1 \wedge dx^2
\ = \ \epsilon^{\mu\nu} d^2 x  \ = \ 
\varepsilon^{\mu\nu} \, \sqrt{g} \, d^2 x  
\ = \ \varepsilon^{\mu\nu} \, \omega \,.  
\end{equation}
It is a simple calculation to verify that:
\begin{equation}
\label{form-wedge-form}
u_1 \wedge u_2 \ = \ \langle  * u_1 , u_2 \rangle \, \omega \,. 
\end{equation}
Given a closed one-form $u$, the vector $v$ associated with the dual form $* u$  is divergenceless:
\begin{equation}
\label{one-form-flow}
du \ = \ 0 \quad \text{then}  \quad   v \ \equiv \ \widehat{* u} \quad
\text{satisfies} \quad \nabla_\mu v^\mu \ = \ 0 \,,  \ \text{and} \quad
 | u | \ = \ | v |  \,. 
\end{equation} 
These identities establish the equivalence of calibrations and flows. 

\medskip
For  the  metric
$ds^2   = E \, dx_1^2  + G  \, dx_2^2$
the Gaussian curvature $K(x)$ is given by
\begin{equation}
\label{gaussian-simple}
K(x) \ = \ - {1\over 2 \sqrt{EG}}  \Bigl[  \partial_1 \Bigl( {\partial_1 G \over \sqrt{EG}} \, \Bigr) +  \partial_2 \Bigl( {\partial_2 E \over \sqrt{EG}}\, \Bigr)  \, \Bigr] \,. 
\end{equation}
The scalar curvature $R$ is twice the Gaussian curvature: $R = 2K$.


\begin{thebibliography}{99}

 
 
 
\bibitem{Witten:1985cc} 
  E.~Witten,
  ``Noncommutative Geometry and String Field Theory,''
  Nucl.\ Phys.\ B {\bf 268}, 253 (1986).
  doi:10.1016/0550-3213(86)90155-0

\bibitem{Giddings:1986wp} 
  S.~B.~Giddings, E.~J.~Martinec and E.~Witten,
  ``Modular Invariance in String Field Theory,''
  Phys.\ Lett.\ B {\bf 176}, 362 (1986).
  doi:10.1016/0370-2693(86)90179-6

\bibitem{Zwiebach:1992ie} 
  B.~Zwiebach,
  ``Closed string field theory: Quantum action and the B-V master equation,''
  Nucl.\ Phys.\ B {\bf 390}, 33 (1993)
  doi:10.1016/0550-3213(93)90388-6
  [hep-th/9206084].
 

\bibitem{strebel}
  K.~Strebel, {\em Quadratic differentials,}  Springer-Verlag Berlin Heidelberg 1984.
  


\bibitem{Saadi:1989tb} 
  M.~Saadi and B.~Zwiebach,
  ``Closed String Field Theory from Polyhedra,''
  Annals Phys.\  {\bf 192}, 213 (1989).
  doi:10.1016/0003-4916(89)90126-7
   
 
\bibitem{Kugo:1989aa} 
  T.~Kugo, H.~Kunitomo and K.~Suehiro,
  ``Nonpolynomial Closed String Field Theory,''
  Phys.\ Lett.\ B {\bf 226}, 48 (1989).
 
\bibitem{Zwiebach:1990ni} 
  B.~Zwiebach,
  ``Consistency of Closed String Polyhedra From Minimal Area,''
  Phys.\ Lett.\ B {\bf 241}, 343 (1990).
  doi:10.1016/0370-2693(90)91654-T
 
 
\bibitem{Zwiebach:1990nh} 
  B.~Zwiebach,
  ``How Covariant Closed String Theory Solves A minimal-area problem,''
  Commun.\ Math.\ Phys.\  {\bf 136}, 83 (1991).
  doi:10.1007/BF02096792

 
  
\bibitem{ahlfors} 
L.~V.~Ahlfors,  {\em  Conformal Invariants:  Topics in Geometric Function Theory},  AMS Chelsea Publishing (1973). 
 
 \bibitem{m_katz}
  M. G. Katz,  {\em Systolic Geometry and Topology,}   Mathematical Surveys and Monographs, Volume 137.  American Mathematical Society 2007.

\bibitem{mberger}
M. Berger, ``A Panoramic View of Riemannian Geometry," 
Springer-Verlag Berlin Heidelberg 2003.

\bibitem{guth}
L.~Guth, ``Metaphors in systolic geometry",  arXiv:1003.4247.
``Systolic inequalities and minimal hyper surfaces", arXiv:0903.5299.

 
 \bibitem{gromov} 
 M. Gromov, ``Filling Riemannian Manifolds,"  
 J.\ Differential Geom. {\bf 18} (1983) 1-147.
 
 \bibitem{bavard}
 C. Bavard,  ``In\'egalit\'es isosystoliques conformes,"
Comment. Math. Helv. {\bf 67} (1992) 146-166. 

\bibitem{calabi}
E. Calabi,  ``Extremal isosystolic metrics for compact surfaces."
pp. 167-204 in {\em Actes
de la Table Ronde de G\'eom\'etrie Diff\'erentielle} (Luminy, 1992), Soc. Math. France, Paris, 1996.  


\bibitem{katz-sabourau}
M.~G.~Katz and S. Sabourau, ``An optimal systolic inequality for 
CAT(0) metrics in genus two,"  Pacific Journal of Mathematics,
Vol. 227, No. 1, 2006, 95- 107.  arXiv:math/0501017.

\bibitem{bryant}
R.~L.~Bryant, ``On extremal with prescribed Lagrangian densities,"
in ``Manifolds and Geometry" (Pisa, 1993), Cambridge University Press, 1996.
arXiv:dg-ga/9406001. 


 \bibitem{boyd}
  S. Boyd and L. Vandenberghe, {\em Convex Optimization}, 
  Cambridge University Press (2004).   
  Available online at: https://web.stanford.edu/~boyd/cvxbook/
  
   \bibitem{headrick-hubeny-refs}
H. Federer, {\em Real flat chains, cochains and variational problems}, Indiana Univ. Math. J. {\bf 24} (1974/75) 351-407; \hfill\break
G. Strang, {\em Maximal flow through a domain}, Math. Programming 
{\bf 26} (1983), no. 2 123-143; \hfill\break
R. Nozawa, {\em Max-flow min-cut theorem in an anisotropic network}, Osaka J. Math. {\bf 27} (1990),
no. 4 805-842; \hfill\break
J. M. Sullivan, {\em A crystalline approximation theorem for hypersurfaces}. ProQuest LLC, Ann Arbor, MI, 1990. Thesis (Ph.D.) Princeton University.

\bibitem{Headrick:2017ucz} 
  M.~Headrick and V.~E.~Hubeny,
  ``Riemannian and Lorentzian flow-cut theorems,''
  arXiv:1710.09516 [hep-th].


\bibitem{Freedman:2016zud} 
  M.~Freedman and M.~Headrick,
  ``Bit threads and holographic entanglement,''
  Commun.\ Math.\ Phys.\  {\bf 352}, no. 1, 407 (2017)
  doi:10.1007/s00220-016-2796-3
  [arXiv:1604.00354 [hep-th]].

 \bibitem{headrick-zwiebach2}
  M.~Headrick and B.~Zwiebach,
  ``Minimal-area metrics on the Swiss cross and punctured torus,''
  arXiv:1806.00450 [hep-th].
  
  \bibitem{headrick-zwiebach3}
  M.~Headrick and B. Zwiebach, ``String diagrams from minimal area metrics of non-positive curvature",  in preparation. 
  
  
  \bibitem{k-s}  
M.~G.~Katz and S.~Sabourau, ``Systolically extremal nonpositively curved 
surfaces are flat with finitely many singularities,"  to appear in 
Journal of Topology and  Analysis.  DOI: 10.1142/S1793525320500144 [arXiv:1904.00730] 

 

\bibitem{Ranganathan:1991qd} 
  K.~Ranganathan,
  ``A Criterion for flatness in minimal area metrics that define string diagrams,''
  Commun.\ Math.\ Phys.\  {\bf 146}, 429 (1992).
  doi:10.1007/BF02097012


    
\bibitem{Wolf:1992bk} 
  M.~Wolf and B.~Zwiebach,
  ``The Plumbing of minimal area surfaces,''
 Journal of Geometry and Physics {\bf 15} (1994) 23-56.
  [hep-th/9202062].


  
   
\bibitem{DHoker:1987hzc} 
  E.~D'Hoker and S.~B.~Giddings,
  ``Unitary of the Closed Bosonic Polyakov String,''
  Nucl.\ Phys.\ B {\bf 291}, 90 (1987).
  doi:10.1016/0550-3213(87)90466-4
  
  
  
\bibitem{Zemba:1988rf} 
  G.~Zemba and B.~Zwiebach,
  J.\ Math.\ Phys.\  {\bf 30}, 2388 (1989).
  doi:10.1063/1.528569
  
  
\bibitem{Sonoda:1989wa} 
  H.~Sonoda and B.~Zwiebach,
  ``Closed String Field Theory Loops With Symmetric Factorizable Quadratic Differentials,''
  Nucl.\ Phys.\ B {\bf 331}, 592 (1990).
  doi:10.1016/0550-3213(90)90086-S



\bibitem{Moosavian:2017qsp} 
  S.~F.~Moosavian and R.~Pius,
  ``Hyperbolic Geometry and Closed Bosonic String Field Theory I: The String Vertices Via Hyperbolic Riemann Surfaces,''
  arXiv:1706.07366 [hep-th].
  
  
     
  \bibitem{berger} 
  M. Berger, ``A l'ombre de Loewner," Ann. Sci. Ecole Norm. Sup. Paris 
  {\bf 5} (1972) 241-260.
\bibitem{Zwiebach:1990ba} 
  B.~Zwiebach,
  Commun.\ Math.\ Phys.\  {\bf 141}, 577 (1991).
  doi:10.1007/BF02102817
 
  \bibitem{bott-tu}
  R. Bott and L. Tu, {\em Differential Forms in Algebraic Topology},
  Springer Verlag, New York (1982). 
 
 \bibitem{hicks}
 N. J. Hicks, {\em Notes on differential geometry,}
 Van Nostrand (1965).
 

\bibitem{singer-thorpe} I. M. Singer and J. A. Thorpe, {\em Lecture notes on elementary topology and geometry},  Undergraduate Texts in Mathematics,
Springer-Verlag New York - Heidelberg - Berlin (1976).

\bibitem{harvey-lawson} R. Harvey and B.H. Lawson, Jr., ``Calibrated geometries'', Acta Math., 148 (1982) 47--157.

\bibitem{MR0097026}
    M. Sion, 
     ``On general minimax theorems'',
   Pacific J. Math.,
    8
(1958),
     171--176.
	
 
 
  \end{thebibliography}
\end{document}